\documentclass[aps,prl,onecolumn,groupedaddress]{revtex4}
\usepackage[pdftex]{graphicx}

\begin{document}
\title{Single electron transport and possible quantum computing in 2D materials}

\author{K. L. Chiu}

\affiliation{CAS Key Laboratory of Quantum Information, University of Science and Technology of China, Hefei, Anhui 230026, China}


\date{\today}

\begin{abstract}
Two-dimensional (2D) materials for their versatile band structures and strictly 2D nature have attracted considerable attention over the past decade. Graphene is a robust material for spintronics owing to its weak spin-orbit and hyperfine interactions, while monolayer 2H-transition metal dichalcogenides (TMDs) possess a Zeeman effect-like band splitting in which the spin and valley degrees of freedom are nondegenerate. Monolayer 1T'-TMDs are 2D topological insulators and are expected to host Majorana zero modes when they are placed in contact with S-wave superconductors. Single electron transport as well as the superconductor proximity effect in these materials are viable for use in both conventional quantum computing and fault-torrent topological quantum computing. In this chapter, we review a selection of theoretical and experimental studies addressing the issues mentioned above. We will focus on: $(1)$ the confinement and manipulation of charges in nanostructures fabricated from graphene and 2H-TMDs (2) 2D materials-based Josephson junctions for possible superconducting qubits $(3)$ the quantum spin Hall states in 1T'-TMDs and their topological properties. We supply the entry-level knowledge for this field by first introducing the fundamental properties of 2D bulk materials followed by the theoretical background relevant to the physics of quantum dots and Josephson junctions. Subsequently, a historical review of experimental development in this field is presented, from graphene nanodevices fabricated on both SiO$_{2}$ and hBN substrate to more recent progress in transport studies of 2H-TMD nanostructures. In the second part of this chapter, we will discuss the properties of 2D material-based Josephson junctions and the observation of quantum spin Hall effect in 1T'-TMDs. We aim to outline the current challenges and suggest how future work will be geared towards developing quantum computing devices in 2D materials.
\end{abstract}

\maketitle
\section{1. Introduction of the family of 2D materials}

Since the 1960s, the density of components on silicon chips has doubled approximately every 18 months, following a trend known as Moore's law after Intel’s cofounder Gordon Moore, who predicted the phenomenon. Silicon-based transistor manufacturing has now reached the sub-10nm scale, heralding the limit of Moore's law and stimulating the development of alternative switching technologies and host materials for processing and storing bits of information. Quantum bits, or 'qubits', are at the heart of quantum computing, an entirely different paradigm in which information is encoded using the superposition states of individual quanta. Ideally, the charge and spin degrees of freedom of a single electron trapped in quantum dots (QDs) are nature candidate of qubits for use in quantum computing operations. To reach this goal, tremendous efforts have been dedicated to study the transport properties of QDs made from semiconductors, such as GaAs and silicon, and, more recently, graphene and other two-dimensional (2D) materials \cite{Schnez2009,Gutinger2008,Guttinger2009,Cho2012,Hong2014,Li2013}. One important parameter in quantum computing is the quality factor defined by $Q=T_{2}^{*}/t_G$, where $T_{2}^{*}$ is the decoherence time and $t_G$ is the gate operation time. A good quantum computing system requires a long decoherence time and a short gate operation time, thus many calculations can be performed before the information is lost. Although the high mobility (clean) and light effective mass (allowing for wider gate separation, hence relatively easy fabrication) in GaAs-based QDs has enabled the rapid development of spin qubits \cite{Hanson2007}, the strong nuclear field limits the spin decoherence time ($T_{2}^{*}$ $\approx$ 10 - 100 ns), making this material less ideal for upscaling. The QDs fabricated in isotopically purified Silicon ($^{28}$Si) do not suffer from the nuclear field and have shown a sufficiently long spin decoherence time ($T_{2}^{*}$ $\approx$ 0.12 ms) \cite{Zwanenburg2013,Veldhorst2015}, but the number of entanglement is hindered by the fabrication difficulty (shorter gate separation required) resulted from the heavy effective mass in silicon. While research on these materials is ongoing, 2D materials such as graphene and transition metal dichalcogenides (TMDs) have attracted considerable attention over the past few years because of their novel electronic properties \cite{Geim2013,Chiu2017a}. Graphene is expected to be a robust material for spintronics owing to its weak spin-orbit and hyperfine interactions. Over the last decade, attempts to confine and manipulate single charges in graphene quantum dots (GQDs) have been widely studied and reported, as noted in several review articles \cite{Stampfer2011,Molitor2011,Guttinger2012,Neumann2013,Chiu2017}. However, early studies of GQDs on SiO$_{2}$ have indicated an absence of spin-related phenomena, such as spin blockade and the Kondo effect. In order to reduce the substrate disorder, which is one of the major sources of fast spin relaxation, recent efforts have been focused on GQDs on atomically flat substrates [e.g., hexagonal boron nitride (hBN)]. Nevertheless, the edge disorder may still play a role hence no significant differences compared with studies on SiO$_2$ have been reported either. Other 2D materials, such as 2H-TMDs, exhibit direct band gap in monolayer form and are promising for switch applications due to the high current on/off rates in their transistors. In addition, the absence of inversion symmetry and the existence of strong spin-orbit coupling in monolayer 2H-TMDs allow the charge carriers to be simultaneously valley- and spin-polarized, providing more degrees of freedom that can be controlled as qubits. On the other hand, 1T' phase TMDs possess a completely different band structure compared to their 2H phase family. They are semimetal in bulk but become 2D topological insulators (TI) in the monolayer form. These insulators hold promise for hosting Majorana zero modes, which are topologically protected and form the core elements for performing fault-tolerant quantum computing at the hardware level. In this chapter, we aim to provide an overview of experimental studies that are relevant to the development of various qubits in 2D materials. We supply the entry-level knowledge for this field by first introducing the fundamental properties of various 2D materials and nanostructures followed by a selection of experimental studies. We discuss the transport properties of graphene nanodevices fabricated on both SiO$_{2}$ and hBN substrates at low temperatures and under high magnetic fields. Our primary focus is the single-electron tunneling regime in transport. In the second part, our focus will be directed to 2H-TMD nanostructures. We review recent developments in the fabrication and understanding of the electronic properties of these 2D nanostructures, including MoS$_{2}$ nanoribbons, WSe$_2$ single quantum dots and MoS$_{2}$ double quantum dots. In the third part, we extend our discussion to 2D material-based Josephson junctions and their potential applications in quantum computing. Finally, we review the quantum spin Hall edge states observed in monolayer 1T'-TMDs, which combined with superconductor could be useful for probing Majorana zero modes. In the summary, we outline how future work should pursue the development of various qubits in 2D materials.

\subsection{1.1 Graphene and hBN}

\begin{figure}
\begin{center}
		\includegraphics[scale=0.72]{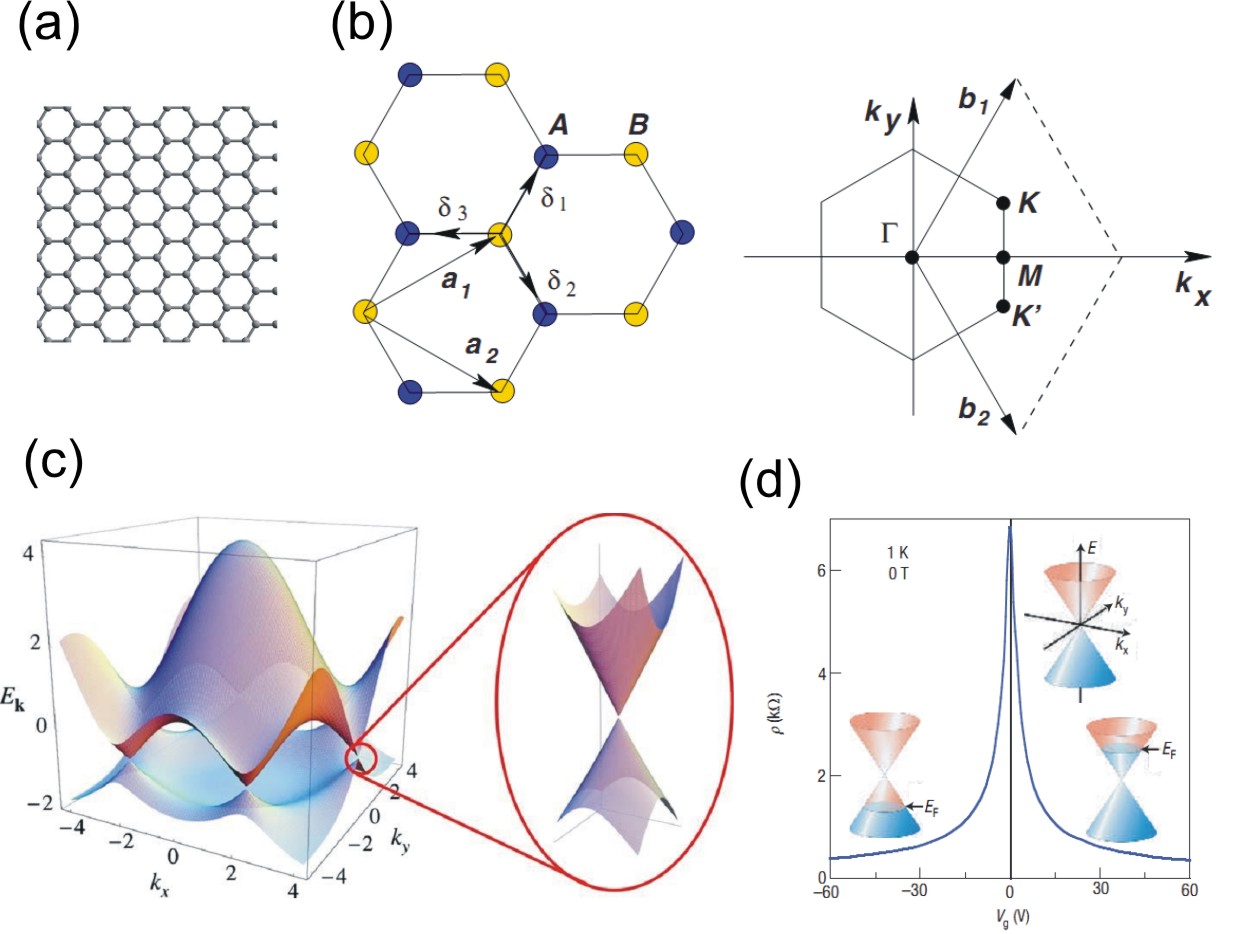}
\caption{(a) Graphene is a honeycomb lattice of carbon atoms. (b) Honeycomb lattice and its Brillouin zone. Left: lattice structure of graphene, made out of two interpenetrating triangular lattices ($\textbf{a}_{1}$ and $\textbf{a}_{2}$ are the lattice unit vectors, and $\vec{\delta_{i}}$, i =1, 2, 3 are the nearest-neighbor vectors). Right: corresponding Brillouin zone. The Dirac cones are located at the K and K' points. (c) The band structure of graphene calculated using a tight-binding model. The zoom in shows the conical dispersion relation around the Dirac point. (d) The ambi-polar electric field effect of graphene. V$_{g}$ is the back gate voltage and $\rho$ is resistivity. By varying V$_{g}$ one can shift the Fermi energy level and therefore determine the type of carriers (either electrons or holes) in graphene. (b, c) adapted with permission from ref. \cite{CastroNeto2009}. Copyright 2009 American Physical Society. (d) adapted with permission from ref. \cite{Geim2007}. Copyright 2007 Nature Publishing Group.}
\label{Fig1}
\end{center}	
\end{figure}

Graphene is a single layer of carbon atoms packed tightly in a honeycomb lattice as shown in Fig. \ref{Fig1}(a). An early study on few-layer graphene can be tracked back to 1948 by G. Ruess and F. Vogt, in which they occasionally observed extremely thin graphitic flakes in transmission electron microscope (TEM) images. However, no one isolated single layer graphene until 2004 when the physicists at the University of Manchester first isolated and spotted graphene on a chosen SiO$_{2}$ substrate \cite{Novoselov2004}. The first line of enquiry stems from graphene's unique gapless bandstructure. The unit cell of graphene consists of two carbon atoms, labeled as A and B sub-lattices, and can be described by the two lattice vectors $\textbf{a}_{1}$ and $\textbf{a}_{2}$, as shown in Fig. \ref{Fig1}(b) (left panel). They include an angle of 60$^{\circ}$ and have a length of $\left|\textbf{a}_{1}\right|$ = $\left|\textbf{a}_{2}\right|$ = $\sqrt{3}$a$_{0}$ $-$ 2.461 $\dot{A}$, where a$_{0}$ is the carbon-carbon bond length (a$_{0}$ = 1.42 $\dot{A}$). The lattice vectors can be determined as $\textbf{a}_{1}$ = $\frac{a_{0}}{2}$ ($3$, $\sqrt{3}$) and $\textbf{a}_{2}$ = $\frac{a_{0}}{2}$ ($3$, -$\sqrt{3}$) and the reciprocal lattice is described by $\textbf{b}_{1}$ = $\frac{2\pi}{3a_{0}}$ ($1$, $\sqrt{3}$) and $\textbf{b}_{2}$ = $\frac{2\pi}{3a_{0}}$ ($1$, -$\sqrt{3}$), as shown in the right panel of Fig. \ref{Fig1}(b). The lattice has high-symmetry points $\Gamma$, $\textit{K}$ and $\textit{M}$, where $\textit{K}$ = $(\frac{2\pi}{3a_{0}}, \frac{2\pi}{3\sqrt{3}a_{0}})$ and $K'$ = $(\frac{2\pi}{3a_{0}}, -\frac{2\pi}{3\sqrt{3}a_{0}})$ are two points at the corners of the hexagonal Brillouin zone \cite{CastroNeto2009}. Around the $\textit{K}$ point, a tight-binding calculation for the bandstructure of this lattice yields a 2D Dirac-like Hamiltonian $\hat{H}_K$ for massless fermions (and around the $K'$ point the Hamiltonian is simply $\hat{H}_{K'}$ $=$ $\hat{H}^{T}_K$): 

\begin{eqnarray}
&& \hat{H}_K \psi (\textbf{r}) = \hbar v_{F}\left( \begin{array}{cc} \ 0 & k_x - ik_y \\ \ k_x + ik_y & 0  \\\end{array} \right)\ \psi (\textbf{r}) = \nonumber -i \hbar v_{F} \left( \begin{array}{cc} \ 0 & \frac{\partial}{\partial{x}} - i\frac{\partial}{\partial{y}} \\ \ \frac{\partial}{\partial{x}} + i\frac{\partial}{\partial{y}} & 0  \\\end{array} \right)\ \psi (\textbf{r}) \\
&&=  -i \hbar v_{F} \vec{\sigma} \nabla \psi (\textbf{r}) = v_{F} \vec{\sigma} \cdot \vec{p} \psi (\textbf{r}) = \textit{E} \psi (\textbf{r})
\label{eqn 1}
\end{eqnarray} where $v_{F}$ is the Fermi velocity, $\vec{\sigma}$ $=$ ($\sigma_{x}$, $\sigma_{y}$) is the 2-D Pauli matrix and $\psi$($\textbf{r}$) is the two-component electron wavefunction. This Hamiltonian gives rise to the most important aspect of graphene's energy dispersion, $E=\hbar$$v_{F}k$, which is a linear energy-momentum relationship at the edge of the Brillouin zone as shown in Fig. \ref{Fig1}(c). The two-component vector part of the wavefunction, which corresponds to the A or B sub-lattices, is the so-called pseudospin degree of freedom, since it resembles the two-component real spin vector. The Pauli matrices $\sigma_x$ and $\sigma_y$ combined with the direction of the momentum leads to the definition of a chirality in graphene ($h = \vec{\sigma} \cdot \vec{p}/2\left|\vec{p}\right|$), meaning the wavefunction component of A or B sub-lattice is polarized with regard to the direction of motion of electrons \cite{CastroNeto2009}. The existence of the $K$ and $K'$ points (as a result of graphene's hexagonal structure), where the Dirac cones for electrons and holes touch each other in momentum space [Fig. \ref{Fig1}(b) and Fig. \ref{Fig1}(c)], is sometimes referred as isospin, and gives rise to a valley degeneracy g$_{\nu}$ $=$ 2 for graphene. The linear dispersion along with the presence of potential disorder leads to a maximum resistivity in the limit of vanishing carrier density (or the so-called Dirac point), as shown in Fig. \ref{Fig1}(d). To change the Fermi level, and hence the charge carrier density, voltage needs to be applied to a nearby gate capacitively coupled to the graphene, which in the case of Fig. \ref{Fig1}(d) is a backgate - a doped Si substrate that is isolated from the graphene by a SiO$_{2}$ insulator layer.

There are rich physics originated from the Dirac nature of the fermions in graphene, such as its electronic, optical and mechanical properties \cite{CastroNeto2009,Basov2014,Amorim2016}. Here, we introduce an important phenomenon in graphene transport, which is relevant to the subjects to be discussed in this chapter: the extreme quantum Hall effect (QHE) that can be observed even at room temperature \cite{Geim2007}. Because the low-energy fermions in graphene are massless, it is obvious that for graphene we cannot apply the results valid for standard semiconductor two-dimensional electron gas (2DEG) systems. Charge carriers in a standard 2DEG have an effective mass, which is related with the parabolic dispersion relation of conduction (valence) band $via$ $E = E_{c} + \frac{\hbar^2k^2}{2m_e^\ast}$ ($E = E_{v} - \frac{\hbar^2k^2}{2m_h^\ast}$), where $E_{c(v)}$ is the conduction (valence) band minimum (maximum) and $m_{e(h)}^\ast$ is the effective mass for the electrons (holes). The band dispersion leads to a constant density of state (DOS) of $\frac{m_{e(h)}^\ast}{\pi\hbar^2}$ for the conduction (valence) band region. In a perpendicular magnetic field, the DOS of electrons in a 2DEG system is quantized at discrete energies given by:

\begin{eqnarray}
E_n &=& \pm\hbar\omega_c(n+1/2), \label{eqn 2}
\end{eqnarray} which is the so-called Landau Level (LL) energy, with $n$ the integer number and $\omega_c = eB/m_e^\ast$ the cyclotron frequency, as sketched in Fig. \ref{Fig2}(a). The resulting Hall plateaus of a 2DEG lie at the conductivity values as follows: 

\begin{eqnarray}
\sigma_{xy} = \nu e^2/h, \label{eqn 3}
\end{eqnarray} where $\nu$ is the filling factor and takes only integer values, as illustrated in Fig. \ref{Fig2}(b). For the QHE in graphene, the 2D massless Dirac equation must be solved in the presence of a perpendicular magnetic field $B$ to find the Landau Level energy $E_{n}$ \cite{Peres2006a,CastroNeto2009}. Thus, the Hamiltonian for graphene now reads:
\begin{equation}
v_{F} \vec{\sigma}\cdot(\vec{p}+\frac{e\vec{A}}{c}) \psi (\textbf{r}) = E \psi (\textbf{r}),
\label{eqn 50}
\end{equation} where momentum $\vec{p}$ in Eq. (\ref{eqn 1}) has been replaced by $\vec{p}+\frac{e\vec{A}}{c}$ and  $\vec{A}$ is the in-plane vector potential generating the perpendicular magnetic field $\vec{B}=B\hat{z}$. The solution of this equation gives rise to the eigenenergy of each Landau level for monolayer graphene: 

\begin{equation}
\ E_{n} = sgn(n) v_{F} \sqrt{2e\hbar B\left|n\right|}
\label{eqn 4}
\end{equation} with the Landau level index $n$ = 0, $\pm$1, $\pm$2, etc, and $sgn(n)$ stands for the sign of $n$. Unlike 2DEG, there will be a Landau level at zero energy ($n$ = 0) separating the positive and negative LLs, and their energies are proportional to $\sqrt{B}$ (instead of $B$ in 2DEG), as sketched in Fig. \ref{Fig2}(c). In addition, the resulting Hall conductivity for monolayer graphene is given by:

\begin{figure}
\begin{center}
		\includegraphics[scale=0.72]{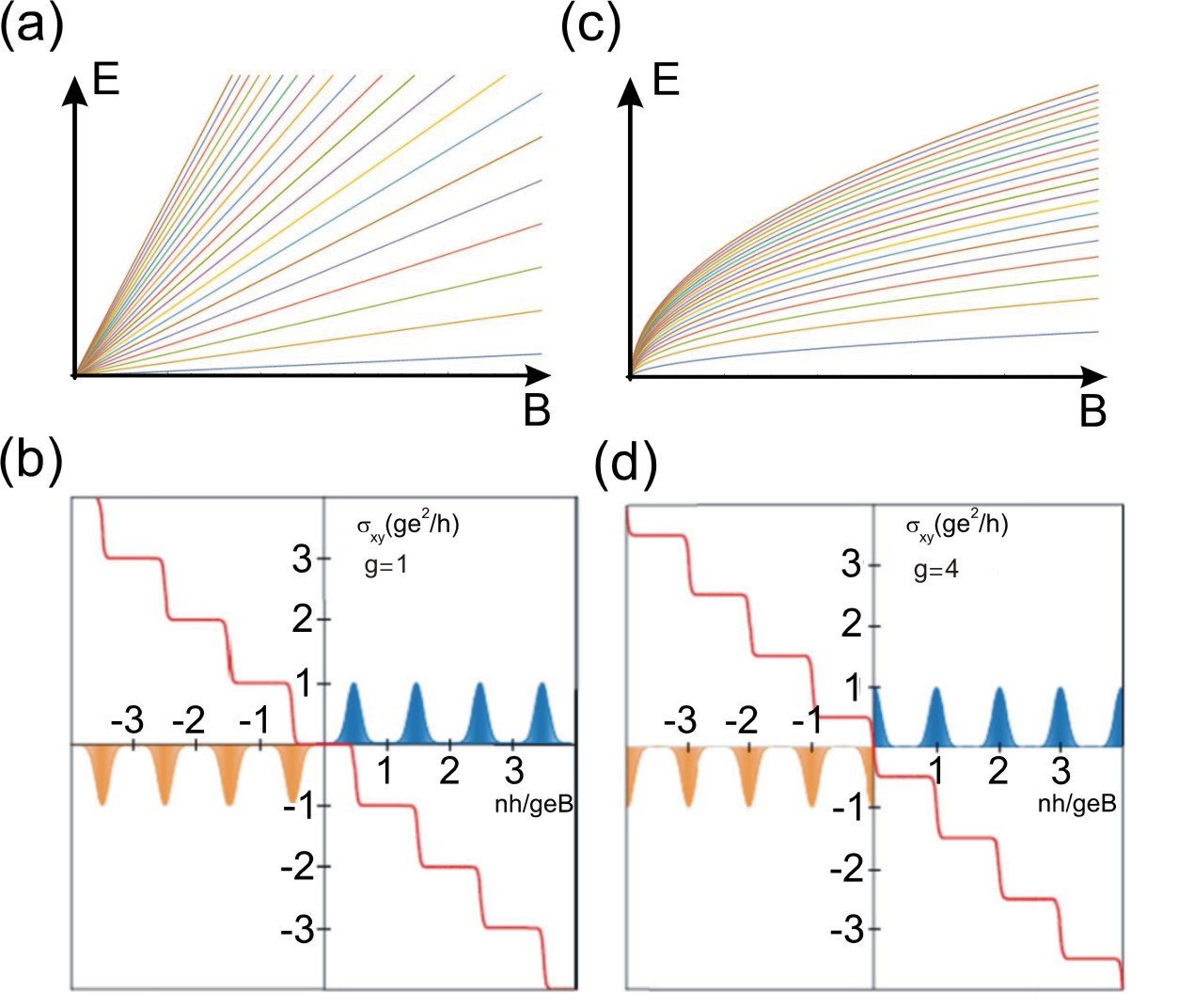}
\caption{(a) Schematic illustration of Landau levels for a standard 2DEG system. (b) Illustration of the integer Quantum Hall Effect (QHE) for a 2DEG system. (c) Same as (a) but for graphene. (d) Same as (b) but for graphene. The QHE plateau $\sigma_{xy}$ lie at half integers of 4e$^{2}$/h. (b, d) adapted with permission from ref. \cite{Novoselov2006}. Copyright 2006 Nature Publishing Group.}
\label{Fig2}
\end{center}	
\end{figure}

\begin{equation}
\sigma_{xy} = 4e^{2}/h (n+\frac{1}{2}) = \nu e^2/h,
\label{eqn 5}
\end{equation}
where $n$ is an integer and the factor 4 is due to the double valley and double spin degeneracy \cite{Geim2007,DasSarma2011,CastroNeto2009}. Note the filling factor now reads: $\nu\equiv 4N/N_{\phi}$ = 4(n+1/2) = $\pm$2, $\pm$6, $\pm$10 etc, where $N$ is the total electron occupancy and $N_{\phi}$ is the magnetic flux divided by the flux quantum $h/e$. This result differs from the conventional QHE found in GaAs heterostructure 2DEGs [Fig. \ref{Fig2}(b)] and is a hallmark of Dirac fermion in monolayer graphene. The quantization of $\sigma_{xy}$ has been observed experimentally \cite{Geim2007}, a sketch of the data is illustrated in Fig. \ref{Fig2}(d). The lowest LL in the conduction band and the highest LL in the valence band merge and contribute equally to the joint level at $E = 0$, resulting in the half-odd-integer QHE. The factor 1/2 in Eq. (\ref{eqn 5}) is due to the additional Berry phase $\pi$ that the electrons, due to their chiral nature, acquire when completing a cyclotron trajectory \cite{Lukanchuk2004,Xue2013}. The observation of the QHE at room temperature is also a consequence of the Dirac nature of the fermions in graphene. Because in graphene $E_{n}$ is proportion to $v_{F}$$\sqrt{B\left|n\right|}$ (where $v_{F}$=10$^{6}$ m/s is the Fermi velocity), at low energy the energy spacing $\Delta E_{n}$ $\equiv$ $E_{n+1}$ $-$ $E_{n}$ between Landau Levels can be rather large. For example, for fields of the order of $B$ = 10 T, the cyclotron energy in a GaAs 2DEG system is of the order of 10 K, however, the same field in graphene gives rise to the cyclotron energy of the order of 1000 K, that is, two orders of magnitude larger.

\begin{figure}
\includegraphics[scale=0.7]{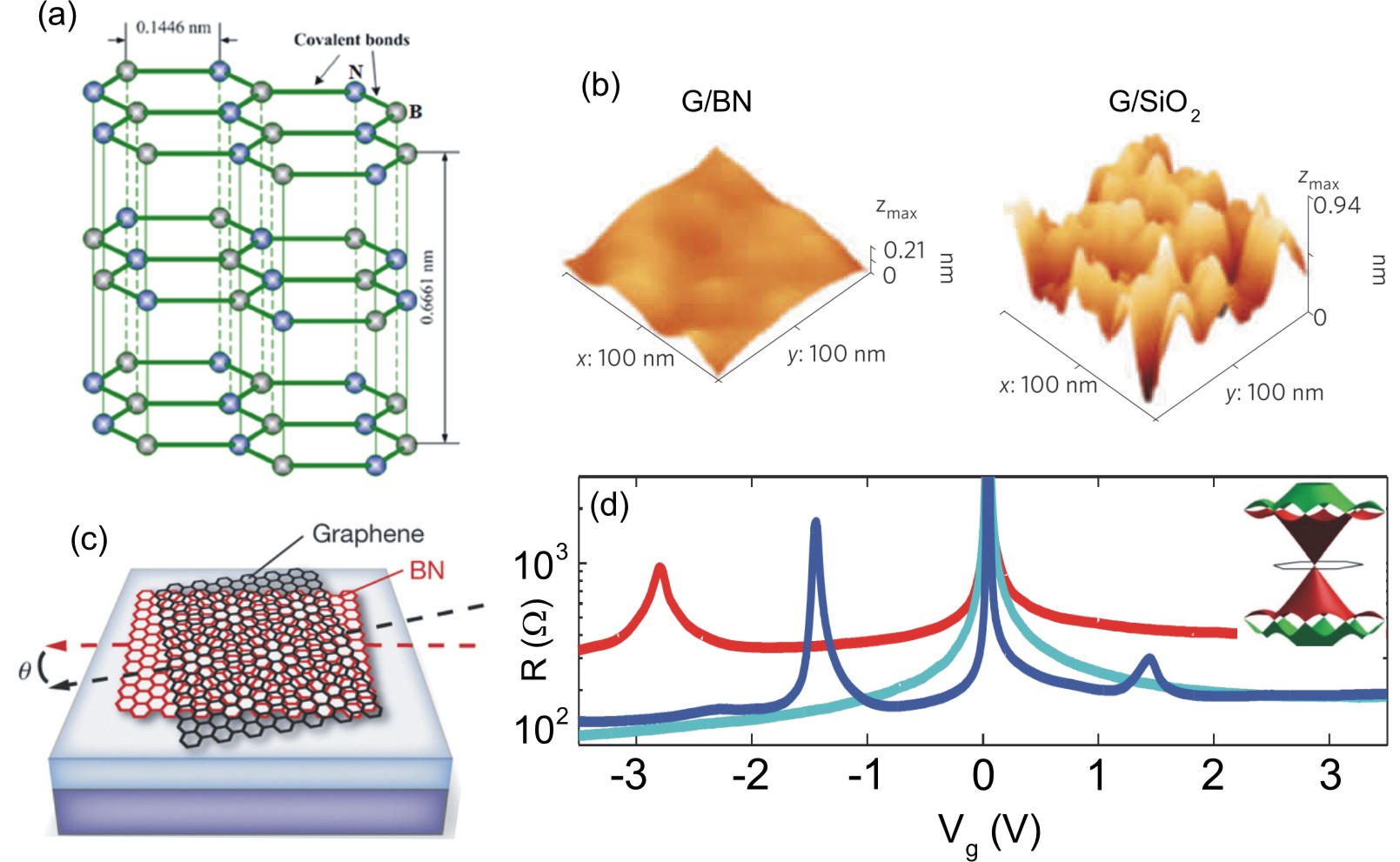}
\caption{(a) Structure of the multilayered Hexagonal Boron Nitride (hBN). (b) STM topographic images of monolayer graphene on hBN (left) and SiO$_{2}$(right) showing the underlying surface corrugations. (c) Schematic of the moir\'e pattern formed from a graphene/hBN stack. The moir\'e wavelength varies with the twist angle $\theta$. (d) Resistance as a function of gate voltage measured from three graphene/hBN stacks (with different moir\'e wavelengths), showing two extra Dirac peaks as a result of the superlattice minibands. Inset shows the band diagram of graphene on hBN. (a) adapted with permission from ref. \cite{Xu2013}. Copyright 2013 American Chemical Society. (b) adapted with permission from ref. \cite{Xue2011}. Copyright 2011 Nature Publishing Group. (c) adapted with permission from ref. \cite{Dean2013}. Copyright 2013 Nature Publishing Group. (d) adapted with permission from ref. \cite{Hunt2013}. Copyright 2013 American Association for the Advancement of Science.}  
\label{Fig3}
\end{figure}

Having briefly introduced graphene, we extend our discussion to Hexagonal Boron Nitride (hBN), which is isostructural to graphene but has boron and nitrogen atoms on the A and B sub-lattices, as shown in Fig. \ref{Fig3}(a). Due to the different onsite energy of A and B sub-lattices, the tight-binding calculation shows that hBN is an insulator with a large band gap of around 6 eV \cite{Golberg2010,Geim2013,Xu2013,Wang2014,Ferrari2015}. Traditionally, hBN has been used as a lubricant or a charge leakage barrier layer in electronic equipments \cite{Xu2013}. More importantly, recent studies have shown the use of hBN thin films as a dielectric layer for gating or as a flat substrate for graphene transistors can improve the electronic transport quality of devices by a factor of ten (or more), compared with the case of graphene on SiO$_{2}$ substrates \cite{R.2010,Dean2013,Ponomarenko2013,Hunt2013}. The high quality of graphene/hBN heterostructures originates from the atomic-level smooth surface of hBN that can suppress surface ripples in graphene. STM topographic images [Fig. \ref{Fig3}(b)] show that the surface roughness of graphene on hBN is greatly decreased compared with that of graphene on SiO$_{2}$ substrates. While graphene on SiO$_{2}$ exhibits charge puddles with diameters of 10$\sim$30 nm, the sizes of charge puddles in graphene on hBN are roughly one order of magnitude larger. The enhanced high mobility of graphene on hBN (up to 10$^{6}$ cm$^{2}$V$^{-1}$s$^{-1}$ reported \cite{Amet2015}) has enabled the studies of many-body physics and phase coherent transport that cannot be accessed in low-mobility samples, such as the observation of the fractional QHE and supercurrent in the quantum Hall regime \cite{Kou2014,Amet2016}.

Due to the similarity in lattice structure, when graphene is stacked on hBN with a small twist angle ($\leq$ 5$^{\circ}$), it can form a superlattice [called the moir\'e pattern, as shown in Fig. \ref{Fig3}(c)] with a wavelength ranging from a few to 14 nm \cite{Yankowitz2012,Hunt2013,Ponomarenko2013,Dean2013}. The superlattice with a relatively large wavelength compared to the bond length of carbon atom introduces additional minibands in graphene's band structure \cite{Hunt2013}. Fig. \ref{Fig3}(d) shows typical transfer curves for three graphene/hBN stacks with different moir\'e wavelengths, in which two extra Dirac peaks, situated symmetrically about the charge neutrality point ($V_{g}$ $=$ 0 V), are observed in all devices. These newly appeared Dirac peaks result from the superlattice minibands, which are away from the original Dirac point of graphene, as shown in the inset of Fig. \ref{Fig3}(d). Such hybrid band structures lend novel transport features to graphene; for example, the observation of the Hofstadter Butterfly spectrum in high magnetic fields \cite{Hunt2013,Ponomarenko2013,Dean2013}.

\subsection{1.2 2H-Transition Metal Dichalcogenides}

2D materials with a hexagonal lattice structure (such as graphene or TMDs with 2H phase) possess valley of energy-momentum dispersion at the corner of the hexagonal Brillouin zone. In graphene, this dispersion at the $K$ and $-K$ points gives rise to a valley degeneracy (note that in this and the subsequent sections we use the notation $-K$ to replace $K'$ for simplicity). The situation is different in 2H-TMDs because of the absence of inversion symmetry, which allows the valley degree of freedom to be accessed independently (valleytronics), although it is still degenerate in energy. 2H-TMDs are semiconductors and have hexagonal lattices of MX$_2$, where M is a transition metal element from group VI (Mo or W) and X is a chalcogen atom (S, Se or Te), as illustrated in Fig. \ref{Fig4}(a). Unlike graphene and hBN, the lattice structure of such a TMD consists of hexagons of M and X, with the M atom being coordinated by the six neighboring X atoms in a trigonal prismatic geometry, as shown in Fig. \ref{Fig4}(b). A key aspect of semiconducting TMDs is the effect exerted by the number of layers on the electronic band structure. Fig. \ref{Fig4}(c) shows the calculated band structure of a 2H-TMD (MoS$_2$), which exhibits a crossover from an indirect gap in the bulk form to a direct gap in the monolayer form as a result of a decreasing interlayer interaction. The photoluminescence (PL) from monolayer MoS$_2$ has shown the quantum yield to be two orders of magnitude larger than that from the multilayer material, providing evidence of such a crossover in the band gap \cite{Splendiani2010,Mak2010}. In the monolayer limit, the conduction and valence band edges are at the $\pm K$ points and are predominantly formed by the partially filled $d$-orbitals of the M atoms and have the following forms:

\begin{figure}
\begin{center}
		\includegraphics[scale=0.42]{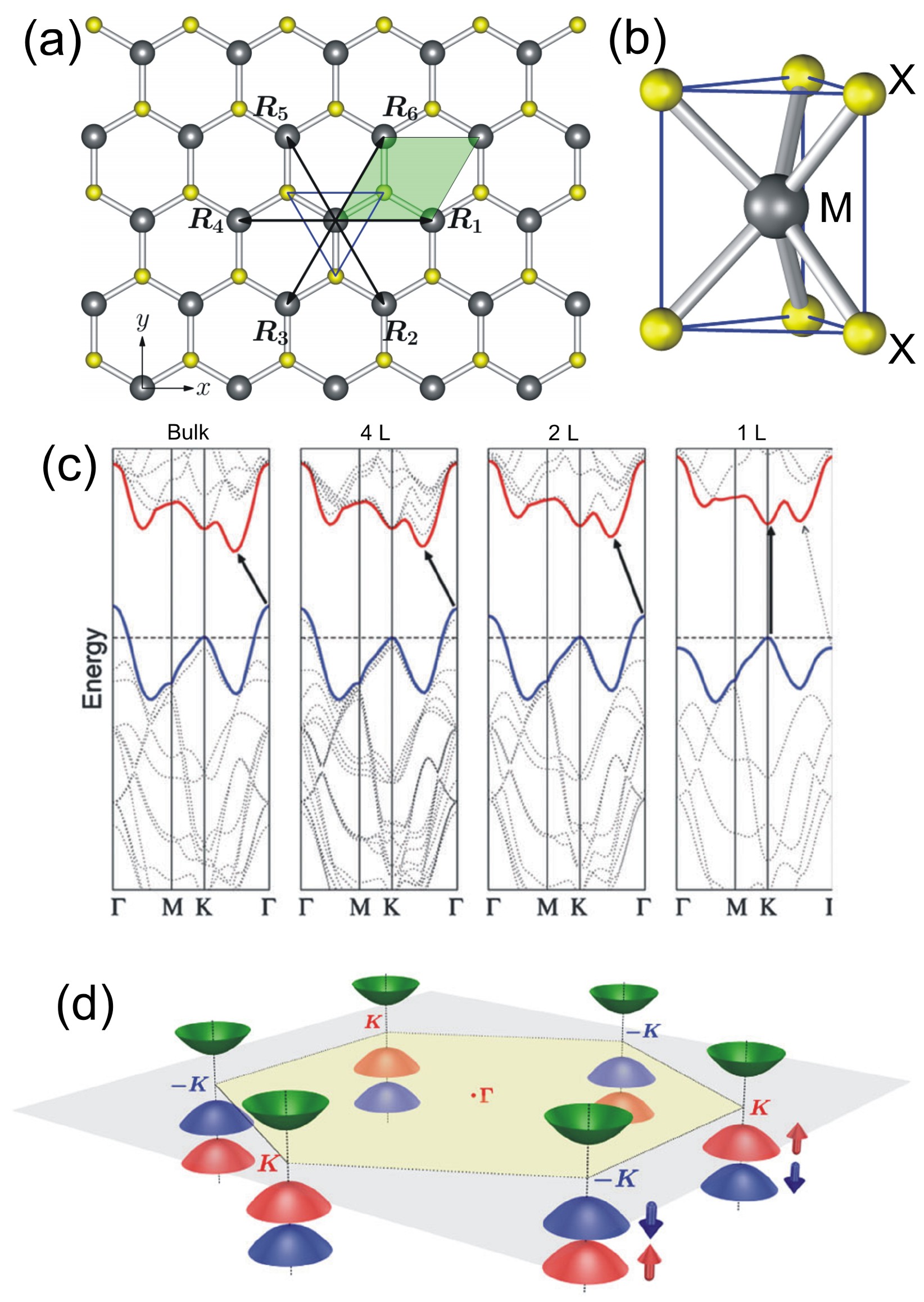}
\caption{(a) Top view of monolayer 2H-MX$_2$. Black balls represent M atoms, and yellow balls represent X atoms. The shadowed diamond region shows the 2D unit cell with lattice constant $a$. $\textbf{R}_1$ - $\textbf{R}_6$ denote the M-M nearest neighbors. (b) Schematic illustration for the structure of trigonal prismatic coordination, corresponding to a side view of the blue triangle in (a). (c) Energy dispersion in bulk, quadrilayer (4L), bilayer (2L) and monolayer (1L) 2H-MoS$_2$, from left to right, showing the transition from an indirect band gap to a direct band gap. (d) Schematic illustration of the band structure at the band edges located at the edges of the Brillouin zone. (a, b) adapted with permission from ref. \cite{Liu2013}. Copyright 2013 American Physical Society. (c) adapted with permission from ref. \cite{Duan2015}. Copyright 2015 Royal Society of Chemistry. (d) adapted with permission from ref. \cite{Xiao2012}. Copyright 2012 American Physical Society.}
\label{Fig4}
\end{center}	
\end{figure}

\begin{eqnarray}
\left|\phi_{c}\right\rangle &=& \left|d_{z^2}\right\rangle \label{eqn 6.1} \\
\left|\phi_{v}^{\tau}\right\rangle &=& \frac{1}{\sqrt{2}}(\left|d_{x^2-y^2} \right\rangle + i\tau\left|d_{xy} \right\rangle), \label{eqn 6.2}
\end{eqnarray}
where $d_{z^2}$, $d_{x^2-y^2}$ and $d_{xy}$ are the $d$-orbitals of the M atom, the subscript $c$($v$) indicates the conduction (valence) band, and $\tau=\pm1$ is the valley index. At the valley points ($\pm K$), a two-band $k\cdot p$ Hamiltonian that takes the form of the massive Dirac fermion model can be used to describe the dispersion at the conduction and valence band edges \cite{Xiao2012}:

\begin{equation}
H = at(\tau k_x \sigma_x + k_y \sigma_y) + \frac{\Delta}{2}\sigma_z - \lambda\tau \frac{\sigma_z -1}{2} \hat{S}_z, 
\label{eqn 7}
\end{equation}
where $\sigma$ denotes the Pauli matrices for the two basis functions given in Eq. (\ref{eqn 6.1}) and (\ref{eqn 6.2}), $a$ is the lattice constant, $t$ is the effective nearest neighbour hopping integral, and $\Delta$ is the band gap. The last term in Eq. (\ref{eqn 7}) represents the spin-orbit coupling (SOC), where 2$\lambda$ is the spin splitting at the top of the valence band and $\hat{S}_z$ is the Pauli matrix for spin. The spin splitting is due to the strong spin-orbit interaction arising from the $d$-orbitals of the heavy metal atoms. The conduction band-edge state consists of $d_{z^2}$ orbitals and remains almost spin-degenerate at the $\pm K$ points, whereas the valence-band-edge state shows a pronounced split. A schematic illustration of the band dispersion at the edges of the hexagonal Brillouin zone is shown in Fig. \ref{Fig4}(d). Note that the spin splitting at the different valleys is opposite because the $K$ and $-K$ valleys are related to one another by time-reversal symmetry.

\begin{figure}
\begin{center}
		\includegraphics[scale=0.5]{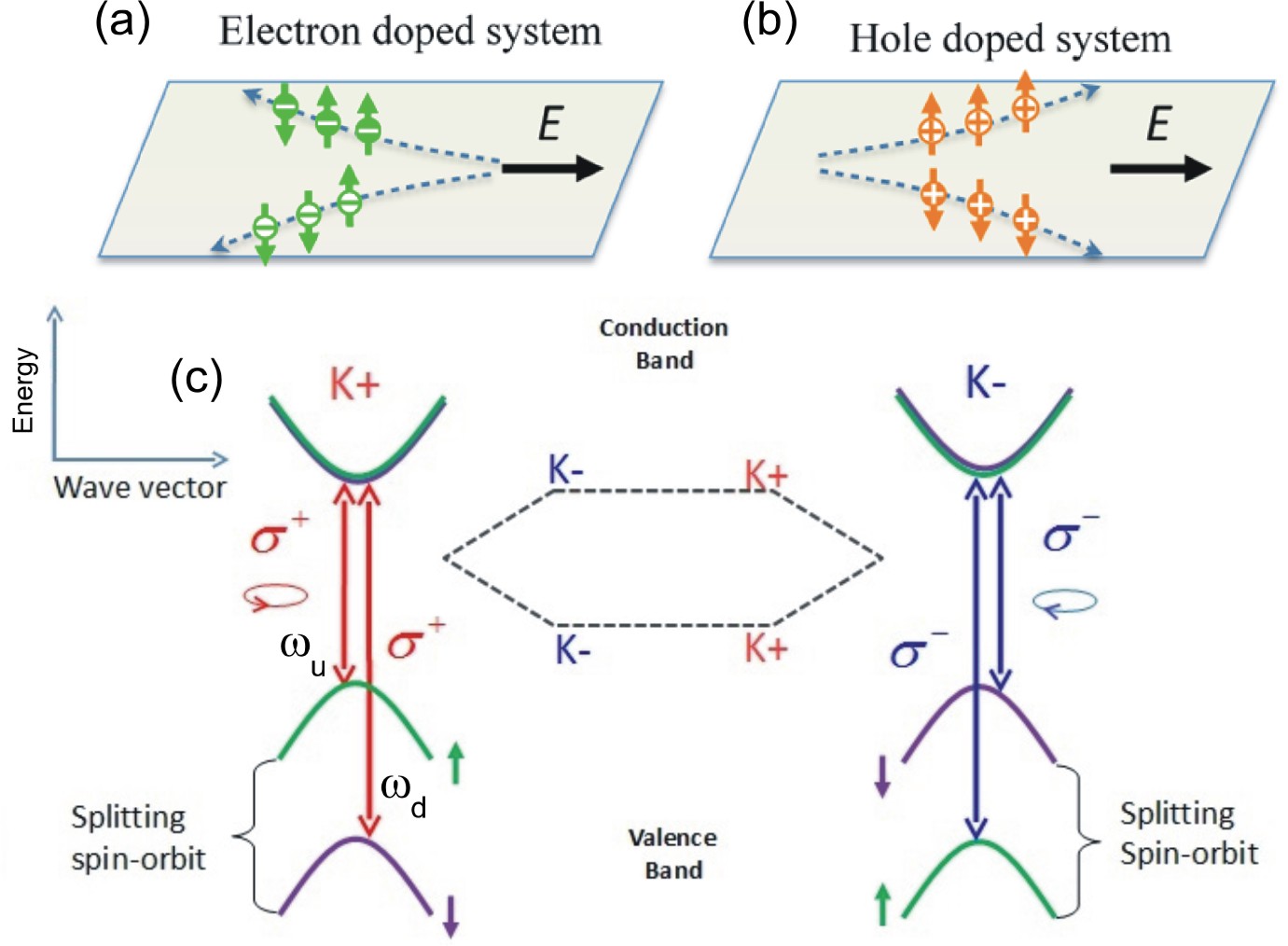}
\caption{Schematic illustration of the valley Hall effect in (a) the electron-doped regime and (b) the hole-doped regime. The electrons and holes in the $+K$ valley are denoted by white "$-$" and "$+$" symbols in dark circles and their counterparts in the $-K$ valley are denoted by an inverted color scheme. (c) Illustrations of the valley- and spin-dependent optical transition selection rules. (a, b) adapted with permission from ref. \cite{Xiao2012}. Copyright 2012 American Physical Society.}
\label{Fig5}
\end{center}	
\end{figure}

Because of the large valley separation in momentum space, the valley index is expected to be robust against scattering by smooth deformations and long-wavelength phonons. To manipulate such a valley degree of freedom for valleytronic applications, measurable physical quantities that distinguish the $\pm K$ valleys are required. The Berry curvature ($\bf{\Omega}$) and the orbital magnetic moment (\textbf{m}) are two physical quantities for $\pm K$ valleys to have opposite values. The Berry curvature is defined as a gauge field tensor derived from the Berry vector potential $\textbf{A}_n$($\textbf{R}$) through the relation $\bf{\Omega}$$_n$($\textbf{R}$)=$\nabla_\textbf{R}$ $\times$ $\textbf{A}_n$($\textbf{R}$), where \textit{n} is the energy band index (in the case of 2H-TMDs and at the $\pm K$ points, $n$ is either the conduction or valence band) and $\textbf{R}$ is the parameter to be varied in a physical system (in the case below $\textbf{R}$ is the wavevector \textbf{k}) \cite{Xiao2010}. The Berry curvature can be written as a summation over the eigenstates as follows \cite{Xu2014}: 

\begin{eqnarray}
\bf{\Omega}_n(k) &=& i\frac{\hbar^2}{m^2}\sum_{i\neq n}\frac{\textbf{P}_{n,i}(\textbf{k}) \times \textbf{P}_{i,n}(\textbf{k})}{\left[E^{0}_{n}(\textbf{k}) - E^{0}_{i}(\textbf{k})\right]^2}
\label{eqn 8}
\end{eqnarray}
Here, $\textbf{P}$$_{n,i}$(\textbf{k}) $\equiv$ $\left\langle u_{n}(\bf{k})\right|$$\bf{\hat{p}}$$\left|u_{i}(\bf{k})\right\rangle$ is the interband matrix element of the canonical momentum operator $\bf{\hat{p}}$, where $u(\bf{k})$ is the periodic part of the Bloch wavefunction, and $E^{0}_{n(i)}(\bf{k})$ denotes the energy dispersion of the $n$($i$)-th band. Upon substituting the eigenfunctions of Eq. (\ref{eqn 7}) into Eq. (\ref{eqn 8}), the Berry curvature in the conduction band is given by:

\begin{eqnarray}
\bf{\Omega_c(k)} &=& -\tau \frac{2a^2 t^2 \Delta'}{(4a^2 t^2 k^2 + \Delta'^2)^{3/2}}
\label{eqn 9}
\end{eqnarray}
where $\tau$ is the valley index and $\Delta'$ $\equiv$ $\Delta - \tau S_z\lambda$ is the spin-dependent band gap. Note that the Berry curvature has opposite signs in opposite valleys, and this also occurs in the conduction and valence bands [$\bf{\Omega_v}$(\textbf{k}) = $-\bf{\Omega_c}$(\textbf{k})]. Here, we write the equations of motion for Bloch electrons under the influence of the Berry curvature and applied electric and magnetic fields \cite{Xiao2010}:
\begin{eqnarray}
\bf{\dot{r}} &=& \frac{1}{\hbar}\frac{\partial E_{n}(\textbf{k})}{\partial \textbf{k}}-\dot{\textbf{k}} \times \bf{\Omega_n(k)} \\ 
\hbar\dot{\textbf{k}} &=& -e\textbf{E} - e\dot{\textbf{r}} \times \bf{B}
\label{eqn 10}
\end{eqnarray}
It can be seen that in the presence of an in-plane electric field, carriers with different valley indices will acquire opposite velocities in the transverse direction because of the opposite signs of their Berry curvatures, leading to the so-called valley Hall effect, as illustrated in Fig. \ref{Fig5}(a) and (b). Here, we note that this result is valid not only for monolayer 2H-TMDs but also for thin films with an odd number of layers because odd numbers of layers also exhibit inversion symmetry breaking, which is a necessary condition for the $\pm K$ valleys to exhibit valley contrast in the Berry curvature.

The valley contrast in 2H-TMDs can also reflect on the optical interband transitions from the top of the spin-split valence-band to the bottom of the conduction band at the $\pm K$ points. The coupling strength with optical fields of $\sigma_\pm$ circular polarization is given by \textbf{\textit{P}}$_{\pm}(\textbf{k})$ $\equiv$ \textbf{\textit{P}}$_{x}(\textbf{k})$ $\pm$ $i$\textbf{\textit{P}}$_{y}(\textbf{k})$, where \textbf{\textit{P}}$_{\alpha}$(\textbf{k}) $\equiv$ $m_0$$\left\langle u_{c}(\bf{k})\right|$$\frac{1}{\hbar}\frac{\partial\hat{H}}{\partial k_\alpha}$$\left|u_{v}(\bf{k})\right\rangle$ is the interband matrix element of the canonical momentum operator ($u_{c(v)}(\bf{k})$ is the Bloch function for the conduction (valence) band, and $m_0$ is the free electron mass). For transitions near the $\pm K$ points and for a reasonable approximation of $\Delta'$ $\gg$ $atk$ (see the parameters in Ref.\cite{Xiao2012}), this expression has the following form \cite{Xiao2012}:

\begin{eqnarray}
\left|\textit{\textbf{P}}_{\pm}(\textbf{k})\right|^2 &=& \frac{m_{0}^{2}a^{2}t^{2}}{\hbar^2}(1\pm\tau)^2 \label{eqn 11}
\end{eqnarray} It is evident that the coupling strength between circularly polarized light and the interband transitions is valley dependent; \textbf{\textit{P}}$_{+}(\textbf{k})$ has a non-zero value in the $+K$ valley, as does \textbf{\textit{P}}$_{-}(\textbf{k})$ in the $-K$ valley. This valley-dependent optical selection rule is illustrated in Fig. \ref{Fig5}(c), where a $\sigma_{+(-)}$ circularly polarized optical field exclusively couples with the interband transitions at the $+(-)K$ valley. Note that spin is selectively excited through this valley-dependent optical selection rule, and consequently, the spin index becomes locked with the valley index at the band edges. For example, an optical field with $\sigma_{+}$ circular polarization and a frequency of $\omega_{d}$($\omega_{u}$) can generate spin-up (spin-down) electrons and spin-down (spin-up) holes in the $+K$ valley, whereas the excitation in the $-K$ valley is precisely the time-reversed counterpart of the above \cite{Xiao2012}.

\subsection{1.3 1T'-Transition Metal Dichalcogenides}

The TMD family has three typical phases, including 2H, 1T and 1T', as shown in Fig. \ref{Figure 34}(a), (b) and (c), respectively. In contrast to 2H structure, the M atoms in the 1T structure are octahedrally coordinated with the nearby six X atoms, resulting in ABC stacking with the $P\bar{3}m1$ space group, as shown in Fig. \ref{Figure 34}(b). 1T-MX$_2$ have very different electronic properties compared to the semiconducting 2H structures. 1T-TMDs are metallic (Fermi level lying in the middle of degenerate d$_{xy,yz,xz}$ single band) and are often unstable in ambient condition, which usually leads to a spontaneous lattice distortion and a doubling periodicity in the x direction \cite{Tang2017}. Eventually they form a 2$\times$1 superlattice structure, i.e., the  1T' structure, consisting of one-dimensional zigzag chains along the y direction, as shown in Fig. \ref{Figure 34}(c). The lattice distortion from the 1T phase to the 1T' phase induces band inversion and causes 1T'-TMDs to become topologically nontrivial \cite{Qian2014}. Fig. \ref{Figure 35}(a) schematically illustrates this topological phase transition in 1T'-WTe$_2$ \cite{Tang2017}. The bulk band starts with a topological trivial phase, then evolves into a non-trivial phase where the energy of the original valeance band (blue) is higher than that of the original conduction band (red), resulting in an inverted bands crossing at a momentum point along the $\Gamma$-Y direction. Finally, a strong spin-orbit coupling lifts the degeneracy and opens up a bulk bandgap as shown in the rightmost panel of Fig. \ref{Figure 35}(a). The actual calculated electronic band structure of 1T'-MX$_2$ (here taking MoS$_2$ as an example) using many-body perturbation theory is shown in Fig. \ref{Figure 35}(b). As can be seen, the band of 1T'-MoS$_2$ shows a gap ($E_g$) of about 0.08 eV, located at $\Lambda$ $=$ $\pm$(0,0.146)${A}^{-1}$. The conduction and valence bands display a camelback shape near $\Gamma$ point and present a large inverted gap (2$\delta$) of about 0.6 eV. To better understand the nature of the inverted bands near $\Gamma$, a low-energy $\textbf{\textit{k}}\cdot\textbf{\textit{p}}$ Hamiltonian for 1T'-MX$_2$, in which the valence band mainly consists of \textit{d}-orbitals of M atoms ($d_{yz}$ and $d_{xy}$) and the conduction band mainly consists of $p_y$-orbitals of X atoms, is written as \cite{Qian2014}:

\begin{figure}
\begin{center}
		\includegraphics[scale=0.5]{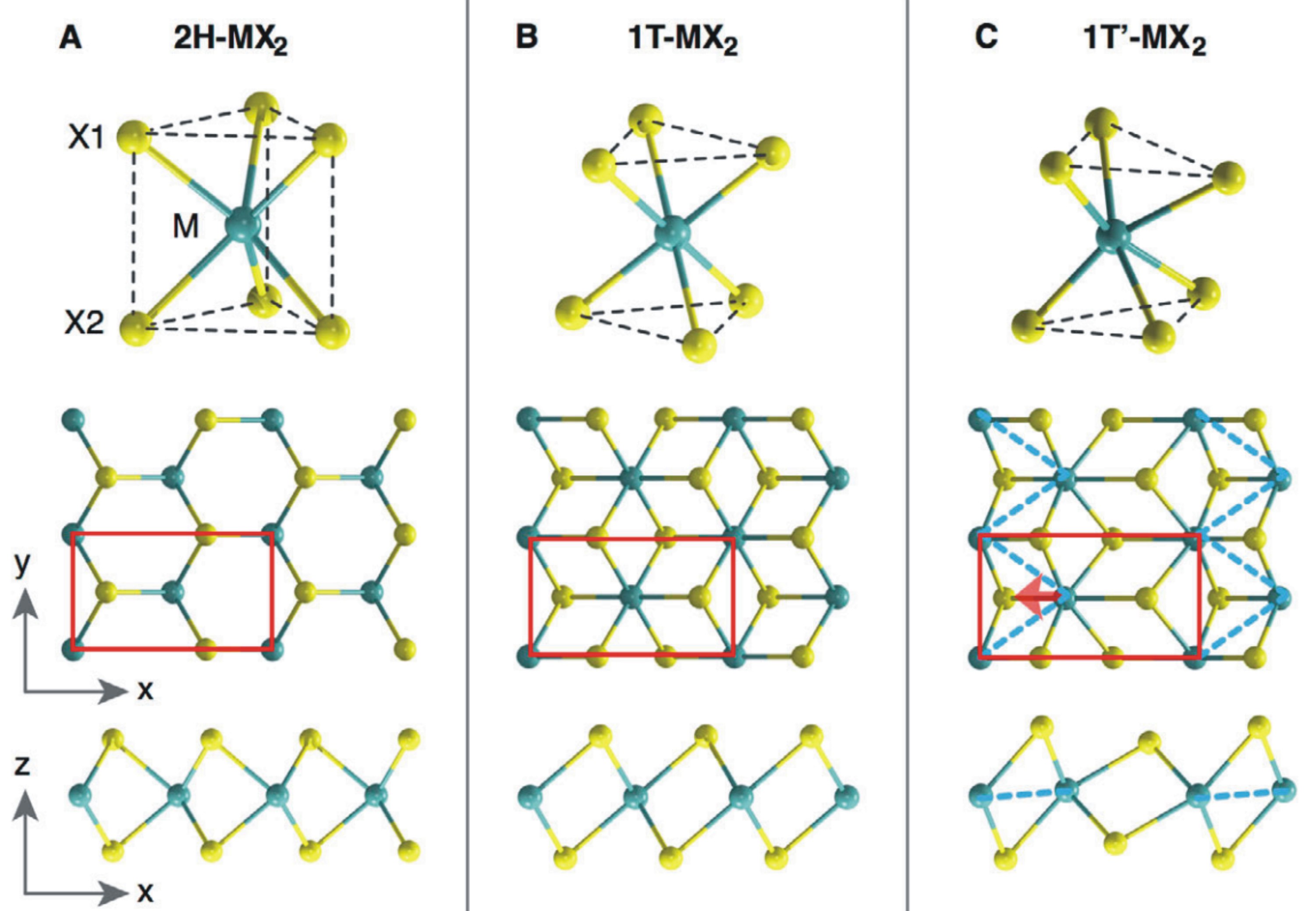}
\caption{Three atomistic structures of monolayer transition metal dichalcogenides MX$_2$, where M stands for (Mo,W) and X stands for (S, Se, Te). (a) 2H-MX$_2$ structure where M atoms are trigonal-prismatically coordinated by six X atoms. (b) 1T-MX2 structure
where M atoms are octahedrally coordinated with the nearby six X atoms. (c) 1T'-MX2, distorted 1T-MX2, where the distorted M atoms
form 1D zigzag chains indicated by the dashed blue lines. The unit cell is indicated by red rectangles. Figures adapted with permission from ref. \cite{Qian2014}. Copyright 2014 American Association for the Advancement of Science.}
\label{Figure 34}
\end{center}	
\end{figure}

\begin{equation}
H = \left( \begin{array}{cccc} \ E_p(k_x,k_y) & 0 & -i\nu_1\hbar k_x & \nu_2\hbar k_y \\ \ 0 & E_p(k_x,k_y) & \nu_2\hbar k_y & -i\nu_1\hbar k_x \\ \ i\nu_1\hbar k_x & \nu_2\hbar k_y & E_d(k_x,k_y) & 0 \\ \ \nu_2\hbar k_y & i\nu_1\hbar k_x & 0 & E_d(k_x,k_y) \\\end{array} \right), 
\label{eqn 56}
\end{equation}
where $E_p$ = $-\delta-\frac{\hbar^2k_x^2}{2m_x^p}$$-$$\frac{\hbar^2k_y^2}{2m_y^p}$, and $E_d$ = $\delta+\frac{\hbar^2k_x^2}{2m_x^d}$$+$$\frac{\hbar^2k_y^2}{2m_y^d}$. Here $\delta$ $<$ 0 corresponds to the $d-p$ band inversion ($E_p$ $>$ $E_d$ near $\Gamma$, see Fig. \ref{Figure 35}(b)). Note that the band inversion arises from the formation of quasi-one dimensional M chains in the 1T' structure, which lowers the metal $d$ orbital below chalcogenide $p$ orbital with respect to the original 1T structure, leading to the band inversion at $\Gamma$ point. By fitting with first-principles band structure in Fig. \ref{Figure 35}(b), parameters in Eq. (\ref{eqn 56}), such as $\delta$, $m_x^p$, $m_y^p$, $m_x^d$ and $m_y^d$, can be estimated \cite{Qian2014}. Since the 1T' structure has inversion symmetry, the Z$_2$ band topology can be determined by the parity of valence bands at four time reversal invariant momenta (TRIM), $\Gamma$, X, Y and R \cite{Chiu2017a,Fu2007a}. Apart from 1T'-MoS$_2$, Qian et al. have calculated the band structures and TRIM of other five 1T'-MX$_2$, including MoSe$_2$, MoTe$_2$, WS$_2$, WSe$_2$ and WTe$_2$ \cite{Qian2014}. Their results suggest that all 1T'-TMDs have Z$_2$ nontrivial band topology resulting from the above $p-d$ band inversion, with inverted band gaps at $\Gamma$ of 1.04, 0.36, 0.28, 0.94, and 1.17 eV, respectively. 1T'-MoSe$_2$, WS$_2$, WSe$_2$ have fundamental gaps of 0.11, 0.12, and 0.12 eV, respectively; while 1T'-MoTe$_2$ and WTe$_2$ are semi-metals due to the increase of valence band maximum at the $\Gamma$ point (although recent experiments indicated that monolayer 1T'-WTe$_2$ is actually a 2D TI, see section 6).

\begin{figure}
\begin{center}
		\includegraphics[scale=0.7]{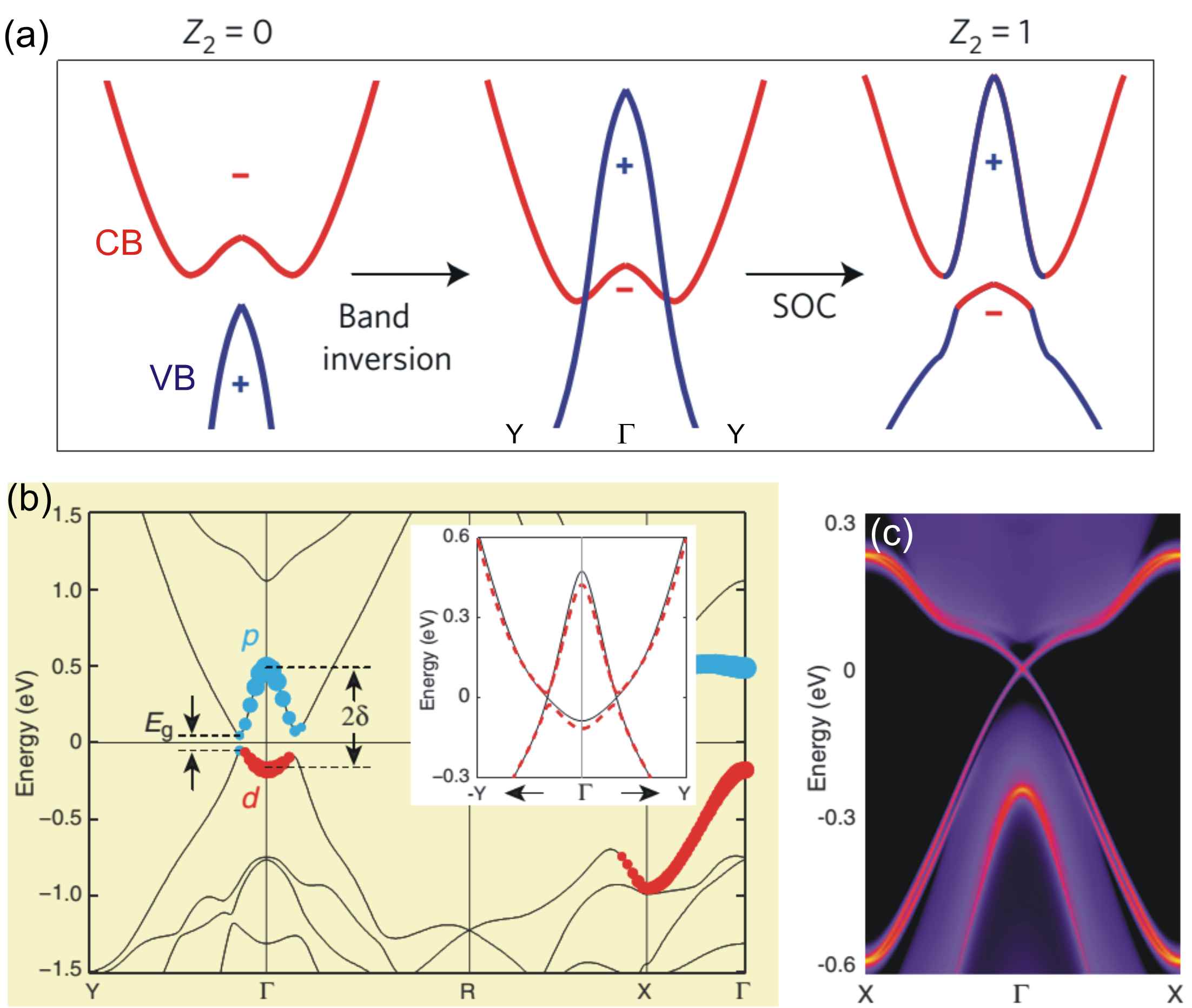}
\caption{Band structures of monolayer 1T'-TMDs. (a) Schematic diagram to show the band evolution from a topologically trivial phase, to a non-trivial phase, and then to a band gap opening due to the spin-orbit coupling. (b) First-principles calculated band structures for monolayer 1T'-MoS$_2$. $E_g$ and 2$\delta$ represent the fundamental gap and inverted gap, respectively. Blue and red dots indicate the major orbital characters in the top valence band and bottom conduction band. The inset compares band structures with (red dashed line) and without (black solid line) spin-orbit coupling. (c) Calculated edge density of states of monolayer 1T'-MoS$_2$. (a) adapted with permission from ref. \cite{Tang2017}. Copyright 2017 Nature Publishing Group. (b, c) adapted with permission from ref. \cite{Qian2014}. Copyright 2014 American Association for the Advancement of Science.}
\label{Figure 35}
\end{center}	
\end{figure}

\begin{figure}
\begin{center}
		\includegraphics[scale=0.5]{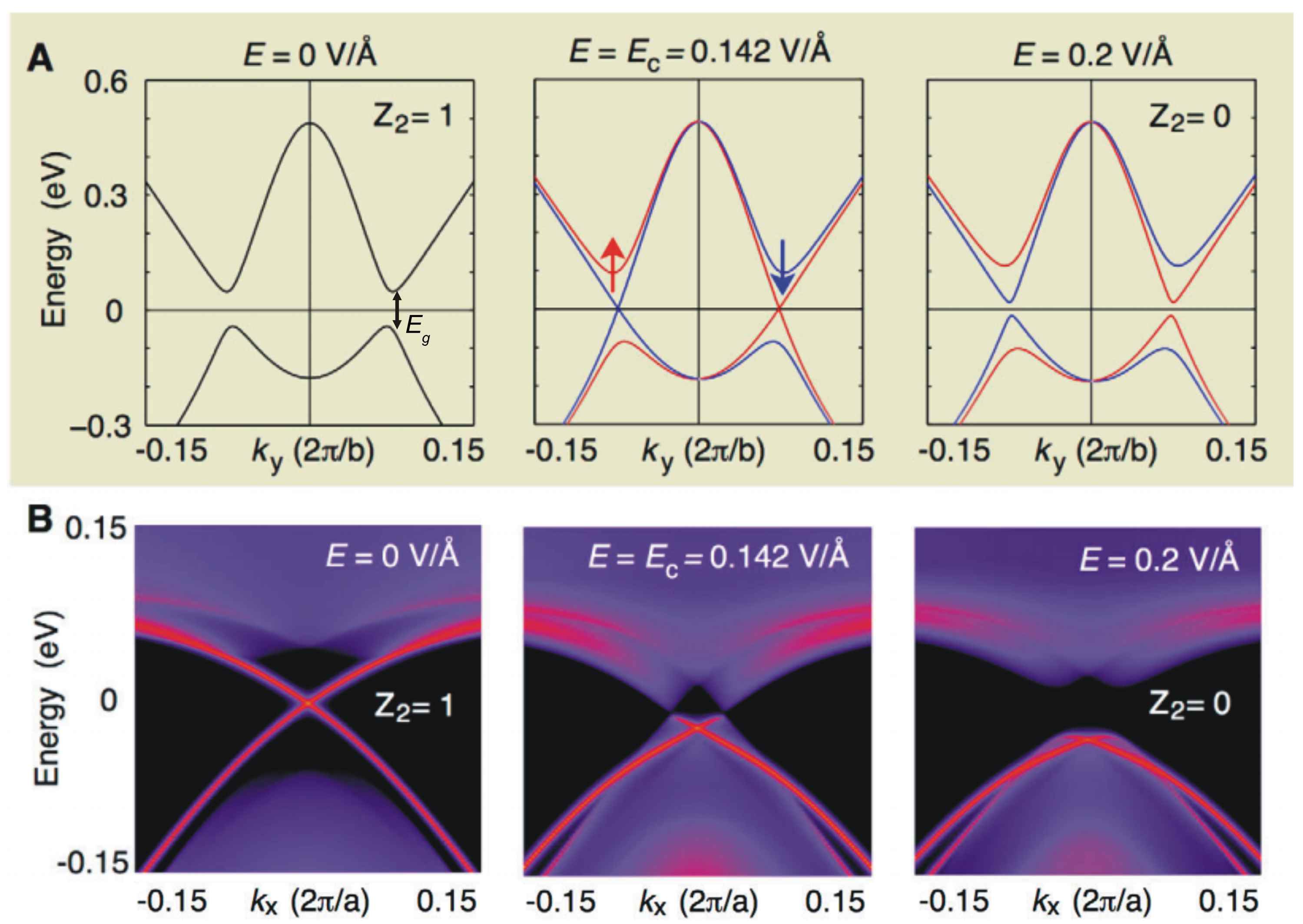}
\caption{Vertical electric field induced topological phase transition in monolayer 1T'-MoS$_2$. (a) Bulk band structure of monolayer 1T'-MoS$_2$ (along Y-$\Gamma$-Y) under different vertical electric field, showing a bandgap closing and reopening around $k_y$ = 0.08 2$\pi$/b. (b) The corresponding edge density of states under different electric fields. Figures adapted with permission from ref. \cite{Qian2014}. Copyright 2014 American Association for the Advancement of Science.}
\label{Figure 36}
\end{center}	
\end{figure}

The topological phase (Z$_2$ = 1) in monolayer 1T'-MX$_2$ makes them a 2D topological insulator which carries helical edge states that are protected from elastic backscattering by time-reversal symmetry (TRS). Fig. \ref{Figure 35}(c) shows the calculated edge density of state of 1T'-MoS$_2$ (similar results are found for other 1T'-MX$_2$) using iterative Green's function and many-body GW theory \cite{Qian2014}. The edge states present a Dirac-like (linear) dispersion located inside the bulk band gap at $\Gamma$, with a high Fermi velocity of $\approx$1$\times$10$^5$m/s. These edge states, as known as quantum spin Hall (QSH) edge states, have the special "spin-filter" property in which upward and downward spins propagate in opposite directions, leading to a phenomenon called spin-momentum locking. Further investigations also indicate that the decay length (from the edge to bulk) of these helical edge states to be as short as 5 nm (50 nm in HgTe quantum wells \cite{Konig2007}), which can greatly reduce scattering with bulk states and hence increase the transport lifetime \cite{Qian2014}. Most interestingly, theory has predicted that the topological phase, hence the existence of helical edge states within the bandgap, can be controlled by gating (vertical electric field) in monolayer 1T'-TMDs \cite{Qian2014}. This tunable topological phase arises from the vertically well-separated planes between chalcogenide's $p$ and metal's $d$ orbitals, which allow a vertical electric field to modify the inverted band. Fig. \ref{Figure 36}(a) displays the first-principles calculated bulk band structures of 1T'-MoS$_2$ under different vertical electric fields from 0 to 0.2 V/A, while Fig. \ref{Figure 36}(b) shows the corresponding edge density of states along X-$\Gamma$-X. The electric field breaks the inversion symmetry and introduces Rashba spin splitting of the original doubly degenerate bands near the fundamental gap $E_g$ [see middle panel of Fig. \ref{Figure 36}(a)]. As the field increases, $E_g$ first decreases to zero at a critical field strength of 0.142 V/A and then reopens [see the rightmost panel in Fig. \ref{Figure 36}(a)]. This gap-closing transition induces a topology change to a trivial phase, leading to the destruction of helical edge states, as shown in Fig. \ref{Figure 36}(b). In addition to the vertical electrical field, Qian et al. also reported that a few percent of in-plane elastic strain can change monolayer 1T'-MoTe$_2$ and WTe$_2$ from semimetals to small-gap QSH insulators by lifting the band overlap. The gate-tunable topological phases are viable for designing all electric-field controlled topological devices, and could be useful for probing Majorana zero modes \cite{Alicea2012}. The quantized conductance of the QSH edge states in 1T'-TMDs has been observed in several literature which will be reviewed in section 6. However, the topological phase transition induced by vertical gating has not been reported up to date. 

In this section, we introduced the fundamental properties of various 2D materials that are to be discussed in the rest of the chapter. Combined with the quantum transport physics presented in the next section, these discussions will serve as a basis for our examination of the experimental studies in the subsequent sections.

\maketitle
\section{2. Theoretical background on quantum transport}

\subsection{2.1 Single quantum dot}
\begin{figure}
\begin{center}
		\includegraphics[scale=0.7]{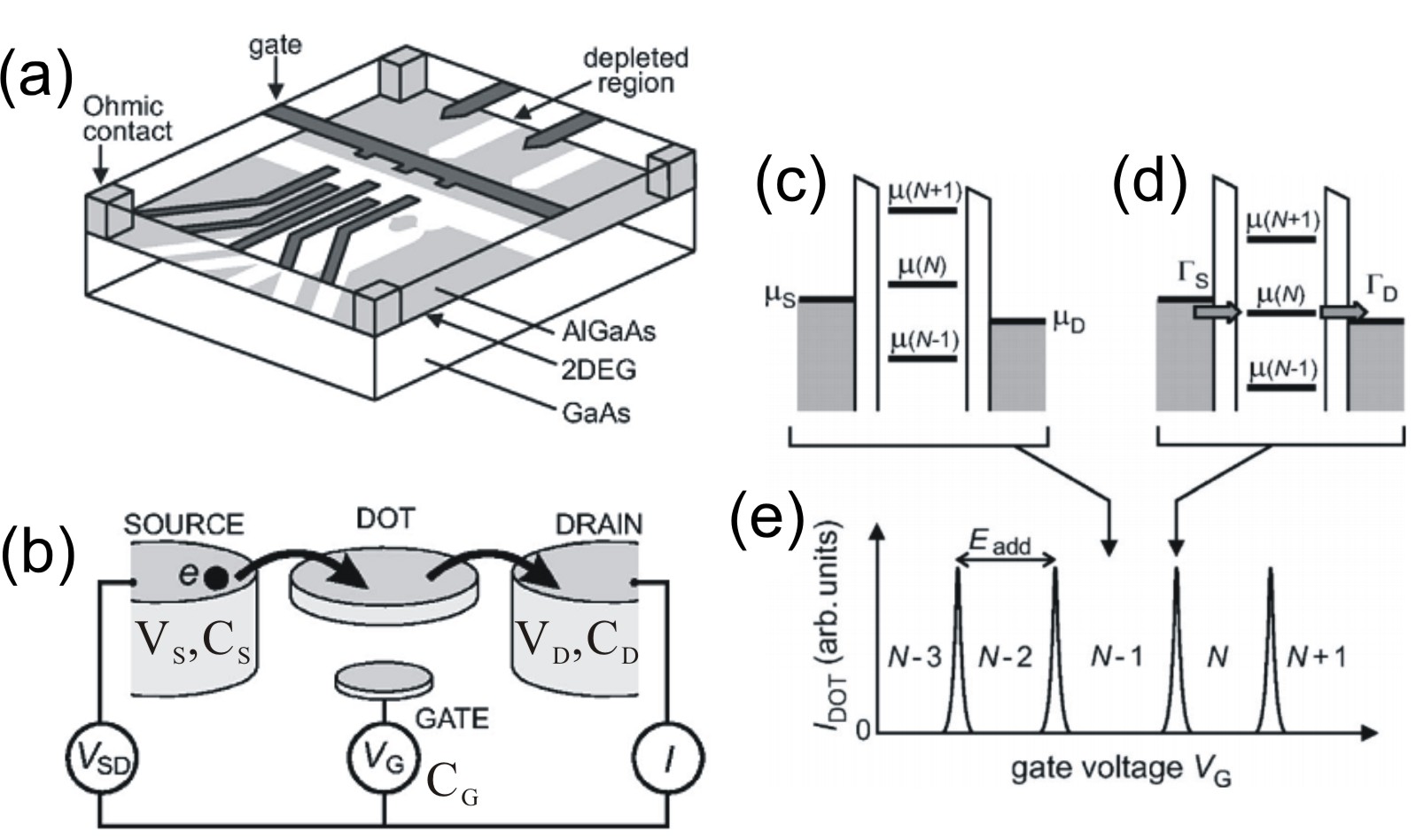}
\caption{(a) Schematic view of a lateral quantum dot device defined by metal surface electrodes on a GaAs/AlGaAs 2DEG system. (b) Electrical network diagram of a single quantum dot. (c) The electrochemical potential energies for a single dot with transport blocked due to Coulomb blockade. (d) With one of the electrochemical potentials lying within the bias window transport through the dot is then permitted. (e) Schematic plot of the current through a single dot against gate voltage showing periodic oscillations with the dot charge. Adapted with permission from ref. \cite{Hanson2007}. Copyright 2007 American Physical Society.}
\label{Fig10}
\end{center}	
\end{figure}

A quantum dot is an artificially structured system that can be filled with only a few electrons or holes \cite{Hanson2007}. The charged carriers in such a system are generally confined in a submicron area, and the confinement potential in all directions is so strong that it gives rise to quantized energy levels that can be observed at low temperatures. The electronic properties of quantum dots are dominated by several effects \cite{Hanson2007}. First, the Coulomb repulsion between electrons on the dot leads to an energy cost called charging energy $E_C$ $=$ $e^{2}/C$, where $C$ is the total capacitance of the dot, for adding an extra electron to the dot. Because of this charging energy, the tunneling of electrons to or from the reservoirs can be suppressed at low temperatures (when $E_C$ $>$ $k_BT$), which leads to a phenomenon called Coulomb blockade. Second, the tunnel barrier resistance $R_{t}$, which describes the coupling of the dot to both the source and drain reservoirs, has to be sufficiently opaque such that the electrons are located either in the source, in the drain, or on the dot. The minimal $R_{t}$ can be estimated using the uncertainty principle, $\Delta E$$\cdot$$\Delta t$ $>$ $h$. From $\Delta E$ $=$ $e^{2}/C$ and $\Delta t$ $=$ $R_{t}C$, the condition $R_{t}$ $>$ $h/e^{2}$ for $R_{t}$ can be found. This means that the energy uncertainty corresponding to the tunneling time can not be greater than the charging energy; otherwise, it would lead to uncertainty in the number of carriers occupying the dot. Third, if the confinement in all three directions is strong enough for electrons residing on the dot to form quantized energy levels $E_{n}$ (often denoted as single-particle energy), the energy spacing $\Delta E$ $=$ $E_{n}$ $-$ $E_{n-1}$ can be observed on top of charging energy if $\Delta E$ $>$ $k_BT$. Because of this discrete energy spectrum $E_{n}$, quantum dots behave in many ways as artificial atoms. Fig. \ref{Fig10}(a) shows an example of a quantum dot formed in a GaAs/AlGaAs 2DEG system, where the dot is defined by a gate-depleted area and is tunnel coupled to the reservoir on each side. Varying the voltages on the surface gates enables several important parameters, such as the number of electrons and the tunnel barrier resistance, to be finely tuned. To understand the dynamics of a single quantum dot, a constant interaction model has been proposed \cite{Hanson2007} and is illustrated in Fig. \ref{Fig10}(b). The model is based on two assumptions. First, the Coulomb interactions among electrons in the dot, and between electrons in the dot and those in the environment, are parameterized by a single, constant capacitance $C$. This capacitance is the sum of the capacitance between the dot and the source $C_{S}$, the drain $C_{D}$ and the gate $C_{G}$: $C$=$C_{S}$+$C_{D}$+$C_{G}$. The second assumption is that the single-particle energy spectrum $E_{n}$ is independent of the Coulomb interaction, therefore of the number of electrons in the dot. Using this model, the total energy of a single dot with $N$ electrons in the ground state is given by \cite{Hanson2007}:
\begin{eqnarray}
U(N) &=&  \frac{\left(-\left|e\right|\left(N-N_0\right)+ C_S V_S + C_D V_D + C_G V_G\right)^2}{2C} + \sum_{n=1}^{N} E_n
\label{eqn 18}
\end{eqnarray}
where -$\left|e\right|$ is the electron charge, $N_0$ is the charge on the quantum dot due to the positive background charge of the donors and $V_S$, $V_D$ and $V_G$ are the voltages of the source, drain and gate respectively. The last term is a sum over the occupied single-particle energy levels $E_{n}$ which depend on the characteristics of the confinement potential.

The electrochemical potential of the dot $\mu \left(N\right)$ is defined as the energy needed to add the $N$-th electron to a dot with $N$-1 occupied electrons \cite{Hanson2007}:
\begin{eqnarray}
\mu \left(N\right) 	 &=& \nonumber U\left(N\right) - U\left(N-1\right) \\
&= & \left(N - N_0 - \frac{1}{2}\right)E_C - \frac{E_C}{e}\left(C_S V_S + C_D V_D + C_G V_G\right) + E_N
\label{eqn 19}
\end{eqnarray}
where 
\begin{eqnarray}
E_{C}=e^{2}/C
\label{eqn 48}
\end{eqnarray} is the charging energy. The addition energy is then given by the energy difference between two successive electrochemical potentials: 
\begin{equation}
E_{add}\left(N\right) = \mu \left(N+1\right) - \mu \left(N\right) \\
= E_C + \Delta E
\label{eqn 20}
\end{equation} 
where $\Delta E = E_{N+1} - E_N$ is the single-particle energy spacing, and is independent of the electron number on the dot (the second assumption).

When the temperature is low enough ($k_{B}$$T \ll \Delta E, E_C$), the transport through the quantum dot depends on whether the dot electrochemical potentials align with bias window, which is defined as the spacing between the electrochemical potentials of the source and drain, i.e., $-eV_{SD}$ $\equiv$ $\mu_S - \mu_D$ $=$ $-eV_{S}-(-eV_{D})$. In the low bias regime where $-eV_{SD} < E_C$, electron tunneling can only happen when the dot electrochemical potential lies in a small bias window, such that $\mu_D < \mu \left(N\right) < \mu_S$ as shown in Fig. \ref{Fig10}(d). When the electrochemical potential is outside the bias window the transport is blocked and no current flows through the dot, which is the Coulomb blockade regime as shown in Fig. \ref{Fig10}(c). When a gate $V_G$ constantly tunes the electrochemical potential of the quantum dot, an on-off current can be observed as peaks with constant spacing ($E_{add}$) between each other as shown in Fig. \ref{Fig10}(e). Each current forbidden regime corresponds to a different electron number on the dot, so in this way the number of electrons on the dot can be varied.

\begin{figure}[!]
\begin{center}
		\includegraphics[scale=0.7]{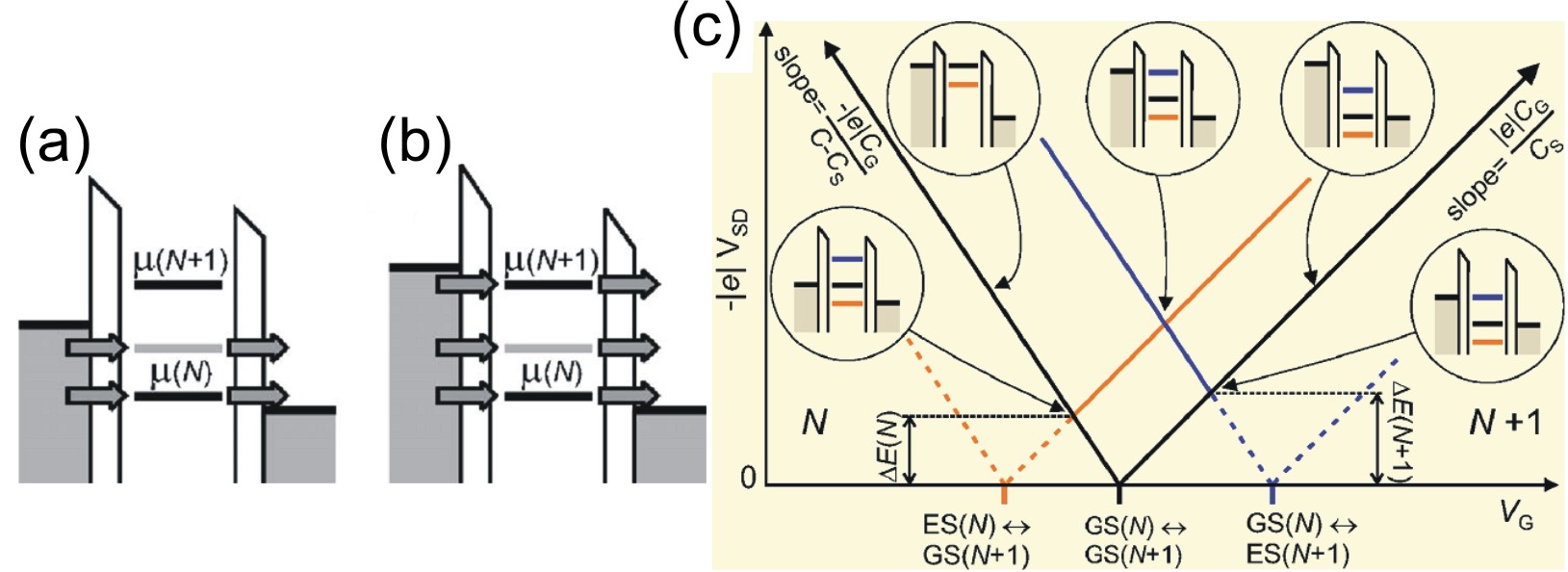}
\caption{ Schematic diagrams of the electrochemical potential levels of a quantum dot in the high bias regime. (a) $V_{SD}$ exceeds $\Delta$E so the electron transport couple to an excited state (the grey level). (b) $V_{SD}$ exceeds the addition energy so the electron transport couples to two successive ground state levels. (c) Differential conductance $dI_{DOT}/dV_{SD}$ through a quantum dot as a function of gate $V_{G}$ and bias voltage -$\left|e\right|V_{SD}$. The insets show different configurations of the dot level with respect to the lead potential in the $V_{SD}-V_{G}$ plane. Adapted with permission from ref. \cite{Hanson2007}. Copyright 2007 American Physical Society.}
\label{Fig11}
\end{center}	
\end{figure}

In the high bias regime where $-eV_{SD} > \Delta E$ and/or $-eV_{SD} > E_{add}$, more dot levels are allowed to lie within the bias window and give rise to multiple tunneling paths as shown in Fig. \ref{Fig11}(a) and (b). Depending on how wide the bias window is, the transition can involve a ground sate and its excited state as shown in Fig. \ref{Fig11}(a), or in an even wider window ($-eV_{SD} >E_{add}$) it can couple to two successive ground states as shown in Fig. \ref{Fig11}(b). From Eq. (\ref{eqn 19}) the electrochemical potential is a function of $V_S$, $V_D$ and $V_G$. Since $\mu_{S(D)}=-eV_{S(D)}$, if we measure the conductance of dot as a function of bias $eV_{SD}$ and gate voltage $V_G$ a spectrum called "Coulomb diamond" is formed as shown in Fig. \ref{Fig11}(c). Since larger biases require a wider spacing in gate voltage for dot levels being pulled out of the window, the V-shape feature can be expected. In Fig. \ref{Fig11}(c) along the left (right) edge of the black V-shape following the slope at $\frac{-\left|e\right|C_{G}}{C-C_{S}}$ ($\frac{\left|e\right|C_{G}}{C_{S}}$), the level of the $N$-electron ground state is aligned with the source (drain) level while the bias window is becoming wider. The black V-shape shows the transition between the $N$-electron ground state and $N+1$-electron ground state, and defines the regimes of blockade (outside the V-shape) and tunneling (within the V-shape). The orange and blue V-shapes shown in Fig. \ref{Fig11}(c) correspond to two different transitions between the dot states which are, the $N$-electon excited state to $N+1$-electron ground state ($ES(N)$$\rightarrow$$GS(N+1)$) and the $N$-electon ground state to $N+1$-electron excited state ($GS(N)$$\rightarrow$$ES(N+1)$). Since the excited state energy $ES(N)$ and $ES(N+1)$ are separated from the ground states $GS(N)$ and $GS(N+1)$ by $\Delta$$E(N)$ and $\Delta$$E(N+1)$ respectively [see Fig. \ref{Fig11}(c)], Coulomb diamond measurements are very useful for studying the excited state spectroscopy in a quantum dot system. The insets shown in Fig. \ref{Fig11}(c) represent a different configuration of dot levels with respect to the source-drain level. Note that the $ES(N)$$\rightarrow$$GS(N+1)$ and $GS(N)$$\rightarrow$$ES(N+1)$ transitions are forbidden outside the black V-shape as $ES(N)$ and $ES(N+1)$ states only exist when the $GS(N)$$\rightarrow$$GS(N+1)$ transition is within the bias window. Finally, the dimension of the Coulomb diamond (current-suppressed region) in the bias direction is a direct measure of $E_{add}$ or the charging energy $E_C$, because beyond the edge of the diamond the bias window is greater than $E_{add}$ and transport is no longer blocked.

\begin{figure}
\begin{center}
		\includegraphics[scale=0.74]{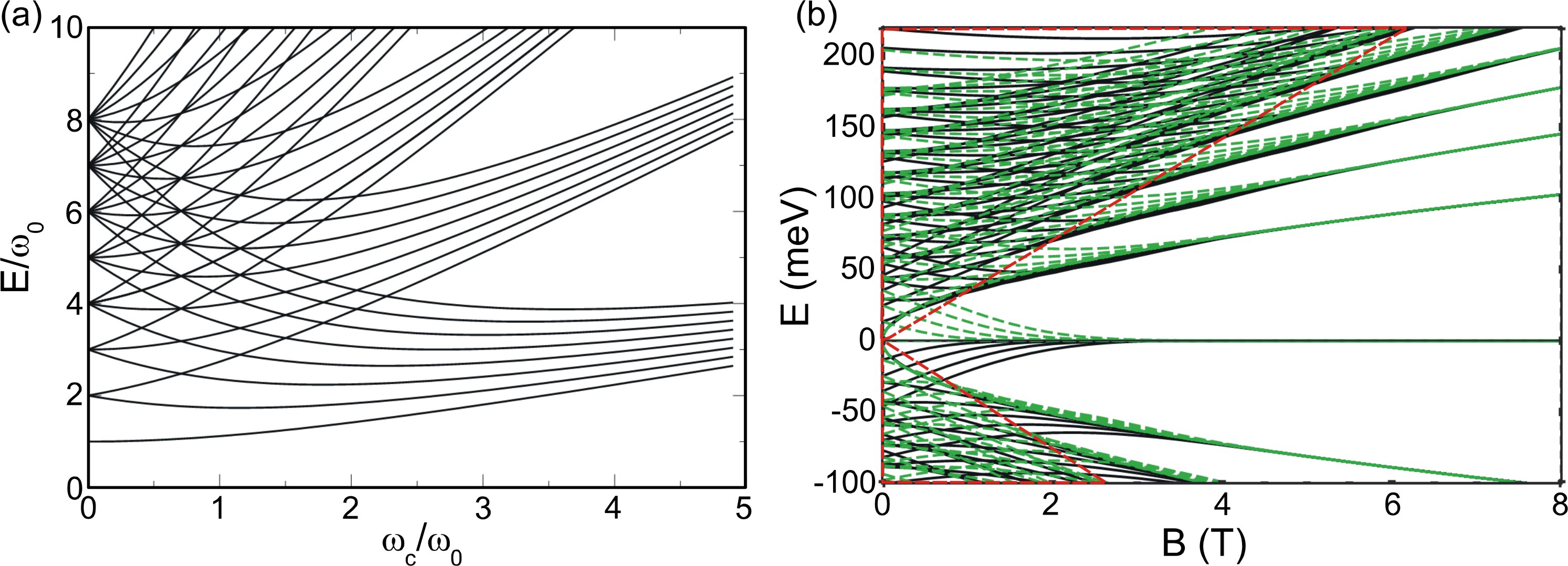}
\caption{(a) Fock-Darwin spectrum of a 2DEG symmetric quantum dot up to a quantum number of $n$ $=$ 7. (b) Energy spectrum of a graphene quantum dot with $R$ $=$ 70 nm for $m$ $=$ -4,...,4 and $n$ $=$ 1,...,6. The energy states for $\tau = +$1 are drawn as solid black lines, and those for $\tau = -$1 are drawn as dashed green lines. The red dashed areas highlight the continuously crossing of different states at low energy, as will be discussed in section 3.1.2. (b) adapted with permission from ref. \cite{Schnez2008}. Copyright 2008 American Physical Society.}
\label{Fig12}
\end{center}	
\end{figure}

Here, we discuss the effect exerted by a magnetic field on the single-particle energy of QDs. The energy spectrum of a 2DEG quantum dot in the presence of a magnetic field is typically solved using a single-particle approximation with a parabolic confinement potential \cite{Fock1928,Kouwenhoven2001}. Such a spectrum is called the Fock-Darwin diagram which describes how 0D levels evolve with respect to an applied perpendicular magnetic field. The symmetric parabolic potential can be approximated as $U(x,y) = \frac{m^\ast}{2}\omega^{2}_{0}(x^2+y^2)$, where $m^\ast$ is the effective mass and $\omega^{2}_{0}$ denotes the strength of the confinement potential. Thus, the Hamiltonian of an electron in the dot can be written as follows: 

\begin{eqnarray}
H = \frac{1}{2m^\ast}(\textbf{p}+e\textbf{A})^2 + \frac{m^\ast}{2}\omega^{2}_{0}(x^2+y^2)
\label{eqn 46}
\end{eqnarray} If we choose the symmetric gauge for the vector potential $\textbf{A}$ $=$ ($-By/2$, $Bx/2$, 0), then the energy spectrum of the Hamiltonian can be solved as follows: 
\begin{eqnarray}
E_{n_+,n_-} = (n_{+} + 1)\hbar\Omega + \frac{1}{2}\hbar\omega_cn_-
\label{eqn 47}
\end{eqnarray}with $\Omega^2$ $\equiv$ $\omega^2_0 + \frac{\omega^2_c}{4}$ where $\omega_c$ $=$ $\frac{\left|eB\right|}{m^\ast}$ is the cyclotron frequency, and with quantum numbers $n_{\pm}$ $=$ $n_x$ $\pm$ $n_y$ where $n_x$, $n_y$ $=$ 0, 1, 2, ..., etc. This spectrum is plotted in Fig. \ref{Fig12}(a). For $B = 0$ the spectrum has a constant level spacing and is simply the spectrum of the two-dimensional harmonic oscillator. In the high-field limit, the spectrum goes over into that of the Landau levels [see Fig. \ref{Fig2}(a)], with the confinement effects of the dot playing an ever-decreasing role. 

In a graphene quantum dot, the Fock-Darwin spectrum is notably different from that in the 2DEG case owing to the existence of a Landau level (LL) at zero energy, which does not shift in energy with increasing magnetic field \cite{Schnez2008,Recher2009}. Together with quantum confinement, the unique linear band dispersion of graphene results in an electron-hole crossover in GQD's magneto-transport \cite{Guttinger2009,Chiu2012}. To solve the Fock-Darwin spectrum for a graphene quantum dot, we start from a free Dirac equation with a circular confinement potential $V(r)$ and include a perpendicular magnetic field, where the symmetric gauge $A$ $=$ $\frac{B}{2}(-y, x, 0)$ $=$ $\frac{B}{2}$($-r$sin$\phi$, $r$cos$\phi$, 0) for the vector potential is used ($\phi$ is the polar angle). Thus, the Hamiltonian now reads (ignoring spin) \cite{Schnez2008}:  

\begin{eqnarray}
H = \nu_F(\textbf{p} + e\textbf{A})\cdot\vec{\sigma} + \tau V(r)\sigma_z,
\label{eqn 48}
\end{eqnarray} where $\vec{\sigma} = (\sigma_x, \sigma_y)$ represents Pauli's matrices and $\tau$ $=$ $\pm$1 is the valley index for $\pm K$. Note that the quantum confinement effect is introduced in the Hamiltonian $via$ a mass-related potential $V(r)$ coupling to the $\sigma_z$ Pauli matrix. We let the mass in the dot to be zero, i.e., $V(r) = 0$ for $r < R$, but let it tend toward infinity at the edge of the dot, i.e., $V(r) = \infty$ for $r > R$. In this way, charge carriers are confined inside the quantum dot which has a radius of $R$. This leads to a boundary condition, which yields the simple relation that $\frac{\psi_2}{\psi_1}$ $=$ $\tau i$ exp[$i\phi$] for circular confinement \cite{Schnez2008}, where $\psi$ $=$ ($\psi_1, \psi_2$) is the eigenfunction of the Hamiltonian. Hence, in the following, we can set $V(r) = 0$; thus, the energy $E$ is related to the wavevector $k$ $via$ $E = \hbar \nu_F k$, and we can determine $k$ using the boundary condition. Following ref. \cite{Schnez2008}, the implicit equation for determining the wavevector $k$ (and therefore the energy $E$) that satisfies the boundary condition is given by:

\begin{eqnarray}
\left(1-\tau\frac{kl_B}{R/l_B}\right)L\left(\frac{k^2l^2_B}{2}-(m+1), m, \frac{R^2}{2l^2_B}\right)+L\left(\frac{k^2l^2_B}{2}-(m+2), m+1, \frac{R^2}{2l^2_B}\right) = 0,
\label{eqn 49}
\end{eqnarray} where $l_B$ $=$ $\sqrt{\hbar/eB}$ is the magnetic length and $m$ is the angular momentum quantum number. The functions $L(a, b, x)$ are generalized Laguerre polynomials, which are oscillatory functions. Hence, there are an infinite number of wavevectors $k_n$ for a given $B$, $m$, and $\tau$ that fulfill Eq. (\ref{eqn 49}). This condition defines the radial quantum number $n$, from which the energy spectrum $E (n, m, \tau)$ $=$ $\hbar \nu_F k_n$ can be plotted, as shown in Fig. \ref{Fig12}(b) for a QD of radius R $=$ 70 nm. Note that $-E (n, m, \tau)$ $=$ $E (n, m, -\tau)$, which gives rise to the electron-hole symmetry in the spectrum. We discuss Eq. (\ref{eqn 49}) under two particular limits. For $B$ $\rightarrow$ 0, Eq. (\ref{eqn 49}) can be written as follows: 

\begin{eqnarray}
\tau J_m(kR) = J_{m+1}(kR),
\label{eqn 50}
\end{eqnarray} where $J_m$ is Bessel function. This relation yields the single-particle energy spectrum and can be used to estimate the energy of the excited states on a graphene dot with $N$ confined charge carriers [$\Delta (N)$ = $\hbar \nu_F$/($d \sqrt{N}$), where $d$ is an effective dot diameter; see ref. \cite{Ponomarenko2008,Schnez2009}]. In addition, there is no state at zero energy under zero magnetic field, which leads to an energy gap separating the states of negative and positive energies. By contrast, at high field, where $R/l_B$ $\rightarrow$ $\infty$, Eq. (\ref{eqn 49}) gives rise to the following:
\begin{eqnarray}
E_m = \hbar \nu_F k_m = \pm\nu_F\sqrt{2e\hbar B(m+1)}
\label{eqn 51}
\end{eqnarray} which are the Landau levels for graphene. Therefore, as the B-field increases, there will be a transition governed by the parameter $R/l_B$, from a regime in which the confinement play an important role ($R \leq l_B$) to the Landau-level regime ($R \geq l_B$). Note that the resonances on both sides of the electron-hole crossover have opposite slopes and merge into the zeroth Landau level. An experimental observation of this effect would constitute clear identification of this crossover, as will be presented in section 3.1.2.

\subsection{2.2 Double quantum dot}
When two single dots are placed in series and separately connected to a source and drain reservoir, a double quantum dot (DQD) with a network of source-dot-dot-drain is formed. To apply the constant interaction model in such a system \cite{Wiel2002}, a schematic diagram of its equivalent electrical network is shown in Fig. \ref{Fig13}(a). In this model, the dots QD1 (QD2) are capacitively coupled to their nearest plunger gate PG1 (PG2) via a capacitance $C_{g1}$ ($C_{g2}$), however, they are also coupled to the further gate PG2 (PG1) through the cross capacitance $C_{g21}$ ($C_{g12}$). The dots themselves also couple to each other through an interdot capacitance $C_{m}$ and to the source and drain reservoir through $C_{S}$ and $C_{D}$ individually. The voltages applied to plunger gate 1, plunger gate 2, source and drain are denoted by $V_{PG1}$, $V_{PG2}$, $V_{S}$ and $V_{D}$ respectively as shown in Fig. \ref{Fig13}(a). The charge and its equivalent voltage on QD1 (QD2) are denoted by $Q_{1(2)}$ and $V_{1(2)}$, also shown in Fig. \ref{Fig13}(a). Based on this model the charge at each dot is given by the vector $\bf{Q} = \bf{CV}$ where $\bf{C}$ is the capacitance matrix, $\bf{Q}$=($Q_{1}$, $Q_{2}$) is the vector of charges and $\bf{V}$=($V_{1}$, $V_{2}$) is the vector of electrostatic potentials. Therefore the components of $\bf{Q}$ are given by \cite{Wiel2002}

\begin{equation}
\left( \begin{array}{cc}
\ Q_{1}+C_{S}V_{S}+C_{g1}V_{g1}+C_{g21}V_{g2} \\
\ Q_{2}+C_{D}V_{D}+C_{g2}V_{g2}+C_{g12}V_{g1}  \\ \end{array} \right)\ =
\left( \begin{array}{cc}
\ C_{1} & -C_{m} \\
\ -C_{m} & C_{2}  \\ \end{array} \right)\
\left( \begin{array}{cc}
\ V_1 \\
\ V_2  \\ \end{array} \right)\
\label{eqn 20}
\end{equation}, where $C_{1(2)}$ = $C_{S(D)}+C_{g1(2)}+C_{g21(12)}+C_{m}$ is the total capacitance of dot 1(2). Making the substitution $Q_{1(2)}=-N_{1(2)}e$, and taking $V_{S}=V_{D}=0$ [in the low bias regime and $N_{1(2)}$ is the electron number in dot 1(2)], Eq. (\ref{eqn 20}) then reads \cite{Wiel2002}:

\begin{figure}[!]
\begin{center}
		\includegraphics[scale=0.60]{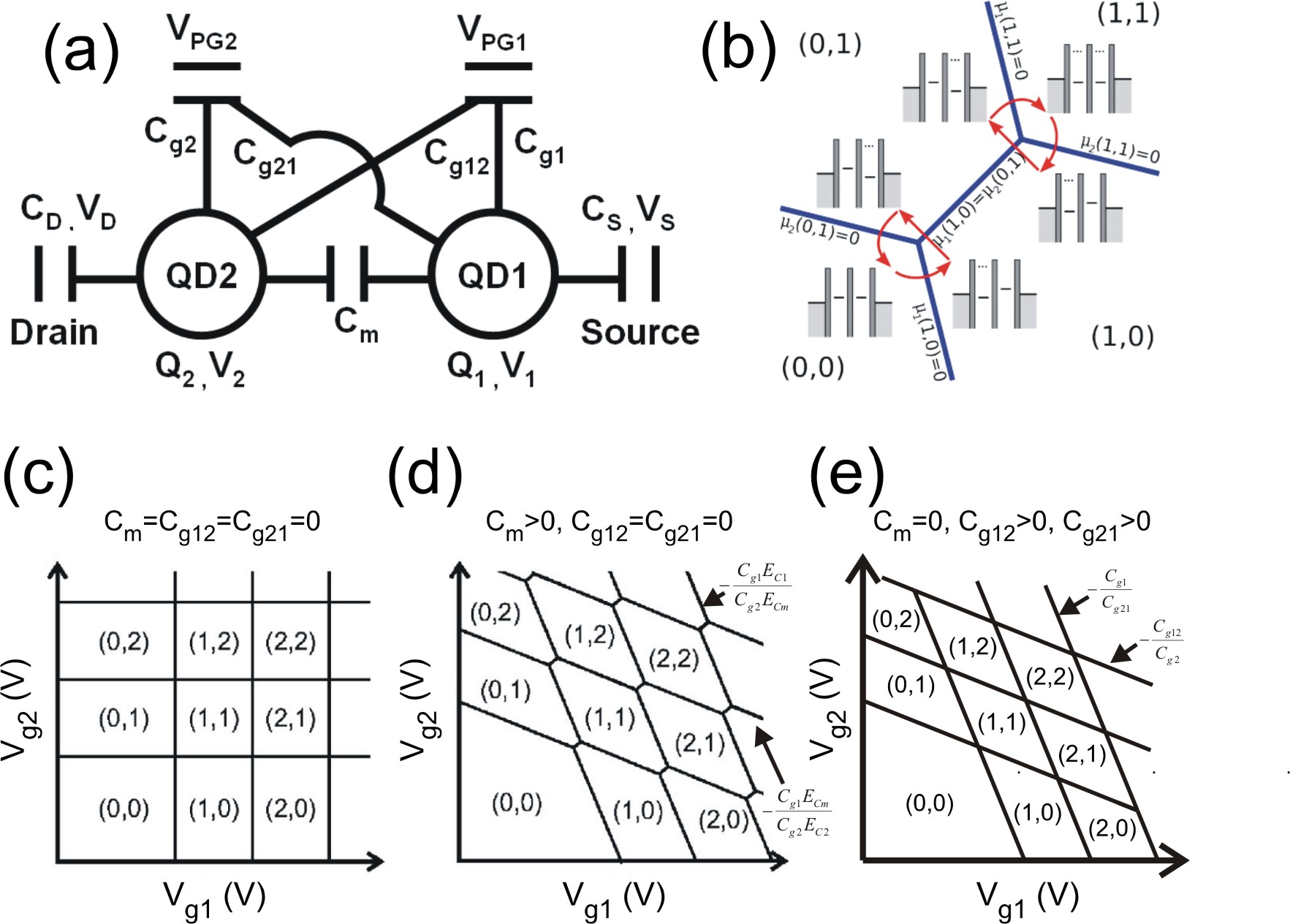}
		\end{center}		
\caption{(a) Electrostatic network model for a double quantum dot considering the cross-capacitance coupling. (b) Charge evolution during conduction at the triple points. The outer numbers in brackets give the stable electron number for that region. The inserts shown are the energy levels of the stable regions at points close to the triple point. (c)-(e) Charge-stability diagrams for a double quantum dot with (c) no interdot and cross-capacitance coupling ($C_{g21}=C_{g21}=E_{Cm}=0$), (d) intermediate interdot coupling but no cross-capacitance coupling ($C_{g21}=C_{g12}=0$, $E_{Cm}\neq0$) and (e) no interdot coupling but intermediate cross-capacitance coupling ($C_{g21}\neq0,C_{g12}\neq0$, $E_{Cm}=0$). (c, d) adapted with permission from ref. \cite{Wiel2002}. Copyright 2003 American Physical Society.}
\label{Fig13}
\end{figure}

\begin{equation}
\left( \begin{array}{cc}
\ V_1 \\
\ V_2  \\ \end{array} \right)\ =\frac{1}{C_{1}C_{2}-C^{2}_{m}} \left( \begin{array}{cc}
\ C_{2} & C_{m} \\
\ C_{m} & C_{1}  \\\end{array} \right)\ \left( \begin{array}{cc}
\ -N_{1}e+C_{g1}V_{g1}+C_{g21}V_{g2} \\
\ -N_{2}e+C_{g2}V_{g2}+C_{g12}V_{g1}  \\ \end{array} \right)\
\label{eqn 21}
\end{equation}

So the total electrostatic energy of such a system is given by \cite{Hanson2007},
\begin{eqnarray}
U\left(N_1,N_2\right) &=& \nonumber \frac{1}{2} \bf{Q}\cdot\bf{V}\\ 
&=& \nonumber \frac{1}{2}N^{2}_{1}E_{C1}+N_{1}N_{2}E_{Cm}+\frac{1}{2}N^{2}_{2}E_{C2}-\frac{1}{2e}C_{g1}V_{g1}N_{1}E_{C1}- \frac{1}{2e}C_{g21}V_{g2}N_{1}E_{C1} \\ \nonumber
&& -\frac{1}{2e}C_{g2}V_{g2}N_{1}E_{Cm}-\frac{1}{2e}C_{g12}V_{g1}N_{1}E_{Cm}-\frac{1}{2e}C_{g1}V_{g1}N_{2}E_{Cm} \\
&&  -\frac{1}{2e}C_{g21}V_{g2}N_{2}E_{Cm}-
\frac{1}{2e}C_{g2}V_{g2}N_{2}E_{C2}-\frac{1}{2e}C_{g12}V_{g1}N_{2}E_{C2}
\label{eqn 22}
\end{eqnarray}
where the charging energies for the dots $E_{C1}$ and $E_{C2}$ and the coupling energy $E_{Cm}$ are given by
\begin{eqnarray}
E_{C1} &=& \frac{e^2}{C_1} \frac{1}{1-\frac{C_{m}^2}{C_1 C_2}} \label{eqn 23}\\
E_{C2} &=& \frac{e^2}{C_2} \frac{1}{1-\frac{C_{m}^2}{C_1 C_2}} \label{eqn 24}\\
E_{Cm} &=& \frac{e^2}{C_m} \frac{1}{\frac{C_1 C_2}{C_{m}^2}-1} \label{eqn 25}
\end{eqnarray}

The electrostatic potentials for the dots are then given by \cite{Hanson2007},

\begin{eqnarray}
\mu_{1}\left(N_1,N_2\right) &=& \nonumber U\left(N_1,N_2\right) - U\left(N_1-1,N_2\right)\\
 &=& \nonumber \left(N_1 - 1\right)E_{C1}  + N_2 E_{Cm} \\ 
&&  - \frac{1}{2e}\left(C_{g1}V_{g1}E_{C1}  +  C_{g2}V_{g2}E_{Cm}+C_{g21}V_{g2}E_{C1}+ C_{g12}V_{g1}E_{Cm}\right) \label{eqn 26}\\
\mu_{2}\left(N_1,N_2\right) &=& \nonumber U\left(N_1,N_2\right) - U\left(N_1,N_2-1\right)\\
 &=& \nonumber \left(N_2 - 1\right)E_{C2}  + N_1 E_{Cm} \\ 
&&  - \frac{1}{2e}\left(C_{g1}V_{g1}E_{Cm}  +  C_{g2}V_{g2}E_{C2}+C_{g12}V_{g1}E_{C2}+ C_{g21}V_{g2}E_{Cm}\right) \label{eqn 27}
\end{eqnarray}

The physical meaning of each term, for example, $(N_{1}-1)E_{C1}$ and $N_{2}E_{Cm}$, stand for the Coulomb statistic energy increases on dot 1 and dot 2 when the $N_{1}$-th electron is added to dot 1. The term $C_{g1}V_{g1}E_{C1}$ in Eq. (\ref{eqn 26}) is the direct coupling energy between PG1 and QD1, while $C_{g2}V_{g2}E_{Cm}$ is the indirect coupling energy between PG2 and QD1 in which PG2 couples to QD2 first then QD2 influences QD1 through the interdot coupling. The last two terms in Eq. (\ref{eqn 26}) shows the cross-coupling effect where $C_{g21}V_{g2}E_{C1}$ is the cross-coupling energy between PG2 and QD1 and $C_{g12}V_{g1}E_{Cm}$ is the indirect cross-coupling energy that PG1 couples to QD2 first and QD2 influences QD1 through the interdot coupling. 
At low temperature ($k_{B}T$ $<$ $e^{2}/C$), the electrical transport through the DQD is only possible in the case where the energy levels in both dots are aligned with the source-drain bias window and this gives rise to the charge-stability diagram as shown in Fig. \ref{Fig13}(b). The outer numbers in brackets ($N_{1}$, $N_{2}$) are the stable electron numbers residing in the dot for that region and the condition for electron transport is met whenever three charge states meet in one point (the so-called triple point). The arrows in Fig. \ref{Fig13}(b) circling each triple point mark the route around the stability diagram that the system takes as electrons shuttle through. The counterclockwise path follows the sequence of charge state $(N_{1}, N_{2})\rightarrow(N_{1}+1, N_{2})\rightarrow(N_{1}, N_{2}+1)\rightarrow(N_{1}, N_{2})$, corresponding to moving an electron to the right. The clockwise path follows the sequence of charge state $(N_{1}+1, N_{2}+1)\rightarrow(N_{1}+1, N_{2})\rightarrow(N_{1}, N_{2}+1)\rightarrow(N_{1}+1, N_{2}+1)$, corresponding to moving a hole to the left. We here try to find a specific slope for $\mu_{1(2)}$ in the $V_{g1}$-$V_{g2}$ plane, along which $\mu_{1(2)}$ will remain constant for a given ($N_{1}$, $N_{2}$). We make the second row of Eq. (\ref{eqn 26}) and Eq. (\ref{eqn 27}) $=$ 0 which gives:

\begin{eqnarray}
V_{g1}\left(C_{g1}E_{C1}+C_{g12}E_{Cm}\right)=-V_{g2}\left(C_{g2}E_{Cm}+C_{g21}E_{C1}\right) \nonumber \\ \Rightarrow \frac{V_{g2}}{V_{g1}}=-\left(\frac{C_{g1}E_{C1}+C_{g12}E_{Cm}}{C_{g2}E_{Cm}+C_{g21}E_{C1}}\right),  (for \  \mu_{1}) \label{eqn 28} \\
V_{g2}\left(C_{g2}E_{C2}+C_{g21}E_{Cm}\right)=-V_{g1}\left(C_{g1}E_{Cm}+C_{g12}E_{C2}\right) \nonumber \\ \Rightarrow \frac{V_{g2}}{V_{g1}}=-\left(\frac{C_{g1}E_{Cm}+C_{g12}E_{C2}}{C_{g2}E_{C2}+C_{g21}E_{Cm}}\right),  (for \  \mu_{2}) \label{eqn 29}
\end{eqnarray}

We discuss the stability diagram for a double quantum dot with three different coupling regimes.
\begin{enumerate}
\item \textbf{No interdot and cross-capacitance coupling}\\
If we do not consider the cross-capacitance and interdot coupling (i.e., $C_{g12}=C_{g21}=E_{Cm}=0$), so that PG1 only influences QD1 and PG2 only influences QD2, Eq. (\ref{eqn 28}) and Eq. (\ref{eqn 29}) now read:
\begin{eqnarray} 
\frac{V_{g2}}{V_{g1}} &=& -\infty,  (for \  \mu_{1}) \label{eqn 30} \\
\frac{V_{g2}}{V_{g1}} &=& 0,  (for \  \mu_{2}) \label{eqn 31}
\end{eqnarray}
The resulting stability diagram is shown as Fig. \ref{Fig13}(c) where the lines for $\mu_{1(2)}$ to stay constant appear as vertical (horizontal) lines.
\item \textbf{Finite interdot but no cross-capacitance coupling}\\
As the interdot coupling or the cross-capacitance coupling opens, the gate PG1(2) has the ability to influence QD2(1). We first consider the case that interdot coupling is finite but the cross-capacitance coupling is weak, so $C_{g21}=C_{g12}=0$, $E_{Cm}\neq0$. In such a case the only way that PG1(2) influences dot 2(1) is to influence dot 1(2) first and through interdot capacitance to tune the other dot indirectly. So now Eq. (\ref{eqn 28}) and Eq. (\ref{eqn 29}) read,
\begin{eqnarray} 
\frac{V_{g2}}{V_{g1}}=-\left(\frac{C_{g1}E_{C1}}{C_{g2}E_{Cm}}\right),  (for \  \mu_{1}) \label{eqn 32} \\
\frac{V_{g2}}{V_{g1}}=-\left(\frac{C_{g1}E_{Cm}}{C_{g2}E_{C2}}\right),  (for \  \mu_{2}) \label{eqn 33}
\end{eqnarray}
The resulting stability diagram is shown as Fig. \ref{Fig13}(d). Instead of appearing as vertical (horizontal) lines, now $\mu_{1(2)}$ has a slope which is determined by the strength of the interdot coupling $E_{Cm}$. The larger $E_{Cm}$ is, the more $\mu_{1(2)}$ deviates from a vertical(horizontal) line. 
\item \textbf{No interdot but finite cross-capacitance coupling}\\
Finally, in the case of no interdot coupling but with cross-capacitance coupling, i.e., $C_{g12}\neq0$, $C_{g21}\neq0$, $E_{Cm}=0$, Eq. (\ref{eqn 28}) and Eq. (\ref{eqn 29}) read:
\begin{eqnarray} 
\frac{V_{g2}}{V_{g1}}=-\frac{C_{g1}}{C_{g21}},  (for \  \mu_{1}) \label{eqn 34} \\ 
\frac{V_{g2}}{V_{g1}}=-\frac{C_{g12}}{C_{g2}},  (for \  \mu_{2}) \label{eqn 35}
\end{eqnarray}
and the resulting stability diagram is shown in Fig. \ref{Fig13}(e) where the slopes are now determined by the ratio between the direct capacitance $C_{g1(2)}$ and the cross-capacitance $C_{g21(12)}$.
\end{enumerate}

\begin{figure}[!]
\begin{center}
		\includegraphics[scale=0.68]{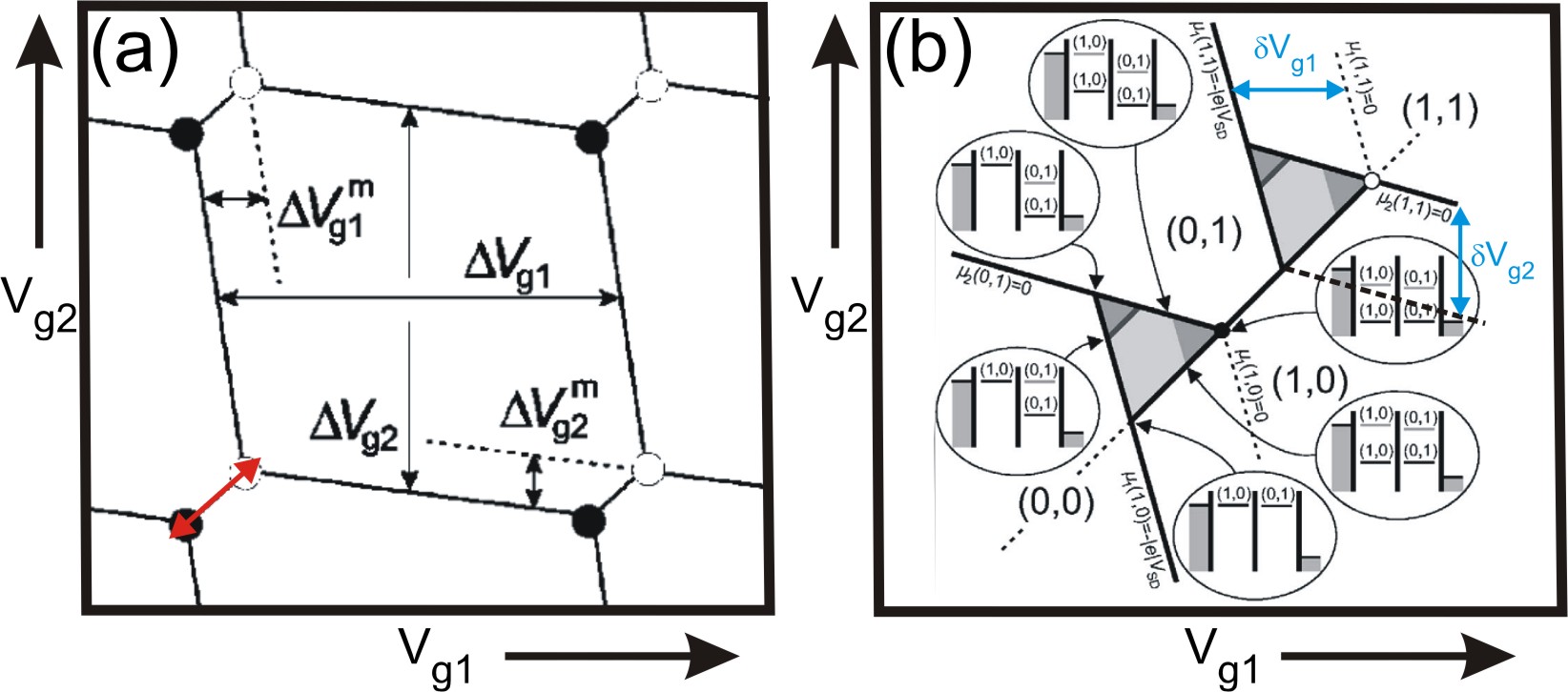}
\caption{(a) An enlarged stability diagram of Fig. \ref{Fig13}(d), with the dimensions for one hexagonal region. (b) Schematic for transport through the double dot at the triple points in the high bias regime. The electrochemical potentials at different points of the triangular regions are also indicated. (a) adapted with permission from ref. \cite{Wiel2002}. Copyright 2003 American Physical Society. (b) adapted with permission from ref. \cite{Hanson2007}. Copyright 2007 American Physical Society.}
\label{Fig14}
\end{center}	
\end{figure}

Usually a double-dot system has a finite interdot and weak cross-capacitance coupling strength so the charge-stability diagram is made up of hexagonal regions of a fixed charge, as shown in Fig. \ref{Fig13}(d) [also an enlarged illustration in Fig. \ref{Fig14}(a)]. The dimensions of the hexagonal regions as indicated in Fig. \ref{Fig14}(a) are given by \cite{Wiel2002}:
\begin{eqnarray}
\Delta V_{g1} &=& e/C_{g1} \label{eqn 36} \\
\Delta V_{g2} &=& e/C_{g2} \label{eqn 37} \\
\Delta V_{g1}^m &=& \Delta V_{g1}\frac{C_m}{C_{2}} \label{eqn 38} \\ 
\Delta V_{g2}^m &=& \Delta V_{g2}\frac{C_m}{C_{1}} \label{eqn 39} 
\end{eqnarray} 

In the high bias regime, the triple points evolve into bias-dependent triangular regions where the two dot levels lie within the bias window as shown in Fig. \ref{Fig14}(b). The dimensions of the triangles are related with the applied bias via \cite{Wiel2002}:
\begin{eqnarray}
\alpha_1 \delta V_{g1} & = & \frac{C_{g1}}{C_1}e \delta V_{g1} = |eV_{SD}| \label{eqn 42} \\
\alpha_2 \delta V_{g2} & = & \frac{C_{g2}}{C_2}e \delta V_{g2} = |eV_{SD}| \label{eqn 43}
\end{eqnarray}
Here $\alpha_{1(2)}$ is the conversion factor between gate voltage and energy which could be extracted from the dimension of the bias triangle. Therefore, the charging energy of the dots and the interdot coupling energy can be found:

\begin{eqnarray}
E_{C1} &=& e^{2}/C_{1} = \frac{\alpha_1e}{C_{g1}} = \alpha_1\Delta V_{g1} \label{eqn 44} \\
E_{C2} &=& e^{2}/C_{2} = \frac{\alpha_1e}{C_{g2}} = \alpha_2\Delta V_{g2} \label{eqn 45} \\
E_{Cm} &=& \alpha_1\Delta V^{m}_{g1} = \alpha_2\Delta V^{m}_{g2} \label{eqn 46}
\end{eqnarray}

\begin{figure}
\begin{center}
		\includegraphics[scale=0.67]{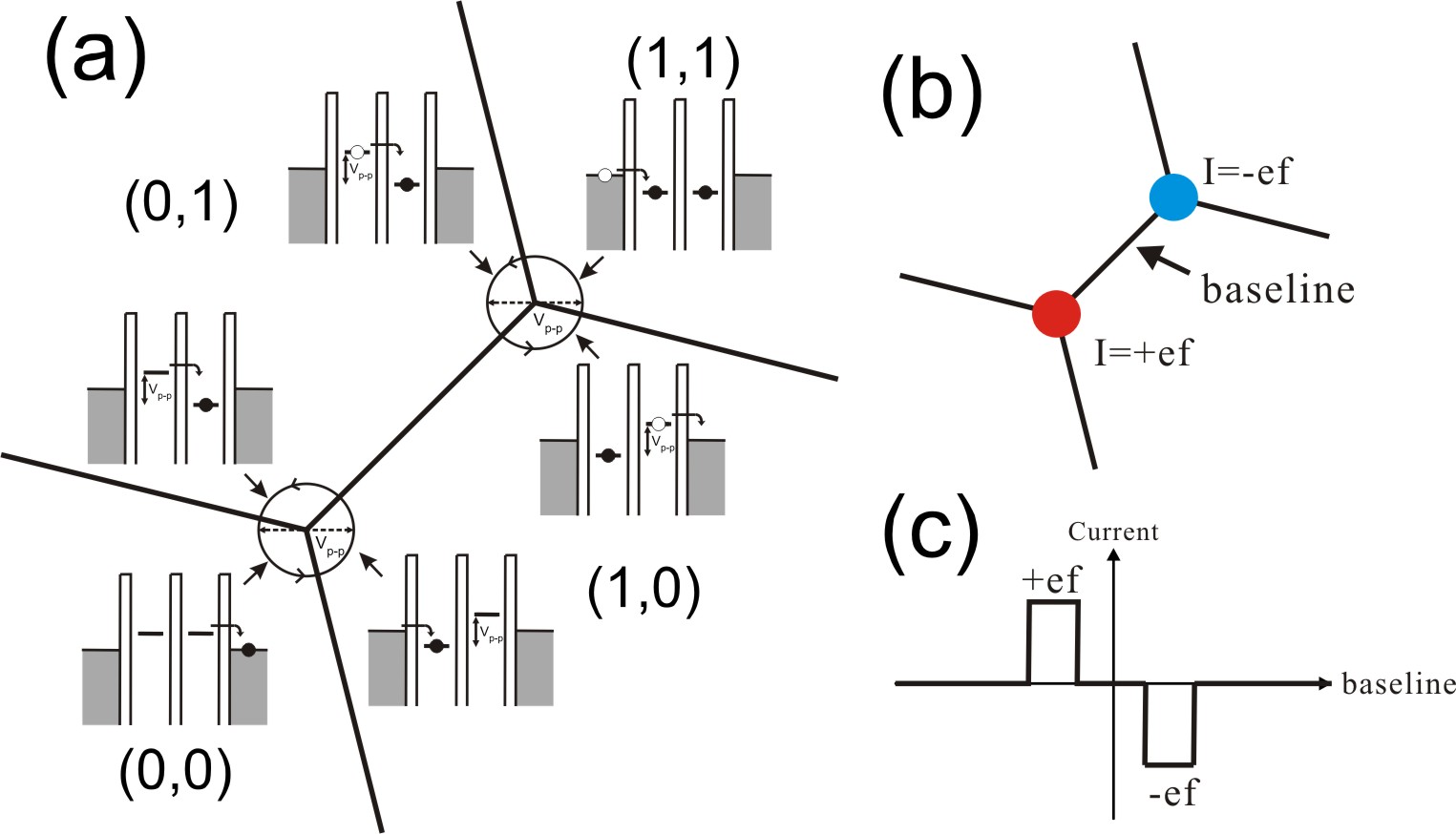}
\caption{(a) The pump loop (counterclockwise close trajectory) and its corresponding charge configuration obtained by modulating the gate voltages by two out-of-phase RF signals, induces one electron to go around the circuit. V$_{p-p}$/2 is the RF amplitude. (b) Schematic for the resulting current map from (a). If the pump is working, one should observe a quantized current $I$ = $\pm ef$ at the location of two nearby triple points. (c) Schematic for the linecut along the baseline in (b). Reproduced from K. L. Chiu, PhD Thesis, 2012.}
\label{Fig15}
\end{center}	
\end{figure}

A phenomenon closely related to the manipulation of the dot levels in double quantum dots is charge pumping. Charge pumping refers to a quantized number $n$ of electrons transferred from the source to the drain at a driven frequency $f$, leading to a total current $I$ $\equiv$ $nef$ even if zero bias is applied ($e$ is the elementary charge). Such quantized charge transport was first demonstrated in single-electron turnstile devices in which an external radio-frequency (RF) signal was applied to linear arrays of tunnel junctions. By doing so electrons could be clocked through each tunnel junction one at a time by exploiting the Coulomb blockade effect \cite{Geerligs1990,Kouwenhoven1991}. When an RF signal is applied to the plunger gates (instead of barriers) of a double quantum dot device, it is also possible to generate an accurate and frequency-dependent quantized current through the device. The AC voltages on both plunger gates with a phase difference between them drives the DQD into different charge states around the triple point. The schematic diagram to illustrate such a pumping mechanism is shown in Fig. \ref{Fig15}(a). Assuming the voltages applied on the plunger gates are AC sinusoidal waves with a phase difference of 90 degrees, it effectively forms a circular pump loop in the stability diagram. The radius of the circle is determined by the amplitude of the sinusoidal wave ($V_{p-p}$/2). When the circular route passes through three charge states around the triple point, it corresponds to shuttling a charge carrier from source reservoir to drain reservoir and generating a current. If the AC amplitude is small enough for the pump loop to just enclose a triple point and the frequency is large enough to produce a measurable pumping current, a current $I=ef$ will follow even when zero source-drain voltage is applied. Depending on the type of triple point that the pumping circle encloses, it generates a different direction of current; i.e., positive current for the electron-transport-type triple point and negative current for the hole-transport-type triple point. So if the pumping is successful the current recorded around two nearby triple points will present a circular shape with equal values but different signs as shown in Fig. \ref{Fig15}(b). This effect can be seen in a linecut along the baseline of triple points, where the current appears as two plateaus as shown in Fig. \ref{Fig15}(c). The experimentally observed quantized pumped current in graphene double dot will be presented in section 3.2.2.

\subsection{2.3 Andreev reflections in ballistic S-N-S Josephson junctions}

Having discussed the fundamental physics related to transport in quantum dots, here we introduce another important topic in this chapter - the proximity effect in Josephson junctions. This section will server as a basis for understanding the superconducting physics in 2D material-based Josephson junctions, as will be discussed in section 5. To start with the proximity effect, we use the following defining line \cite{Klapwijk2004}: If a normal metal N is deposited on top of a superconductor S, and if the electrical contact between the two is good, Cooper pairs can leak from S to N. In such a way, the normal metal acquires some superconducting-like properties at a low temperature. This proximity effect is a well-known phenomenon in superconductivity for over 50 years, and is still attracting enormous interests owing to its rich physics underneath. The key mechanism responsible for the proximity effect, the Andreev reflection, offers phase correlations in a system without interacting electrons at mesoscopic scales, is the main topic to be introduced below. Andreev reflection is a microscopic process that happens in a S-N-S junction, in which single particles in the normal region cannot enter the superconductor and therefore experience a special type of reflection at each S-N interface. This process results in Andreev bound states, which are capable of carrying superconducting current across the normal region. Thus, one can say Andreev reflection and the proximity effect are intimately connected and not two distinct phenomena. In the following, we adopt the discussions in ref. \cite{Wang2016} to illustrate the process of Andreev reflections in a S-N-S junction [Fig. \ref{Fig39}]. Assuming an incident electron with energy $E_k$ and spin $\sigma$ ($E_k-E_F$ $\leq$ $\Delta$, where $E_F$ is the Fermi energy of the normal metal and $\Delta$ is the energy gap of the superconductor) is moving toward the N/S$_1$ interface. The incoming electron would grab an electron with a spin and momentum that is opposite to its own, thereby forming a Cooper pair that can propagate freely into the superconductor S$_1$. In order to conserve the momentum, spin, and charge, this process leaves behind an empty electronic state (hole) with the opposite spin -$\sigma$ and wave vector $-k$ as shown in the Fig. \ref{Fig39} by the dashed red arrow. The bounced hole follows the time-reversed trace of the incoming electron and eventually hit the N/S$_2$ interface, where another Andreev reflection takes place. The hole will pass through the N/S$_2$ interface with another hole excitation that pairs with it (the Cooper pair of hole-like). A Cooper pair of electrons in superconductor S$_2$ is thus annihilated, resulting in an electron with momentum $k$ and spin $\sigma$ (identical to the electron we started with), ejected into the normal metal and completes a roundtrip. In completing each roundtrip, one Cooper pair in the left superconductor is annihilated while another one in the right superconductor is created, leading to a transfer of Cooper pair from left to right. This microscopic process gives rise to the transport of supercurrent across the S-N-S junction, in which current flows without dissipation from one superconductor to another by passing through a normal metal that is not inherently superconducting. Andreev reflections in a S-N-S junction can lead to a series of bound states, known as Andreev bound state (ABS), whose energy strongly depends on the length of junction and the phase acquired in each roundtrip. To understand the mechanism of ABS, let us consider the phase an electron would acquire through the Andreev reflection that convert the incoming electron into a hole at the N/S$_1$ interface:

\begin{equation}
\ \phi_{e\rightarrow h} = \phi_1 + arccos(\frac{E}{\Delta}),
\label{eqn 58}
\end{equation} where $\phi_1$ denotes the phase of the superconductor S$_1$ and $E$ is the excitation energy of electron measured with respect to $E_F$ ($E$ = $E_k-E_F$). The first phase term arises from the requirement that particles absorbed by the superconductor must be in phase with the macroscopic wave function that describes the condensate. The second phase term $arccos(\frac{E}{\Delta})$ comes from the reflection probability amplitude, which depends on the relative strength of the excitation energy $E$ and the barrier $\Delta$ \cite{Blonder1982}. Similarly, the phase a hole acquired through the Andreev reflection at the N/S$_2$ interface can be written as:

\begin{equation}
\ \phi_{h\rightarrow e} = -\phi_2 + arccos(\frac{E}{\Delta})
\label{eqn 59}
\end{equation} In addition to the phase associated with Andreev reflections, we also need to consider the dynamic phase $\phi_{e(h)} = \int k_{e(h)}\cdot d\vec{l}$ that electron (hole) acquired when traveling between the superconductors. In a one-dimensional ballistic case (no scattering), the phase accumulated by an electron traveling from S$_2$ to S$_1$ can be simply written as:

\begin{equation}
\ \phi_{e} = k_e \cdot L = \sqrt{\frac{2m}{\hbar^2}(E+E_F)}\cdot L
\label{eqn 60}
\end{equation}, where $L$ is the length of the normal metal region. Similarly, the dynamic phase a hole acquired when traveling from S$_1$ to S$_2$ is:

\begin{equation}
\ \phi_{h} = k_h \cdot (-L) = \sqrt{\frac{2m}{\hbar^2}(E_F-E)}\cdot (-L)
\label{eqn 61}
\end{equation} As $E \ll E_F$, the total dynamic phase can be approximated as:

\begin{figure}
\includegraphics[scale=0.53]{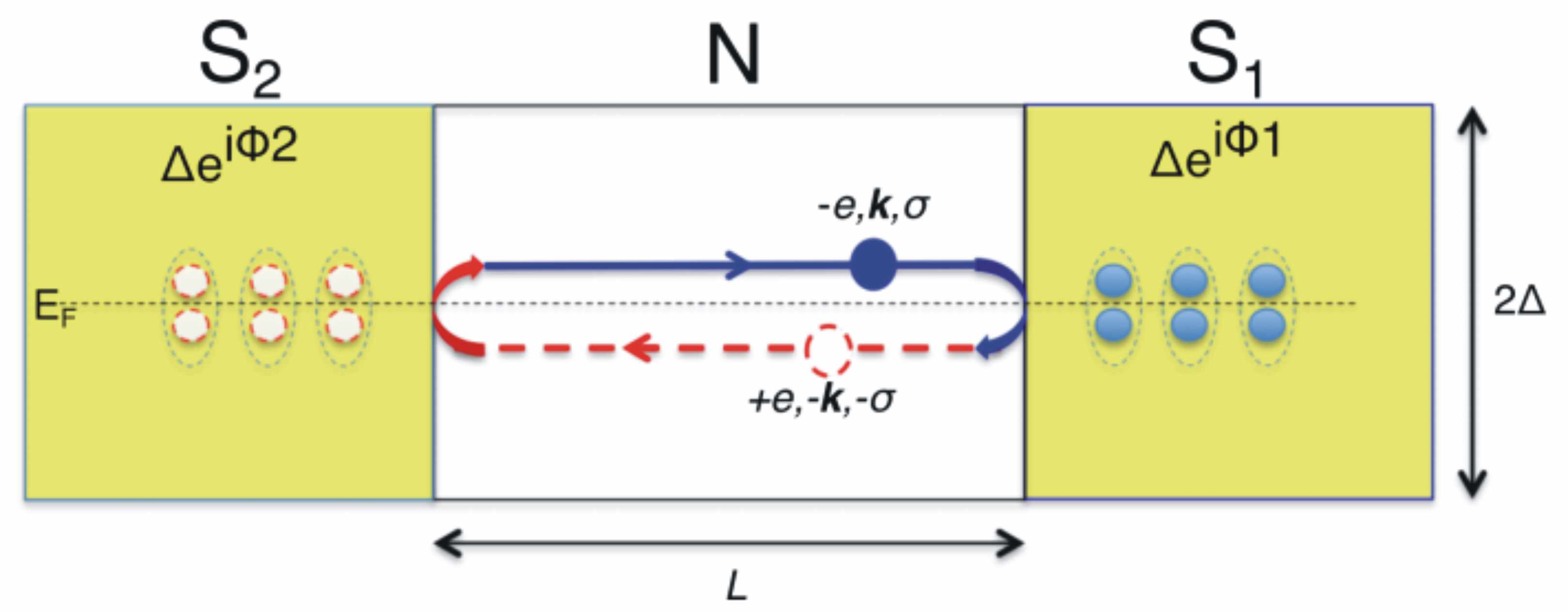}
\caption{An S-N-S junction and Andreev roundtrips responsible for transporting Cooper pairs across a ballistic normal metal weak-link. Reproduced with permission from ref. \cite{Wang2016}.}  
\label{Fig39}
\end{figure}

\begin{equation}
\ \phi_{e} + \phi_{h} = k_F L \frac{E}{E_F}
\label{eqn 62}
\end{equation} Therefore, the ABS energy is determined by requiring the total phase accumulated in the Andreev process (roundtrip) to be multiples of 2$\pi$ (i.e., $\phi_{e\rightarrow h}$+$\phi_{h\rightarrow e}$+$\phi_{e} + \phi_{h}$ = 2$\pi n$), which can be written as followed \cite{Kulik1969}:  

\begin{equation}
\ 2\ arccos(\frac{E}{\Delta}) \pm (\phi_{1}-\phi_{2}) + k_F L \frac{E}{E_F} = 2\pi n,
\label{eqn 63}
\end{equation} where $n$ is an integer and $\pm$ accounts for the two possible directions the roundtrip can take. Note that in the above expressions, we have limited our discussion to ballistic normal metal, meaning that no scattering event (which usually introduces extra phases) takes place within the roundtrip. As can be seen now, the ABS energy depends on the phase difference between the two superconductors and the junction length. The term $k_F L \frac{E}{E_F}$ provides a criteria to estimate the short and long-junction limits. The junction is "short" if this term is negligible, so that the phase depends almost completely on the Andreev reflection and is insensitive to the dynamic phase associated with geometry. The short junction limit corresponds to a condition $L \ll \xi$, where $\xi$ = $\sqrt{\hbar D/\Delta}$ is the superconducting coherence length, with $D$ = $\nu_F l_e/2$ the the Einstein diffusion coefficient, $\nu_F$ the Fermi velocity and $l_e$ the mean free path in normal metal \cite{Bretheau2017}. In the short-junction limit, the bound state energy can be solved from Eq. (\ref{eqn 63}) as \cite{Klapwijk2004a}: 

\begin{equation}
\ E^{\pm}_{L \ll \xi} (\Delta\phi) = \pm\Delta \ cos(\frac{\Delta\phi}{2}), \ -\pi \leq \Delta\phi \leq \pi
\label{eqn 64}
\end{equation} where $\Delta\phi$ = $\phi_1 - \phi_2$ is the phase difference between two superconductors. Note that only a pair of ABS exist within the superconducting gap $\Delta$. In the opposite limit, where $L$ is much larger than the phase coherence
length $\xi$, the phase acquired in each roundtrip is dominated by the dynamic phase associated with $L$ and the ABS energy in this long junction limit can be written as \cite{Klapwijk2004a}:

\begin{equation}
\ E^{\pm}_{L \gg \xi} (\Delta\phi) = \frac{\hbar\nu_F}{L}\left[(n-\frac{1}{2})\pi\mp\frac{\Delta\phi}{2} \right], \ -\pi \leq \Delta\phi \leq \pi
\label{eqn 65}
\end{equation} Due to the weaker quantum confinement in a long junction as compared to that in a short junction, the energy spacing between bound states decreases with $L$ and thus multiple bound states (denoted by the index $n$) can be accommodated within the gap. Here, we emphasize again that the above formulas are based on the assumption of a ballistic (scattering-free) normal metal in a S-N-S junction. However, the general results presented here (especially the oscillation behavior with phase $\Delta\phi$) still qualitatively cover the experimental results presented in this chapter. For ABS considering the scattering from defects, more information can be found in ref. \cite{Bagwell1992}. The phase dependent ABS energy in Eq. (\ref{eqn 64}) and Eq. (\ref{eqn 65}) provide a way to study the ABS spectrum in devices capable of varying magnetic flux, as will be discussed in section 5.

\ \\ 
In summary, we have introduced the relevant physics useful to understand the transport properties in quantum dots and Josephson junctions. In the subsequent sections, we will review a series of experimental studies relevant to developing qubits in 2D materials. Following the development of spin qubits, we will discuss single-electron transport properties of various graphene nanostructures in section 3, while the same properties of 2H-TMDs nanostructures will be reviewed in section 4. For potential use in superconducting qubits, we investigate the Josephson effects of 2D material-based S-N-S junctions in section 5. In section 6, we provide recent studies on QSH edge states in 1T'-TMDs and discuss their potential applications in probing Majorana zero modes.  

\maketitle
\section{3. Single Electron Transport in Graphene}
In this section, we will review the early development of graphene nanostructures fabricated on SiO$_{2}$/Si substrates. After briefly introducing graphene nanoribbons and their function as tunnel barriers, we will focus mainly on graphene quantum dots and their transport properties.  

\begin{figure}
\begin{center}
		\includegraphics[scale=0.7]{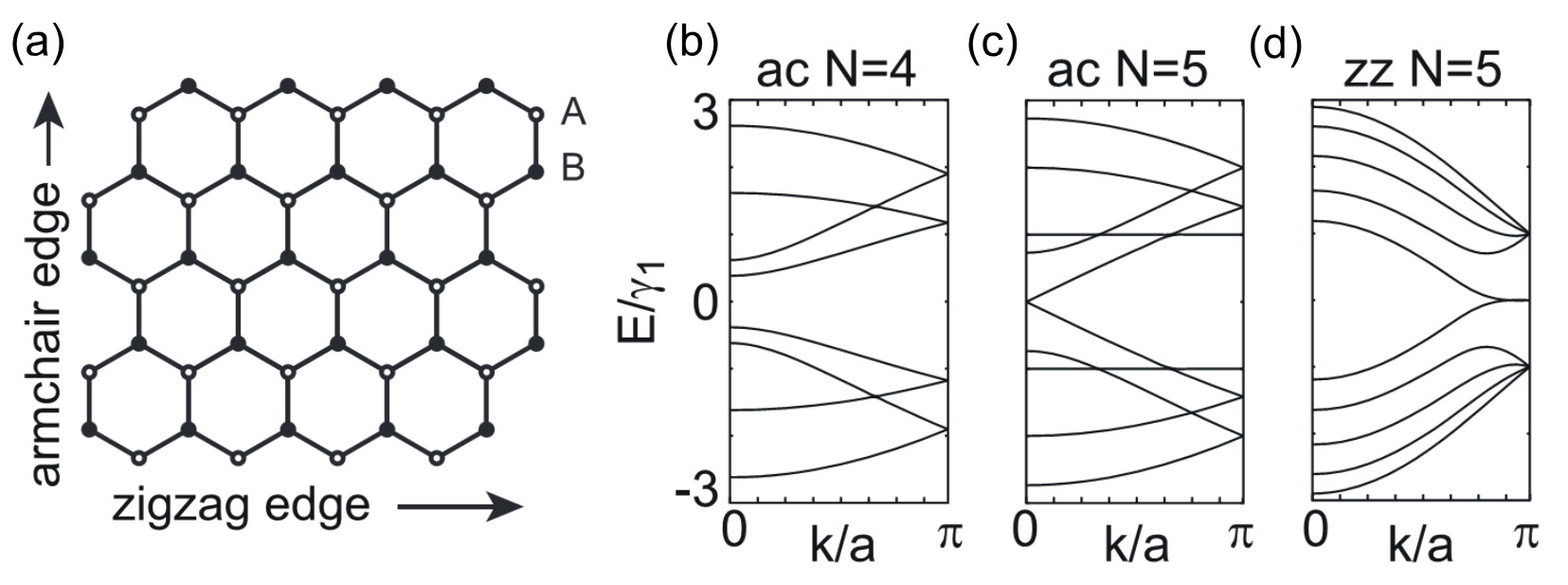}
\caption{(a) Lattice of a zigzag (armchair) graphene nanoribbon by extension in the x-(y-) direction. (b-d) Tight-binding calculations of the nanoribbon subbands for (b) an $N$ = 4 semiconducting armchair (ac) nanoribbon, (c) an $N$ = 5 metallic armchair (ac) nanoribbon and (d) an $N$ = 5 metallic zigzag (zz) nanoribbon. $N$ denotes the number of the dimer (carbon-site pair) lines for the armchair ribbon and the number of the zigzag lines for the zigzag ribbon, respectively. Adapted with permission from ref. \cite{Guttinger2012}. Copyright 2012 IOP Publishing Ltd.}
\label{Fig16}
\end{center}	
\end{figure}

Although graphene is a superb conductor which offers advantages in terms of sensing and analog electronics, its gapless bandstructure hinders its use in logic circuit applications. Owing to the absence of a band gap, the current in graphene cannot be completely turned off, leading to low on/off ratios that are insufficient for switches \cite{Geim2007}. Engineering band gaps in graphene is thus a major challenge that must be addressed to enable the use of graphene-based transistors in digital electronics. First-principle calculations predict that cutting graphene into one-dimensional nanoribbons can open up a scalable band gap $E_{g}$ = $\alpha$/$w$, where $w$ is the nanoribbon width and $\alpha$ is in the range of 0.2 eV$\cdot$nm to 1.5 eV$\cdot$nm, depending on the model and the crystallographic orientation of the edges \cite{Lin2008a,Stampfer2011}. Similar results are also obtained from tight-binding calculations \cite{Nakada1996,Wakabayashi1999}. A GNR can have two possible types of edge terminations, namely, armchair and zigzag edges, as shown in Fig. \ref{Fig16}(a). These two edge types correspond to different boundary conditions, from which the energy band dispersion can be found. The tight-binding calculated energy band structures for armchair GNRs (of two different ribbon widths) and zigzag GNRs are shown in Fig. \ref{Fig16}(b) - (d), where $N$ denotes the number of dimer (carbon-site pair) lines (for the armchair ribbons) or the number of zigzag lines (for the zigzag ribbons). The band dispersion for an armchair nanoribbon with $N$ = 3$m$ $-$ 2 dimers exhibits a band gap (semiconducting), whereas for an armchair nanoribbon with $N$ = 3$m$ $-$ 1 dimers, the dispersion is metallic ($m$ is an integer). For semiconducting ribbons, the direct gap decreases with increasing ribbon width and tends toward zero in the limit of very large $N$. Zigzag nanoribbons always exhibit metallic behavior [Fig. \ref{Fig16}(d)] regardless of how the width ($N$) is varied. The predicted existence of band gaps in GNRs has motivated an experimental effort to establish whether nanostructuring graphene is a feasible route for preparing graphene-based switches \cite{Han2007,Todd2008,Molitor2009a,Bai2010,Connolly2011,Wang2011,Jiao2010,Wei2013b}. GNRs can be fabricated by means of O$_{2}$ plasma etching using physical masks \cite{Han2007,Todd2008,Molitor2009a,Bai2010,Connolly2011}, unzipping carbon nanotubes \cite{Wang2011,Jiao2010,Wei2013b}, gas phase etching \cite{Wang2010a} or functionalization \cite{Withers2011,Lee2011}. Such devices have been tested for their transport properties at various temperatures, and the general results will be discussed below.

\begin{figure*} 
\includegraphics[scale=0.9]{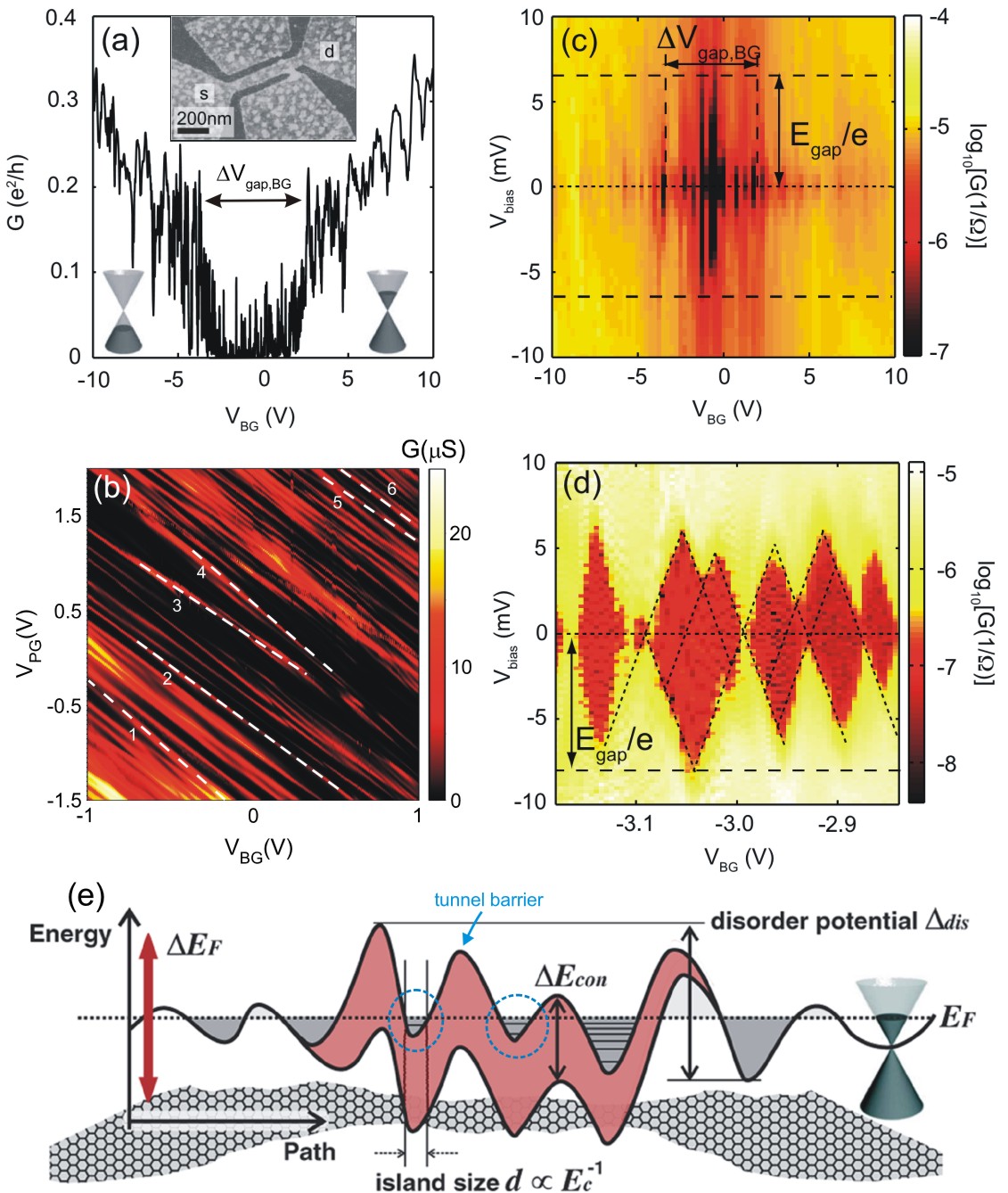}
\caption{(a) Conductance through a nanoribbon (shown in the insert) as a function of the back-gate voltage $V_{BG}$, recorded under an applied bias voltage of $V_{bias}$ = 300 $\mu$V and at a temperature of $T$= 2 K. Insert: Atomic force microscopy image of a graphene nanoribbon ($w$ = 85 nm, $l$ = 500 nm) etched using O$_{2}$ plasma. (b) Conductance as a function of $V_{BG}$ and the plunger-gate voltage $V_{PG}$ of another GNR device, showing a variation in the slopes of the Coulomb resonances (indicated by dashed lines and numbers). (c) Logarithmic conductance as a function of $V_{bias}$ and $V_{BG}$ at $T$= 2 K for the device shown in (a), with indication of the extent of the transport gap $\Delta V_{gap,BG}$ in the back-gate direction and bias-gap $E_{gap}$ in the bias direction. (d) Zoom-in of the region of suppressed conductance depicted in (c). (e) Schematic illustration of the formation of localized states induced by disorders. $\Delta_{dis}$ characterizes the strength of the charge-neutrality point fluctuation and $\Delta E_{con}$ is a confinement gap induced by local constriction. $\Delta E_{F}$ denotes the Fermi energy spacing that the transport gap has to overcome. (a, c, d) adapted with permission from ref. \cite{Molitor2009a}. (b) reproduced from K. L. Chiu, PhD Thesis, 2012. (e) adapted with permission from ref. \cite{Stampfer2009}. Copyright 2009 American Physical Society.}  
\label{Fig17}
\end{figure*}

Fig. \ref{Fig17}(a) shows the conductance of an O$_{2}$ plasma etched GNR [inset of Fig. \ref{Fig17}(a)] as a function of the voltage applied to the back-gate. This back-gate sweep shows a typical V-shape, with a region around 0 V separating the hole- from electron-transport regime where the conductance is strongly suppressed. In contrast to the prediction of energy gaps in clean GNRs (i.e., without considering bulk disorder and edge roughness), where transport should be completely pinched-off, this gap exhibits a large number of conductance peaks reminiscent of Coulomb blockade resonances in quantum dots. The nature of these resonances can be interrogated by varying the potential of the GNR. Fig. \ref{Fig17}(b) shows the conductance as a function of both back-gate and plunger-gate (an in-plane gate close to the GNR) voltages within the transport gap. The conductance resonances exhibiting a range of relative lever arms indicated by dashed lines are present over a wide range of $V_{BG}$ and $V_{PG}$ voltages. One explanation for this behavior draws on its similarity to a series of charge islands (or QDs), each coupled to the plunger-gate through different capacitive coupling strength, assuming the lever arm of the back-gate to the charge islands is nearly constant all over the GNR. More information about such localized states in the GNR can be gleaned by the Coulomb diamond measurements (see section 2.1), in which the differential conductance as a function of back-gate voltage and source-drain bias is recorded, as shown in Fig. \ref{Fig17}(c). Within this picture, the extent in bias voltage of the diamond-shaped regions of suppressed current [see $E_{gap}$/$e$ in Fig. \ref{Fig17}(c) and its zoom-in in Fig. \ref{Fig17}(d)] is a direct indication of the charging energy of the dots (see section 2.1), which fluctuates strongly with $V_{BG}$ and extends to $\approx$ 8.5 meV. The overlapping diamonds in Fig. \ref{Fig17}(d) resembles the behavior of a QD network \cite{Dorn2004}, supporting the notion that multiple QDs form along the GNR. In addition, the gap in Fermi energy $\Delta E_{F}$ corresponding to the transport gap $\Delta V_{gap,BG}$ can be estimated using $\Delta E_{F}$ $\approx$ $\hbar \nu_{F}$$\sqrt{2\pi C_{g}\Delta V_{gap,BG}/\left|e\right|}$, where $C_{g}$ is the back-gate capacitance per area and $\nu_{F}$ is the Fermi velocity in graphene \cite{Stampfer2009,Guttinger2012}. This leads to an energy gap $\Delta E_{F}$ $\approx$ 110 - 340 meV which is significantly larger than the observed $E_{gap}$ (8.5 meV) and the band gaps $\Delta E_{con}$ ($\leq$ 50 meV) estimated from calculations of a GNR with width $W$ = 45 nm \cite{Guttinger2012}.

A schematic model shown in Fig. \ref{Fig17}(e) is able to qualitatively explain the findings described above \cite{Stampfer2009}. This model consists of a combination of quantum confinement energy gap $\Delta E_{con}$ (the intrinsic band-gap of a clean GNR) and strong bulk and edge-induced disorder potential fluctuation $\Delta _{dis}$. The confinement energy $\Delta E_{con}$ alone can neither explain the observed energy scale $\Delta E_{F}$, nor the dots formation in the GNR. However, superimposing a fluctuation in the disorder potential ($\Delta _{dis}$) can result in tunnel barriers separating different localized states (i.e., puddles or QDs), as shown in in Fig. \ref{Fig17}(e). Therefore, transport in such a system is described by a percolation between the puddles [in Fig. \ref{Fig17}(e) the dashed circles indicate the puddles, whereas the blue arrow indicates the tunnel barrier]. Within this model, $\Delta E_{F}$ depends on both the confinement energy gap and the disorder potential fluctuation, and can be approximated using the relation $\Delta E_{F}$ = $\Delta _{dis}$ $+$ $\Delta E_{con}$. $\Delta _{dis}$ can be estimated from the bulk carrier density fluctuations $\Delta n$ (due to substrate disorder) using $\Delta _{dis}$ = $\hbar \nu_{F}$$\sqrt{4\pi \Delta n}$, where $\Delta n$ $\approx$ $\pm$2 $\times$ 10$^{11}$ is extracted from ref. \cite{Martin2008}. This in turns gives $\Delta E_{F}$ = $\hbar \nu_{F}$$\sqrt{4\pi \Delta n}$ $+$ $\Delta E_{con}$ $\approx$ 126 meV \cite{Guttinger2012}, which is comparable to the experimental value (110 - 340 meV). The energy gap in the bias direction ($E_{gap}$) is not directly related with the magnitude of the disorder potential but rather with its spatial variation. When the Fermi energy (or said $V_{BG}$) lies in the center of the transport gap, the smaller localized states are more likely to form, giving rise to the larger charging energies (larger Coulomb diamonds). By contrast, when the Fermi energy is tuned away from the charge-neutrality point, the size of the relevant diamonds gets generally smaller due to the merging of individual puddles. 

Although the localized states in GNRs pose additional complications, their tunability in resistances still allows them to be used as tunnel barriers for transport in GQDs. While a large number of studies on GNRs have been reported in the field; however, in this section, we will focus primarily on GQDs in which GNRs are used as tunnel barriers. Further discussion of the transport properties of GNRs can be found in ref. \cite{Bischoff2015b}. 

\subsection{3.1 Graphene single quantum dots on SiO2/Si substrates}

\begin{figure}
\includegraphics[scale=1.15]{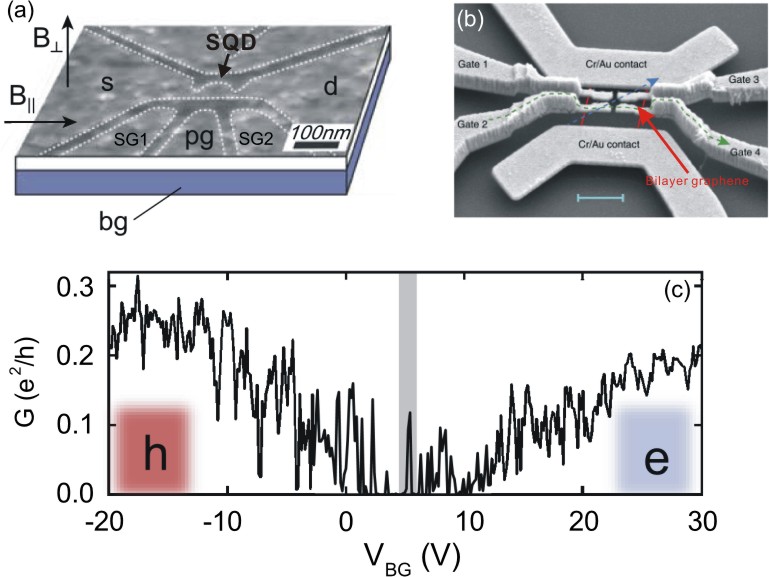}
\caption{(a) Atomic Force Microscope image of a graphene single quantum dot ($\approx$ 50 nm wide and $\approx$ 80 nm long) etched by O$_{2}$ plasma. (b) Scanning Electron Micrograph of a suspended bilayer GQD device. Bilayer graphene (highlighted by red line) is suspended between two electrodes below local top gates. Scale bar, 1 $\mu$m. (c) Source-drain conductance as a function of back-gate voltage $V_{BG}$ at bias $V_{b}$ = 4 mV measured from the device shown in (a). (a, c) adapted with permission from ref. \cite{Guttinger2009}. Copyright 2009 American Physical Society. (b) adapted with permission from ref. \cite{Allen2012}. Copyright 2012 Nature Publishing Group.}  
\label{Fig18}
\end{figure}

Owing to the expected long spin relaxation time, graphene quantum dots (GQDs) are considered to be a viable candidate for preparing spin qubits and spintronic devices \cite{Trauzettel2007}. Over the past decade, GQDs have proven to be a useful platform for confining and manipulating single electrons \cite{Gutinger2008,Volk2013,Guttinger2009,Chiu2012,Guttinger2010,Liu2010,Volk2011,Connolly.2013a}. In this section, we will review a few relevant transport experiments performed on graphene single quantum dots (GSQDs) fabricated on SiO$_{2}$/Si substrates. These include the Coulomb blockade at zero field, Fock-Darwin spectrum, spin states and charge relaxation dynamics, as will be discussed below. 

\subsection{3.1.1 Coulomb blockade at zero field}

GQDs can be formed by etching isolated islands connected to source and drain graphene reservoirs via nanoconstrictions that are resistive enough to act as tunnel barriers \cite{Gutinger2008,Volk2013,Guttinger2009}. An example of such a device is shown in Fig. \ref{Fig18}(a), in which in-plane graphene side and plunger gates (SG1, SG2, PG) are used to locally tune the potential of the tunnel barriers and the 50 nm diameter dot, while the doped-silicon back-gate (BG) is used to adjust the overall Fermi level. Another way to define a GQD is to induce a band-gap in bilayer graphene by applying an electric field perpendicular to the layers; in this way, charges are confined in an island defined by top gate geometry \cite{Goossens2012,Allen2012}. Such a structure can be seen in Fig. \ref{Fig18}(b), where a bilayer graphene is suspended between two Cr/Au electrodes and sits below suspended local top gates that are used to break interlayer symmetry. Graphene quantum dots can also be formed from the disorder potential \cite{Zhang2009,Amet2012}, strain engineering \cite{Klimov2012} and gated GNRs \cite{Liu2010}, in all of which Coulomb blockade can be observed. 

\begin{figure}[!t]	
\includegraphics[scale=1.2]{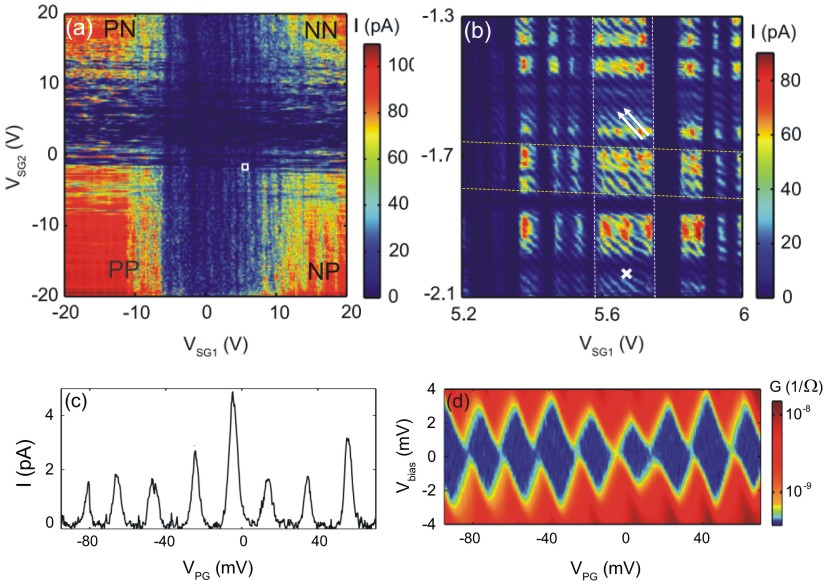}
\caption{(a) Current through a GQD (diameter $\approx$ 180 nm) as a function of two side-gate voltages $V_{SG1}$ and $V_{SG2}$. (b) Current as a function of $V_{SG1}$ and $V_{SG2}$ in the range indicated by the white square in (a). (c) Current as a function of plunger gate voltage $V_{PG}$ at $V_{SG1}$ = 5.67 V and $V_{SG2}$ = -2.033 V [the white cross in (b)]. (d) Coulomb diamonds associated with the Coulomb resonances in (c). Adapted with permission from ref. \cite{Molitor2011}.}  
\label{Fig19}
\end{figure}

Fig. \ref{Fig18}(c) shows the back-gate sweep (conductance as a function of back-gate voltage) of the device shown in Fig. \ref{Fig18}(a). The measurement shows a transport gap ranging from 0 $\leq$ $V_{BG}$ $\leq$ 10 V, in which current is suppressed except for multiple sharp Coulomb resonances, separating hole- from electron-transport regime. The transport gap resulting from the GNR tunnel barriers can be lifted using the side-gate voltage. Fig. \ref{Fig19}(a) shows the current measurements of another GQD (diameter $\approx$ 180 nm) as a function of its side-gate voltages $V_{SG1}$ and $V_{SG2}$ at a fixed back-gate voltage within the transport gap. There is a cross-like region of suppressed current separating four large conductance regions, which correspond to different doping configurations of the constrictions, labeled as NN, NP, PP and PN at the corners of the diagram, respectively. For example, keeping $V_{SG1}$= -20 V constant and sweeping $V_{SG2}$ from -20 V to $+$20 V keeps constriction 1 in the $p$-doped regime whereas constriction 2 is tuned from $p$-doped to $n$-doped (PP to PN transition). In order to observe single electron transport, it is necessary to operate in a region of gate space where both tunnel barriers are resistive (i.e., within the center of the cross-like current suppressed regime). Fig. \ref{Fig19}(b) shows the case with the Fermi energy located at the edge of the transport gap for both constrictions [marked by the white square in Fig. \ref{Fig19}(a)]. The measurement shows broaden vertical and horizontal resonances [white and yellow dashed lines in Fig. \ref{Fig19}(b)], which correspond to resonant transmission through the localized states in the left and right constrictions, tuned with the respective side-gate. The fact that those lines are almost perfectly vertical and horizontal indicates that the side-gate only influences its adjacent constriction. A closer inspection of Fig. \ref{Fig19}(b) shows a series of diagonal lines (indicated by arrows), which correspond to the Coulomb blockade resonances from the central quantum dot, where both side gates are expected to have a similar lever arm. These 0D Coulomb resonances can be unambiguously resolved as a series of well-defined and regular peaks, as shown in Fig. \ref{Fig19}(c), by sweeping a plunger gate voltage $V_{PG}$ with sides gates fixed at $V_{SG1}$ = 5.67 V and $V_{SG2}$ = −2.03 V [the white cross in Fig. \ref{Fig19}(b)]. A Coulomb diamond measurement of these resonances further confirms their origin. A charging energy $E_{C}$ $\approx$ 3.2 meV is extracted from the vertical extent of the Coulomb diamonds shown in Fig. \ref{Fig19}(d), in reasonable agreement with the dot diameter if the Disc plate capacitance model $E_{C}$=$e^{2}/8\epsilon\epsilon_{0}r$, where $r$ is the radius of the quantum dot, is used \cite{Molitor2011}. In the following sections, we discuss how these Coulomb blockade peaks evolve with the applied perpendicular and in-plane magnetic fields.

\subsection{3.1.2 Electron-hole crossover in perpendicular magnetic field}

\begin{figure*}[!t]	
\includegraphics[scale=0.7]{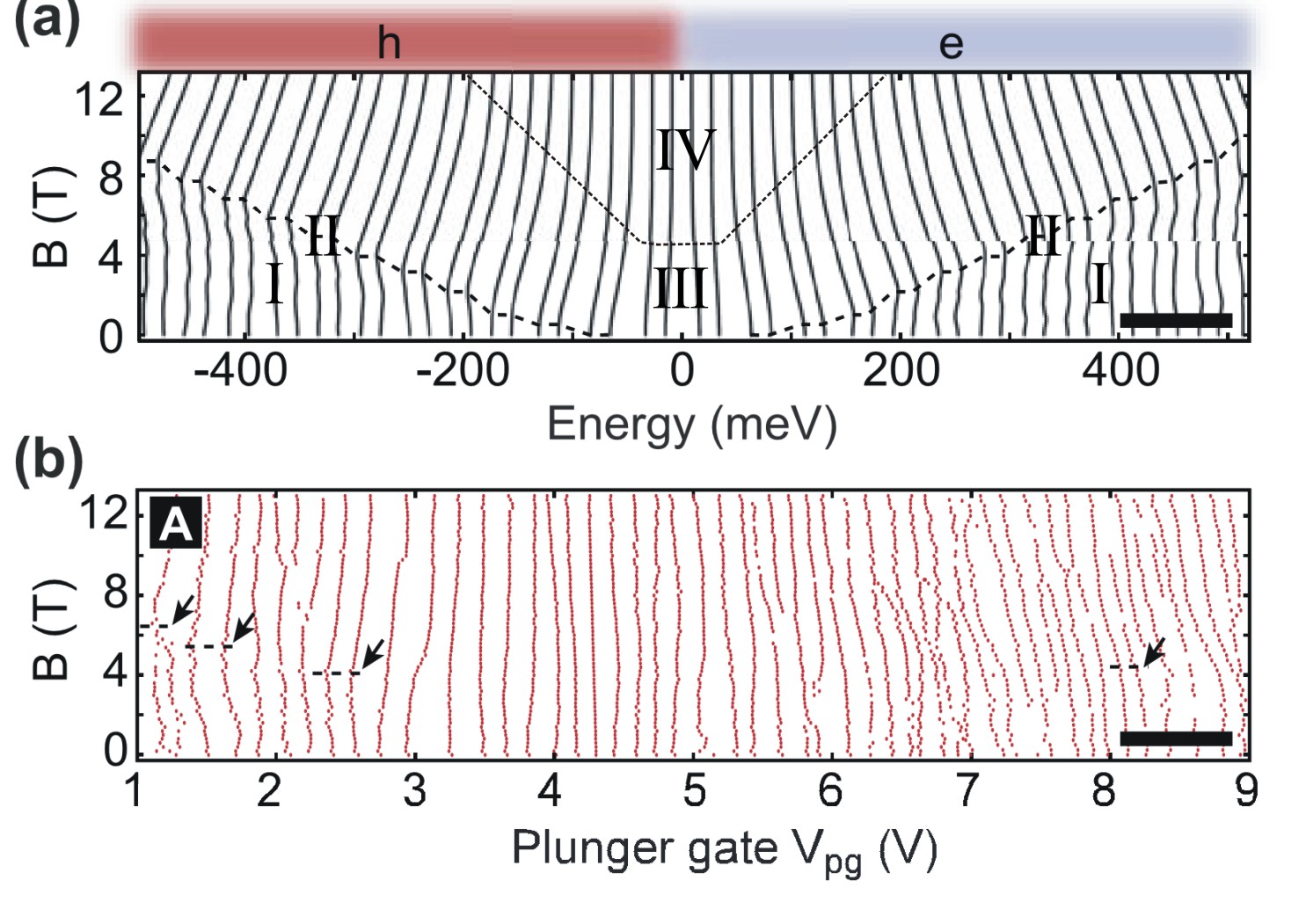}
\caption{(a) The Fock-Darwin spectrum of a 50 $\times$ 80 nm GQD calculated by assuming a constant charging energy and spin degenerate states. The dashed line (regime II) indicates filling factor $\nu$=2 above which all eigenstates continuously evolve into the zero-energy Landau level. The captions I-IV denote different regimes mentioned in the content. (b) Coulomb peak position as a function of perpendicular magnetic field, measured from the device shown in Fig. \ref{Fig18}(a). The arrows indicate the filling factor $\nu$=2 kinks. Adapted with permission from ref. \cite{Guttinger2009}. Copyright 2009 American Physical Society.}  
\label{Fig20}
\end{figure*}

In section 2.1, we have shown the calculated Fock-Darwin spectrum of a graphene quantum dot. Here, we consider a more practical case where a charging energy is included in the spectrum. Fig. \ref{Fig20}(a) shows a tight-binding simulated Fock-Darwin spectrum of a 50 $\times$ 80 nm GQD, where a constant charging energy $E_{C}$=18 meV have been added to each single-particle level spacing ($\approx$4 meV in average). Several key features seen from the spectrum are summarized in the following. At low $B$-field, the 0D levels fluctuate but stay at roughly the same energy, as can be seen in the regime I of Fig. \ref{Fig20}(a). This fluctuation of the Coulomb blockade resonances at low $B$ is due to the continuously crossing of different unfilled states at low energy, as seen in Fig. \ref{Fig12}(b) (red dashed line highlighted regimes). This situation changes when the second lowest LL (LL$_1$) is full, at which point the levels show a kink (regime II) indicating that the electrons (or holes) start to condense into the lowest Landau level (i.e., LL$_0$ at energy $E_{0}$), and the $B$-field onset of this kink increases with increasing number of particles in the quantum dot. Beyond this $B$-field, the levels tend to move towards the charge-neutrality point (regime III), meaning the hole levels move to higher energies while the electron levels move to lower energies. At large enough $B$-field, eventually the levels stop moving and stay roughly at the same energy again (regime IV), indicating the full condensation of electrons/holes into the lowest LL. The Fock-Darwin spectrum of the GQD in Fig. \ref{Fig18}(a) has been studied experimentally by tracking the position of Coulomb peaks under the influence of perpendicular magnetic fields, as shown in Fig. \ref{Fig20}(b). Comparing the numerical simulation and the experimental data [Fig. \ref{Fig20}(a) and (b)], one can find the same qualitative trend of states running toward the center ($E_{0}$). The arrows in Fig. \ref{Fig20}(b) indicate the kinks beyond which all the levels start to fall into the lowest Landau level. These kinks in the magnetic-field dependence of Coulomb resonances can be used to identify the few-carrier regime in graphene quantum dots. The opposite energy shift for electrons and holes in the Fock-Darwin spectrum also provides a method to estimate the charge neutrality point in GQDs \cite{Chiu2012}, but the precise first electron to hole transition is difficult to identify. This can be attributed to the formation of localized states near the Dirac point, which exhibit a weak magnetic-field dependence that alters the spectrum. It is also worth noting that the parasitic magnetic resonances in the tunnel barrier GNRs can also alter the magnetotransport in the GQD \cite{Chiu2012}, which complicates a direct comparison with the simulated Fock-Darwin spectrum.

\subsection{3.1.3 Spin states in in-plane magnetic field}

\begin{figure}	
\includegraphics[scale=0.8]{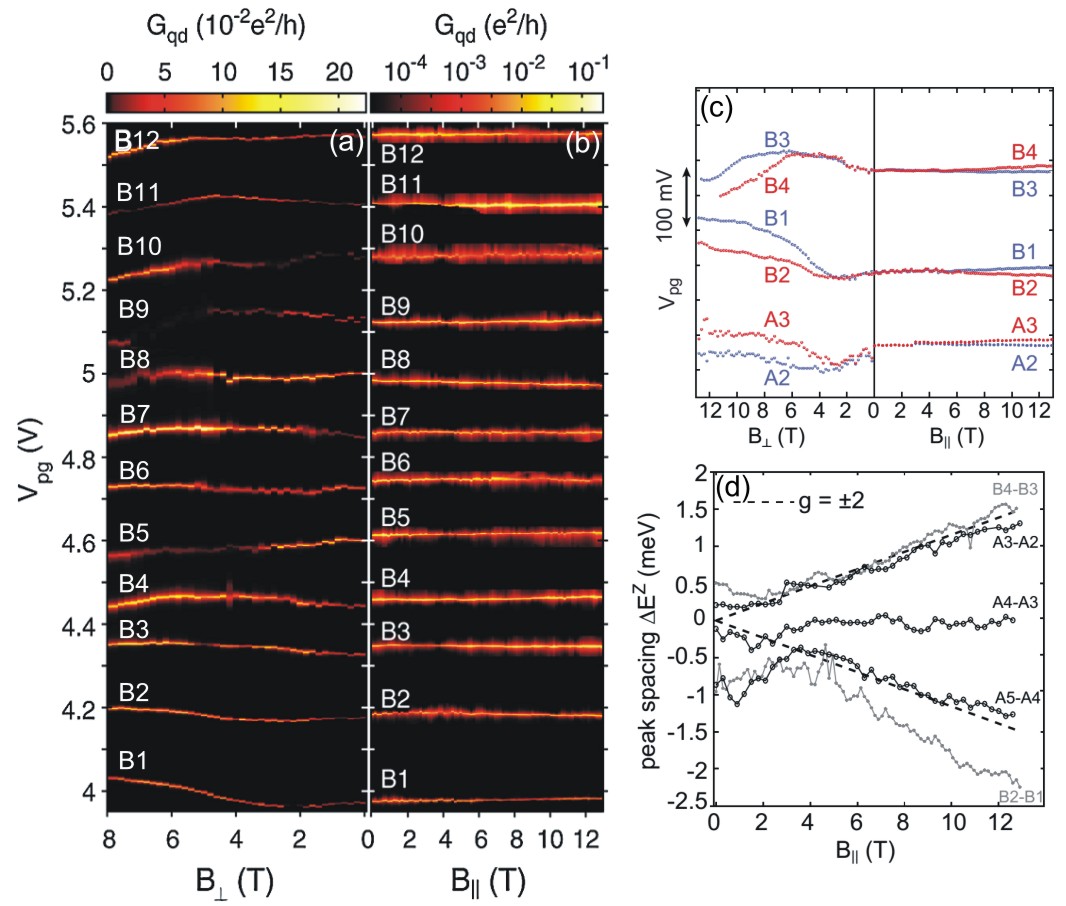}
\caption{(a) Coulomb peaks as a function of perpendicular magnetic field recorded at $V_{b}$=100 $\mu$V, measured from the device shown in Fig. \ref{Fig18}(a). (b) The same Coulomb peaks in (a) but measured in parallel (in-plane) magnetic field. (c) Comparing the evolution of three peak pairs in perpendicular (left) and parallel (right) magnetic field. The peak positions are extracted by fitting the data in (a) and (b), and are offset in $V_{PG}$ voltage such that the pairs coincide at $B$=0 T. (d) Peak spacing as a function of in-plane magnetic field for the three pairs in (c). The dashed lines represent the Zeeman splitting $\Delta E$$^{Z}$=$\pm$$\left|g\right|$$\mu_{B}B$ for a $g$-factor $\left|g\right|$=2. Adapted with permission from ref. \cite{Guttinger2010}. Copyright 2010 American Physical Society.}  
\label{Fig21}
\end{figure}
Perpendicular magnetic fields strongly affect the component of the electron wavefunctions in a QD, resulting in the Fock-Darwin spectrum. In-plane magnetic fields, on the other hand, leave the orbital component unaffected, making it possible to explore Zeeman splitting of QD states \cite{Folk2001,Lindemann2002,Guttinger2010}. It is critical to perfectly align the sample plane to the magnetic field to reduce the perpendicular components, which can be technically difficult. However, this problem can be minimized if one can analyze spin pairs, i.e., two subsequently filled electrons occupying the same orbital state with opposite spin orientation. In this case, the orbital contributions can be significantly reduced by subtracting the positions of individual peaks sharing the same orbital shift in perpendicular magnetic field. Potential spin pairs can be identified by tracking the evolution of two subsequent Coulomb peaks with increasing perpendicular magnetic field, as shown in Fig. \ref{Fig21}(a). For example, the lowest two peaks (B1 and B2) and the following two (B3 and B4) are identified as potential spin pairs due to their similar peak evolution. Fig. \ref{Fig21}(b) shows a measurement of the same peaks in Fig. \ref{Fig21}(a) but with increasing in-plane magnetic fields after the sample is carefully rotated into an orientation parallel to the applied $B$-field. The peaks show a small energy shift with in-plane $B$-field, indicating the orbital effect is negligible. In order to analyze the movement of the peaks in more details, Fig. \ref{Fig21}(c) show the fit of the data selected from Fig. \ref{Fig21}(a) and (b), in which two adjacent peaks (a spin pair) are plotted with suitable offsets in $V_{pg}$ such that pairs coincide at \textit{B}=0 T. As can be seen from the left panel of Fig. \ref{Fig21}(c), the orbital states of each pair have approximately the same $B_{\bot}$ dependence, hence spurious orbital contributions (from slight misalignment) to the peak spacing in $B_{\left|\right|}$ are limited, resulting in a resolvable Zeeman splitting [the right panel of Fig. \ref{Fig21}(c)]. The energy scale of the Zeeman splitting for the spin pairs in Fig. \ref{Fig21}(c) and for two additional peak spacings [A3-A4 and A5-A4, not shown in Fig. \ref{Fig21}(c)] are plotted in Fig. \ref{Fig21}(d). The spin differences between three successive spin ground states take the integer
values $\Delta^{(2)}$ = 0, $\pm$1, ....[e.g., for two successive states, the spin difference can be 1/2 (-1/2) for adding a spin-up (spin-down) electron or 3/2 (-3/2) for adding a spin-up (spin-down) electron while flipping another spin from down (up) to up (down)]. Therefore, apart from the slight deviation of B2-B1, all spin pairs in Fig. \ref{Fig21}(d) follow the relation $\Delta E$$^{Z}$ = $\Delta^{(2)}$$g$$\mu_{B}B$ and a $g$-factor value of approximately 2 can be extracted. The study of Zeeman splitting on spin pairs enables the extraction of the spin-filling sequence in a GQD, which follows an order of $\downarrow\uparrow\uparrow\downarrow\downarrow\uparrow\uparrow\downarrow$ (data not shown) \cite{Guttinger2010}. It is deviated from a sequence of $\uparrow\downarrow\uparrow\downarrow$ observed in the low carrier regime of carbon nanotube quantum dots \cite{Buitelaar2002,Cobden2002}. This phenomenon has been attributed to the exchange interaction between the charge carriers in graphene, which is comparable to the single-particle energy spacing in GQDs and can therefore lead to a ground-state spin polarization \cite{Guttinger2010}. The spin states in GQDs can in principle be considered as a candidate of spin qubits. However, the spin related transport in graphene has shown to suffer from the extrinsic perturbations \cite{Tombros2007,Han2010a,Han2011}. We will address this issue again in section 3.3, where transport properties of GQDs on less disordered substrate will be discussed.

\subsection{3.1.4 Charge relaxation time}

\begin{figure*}[!t]	
\includegraphics[scale=0.72]{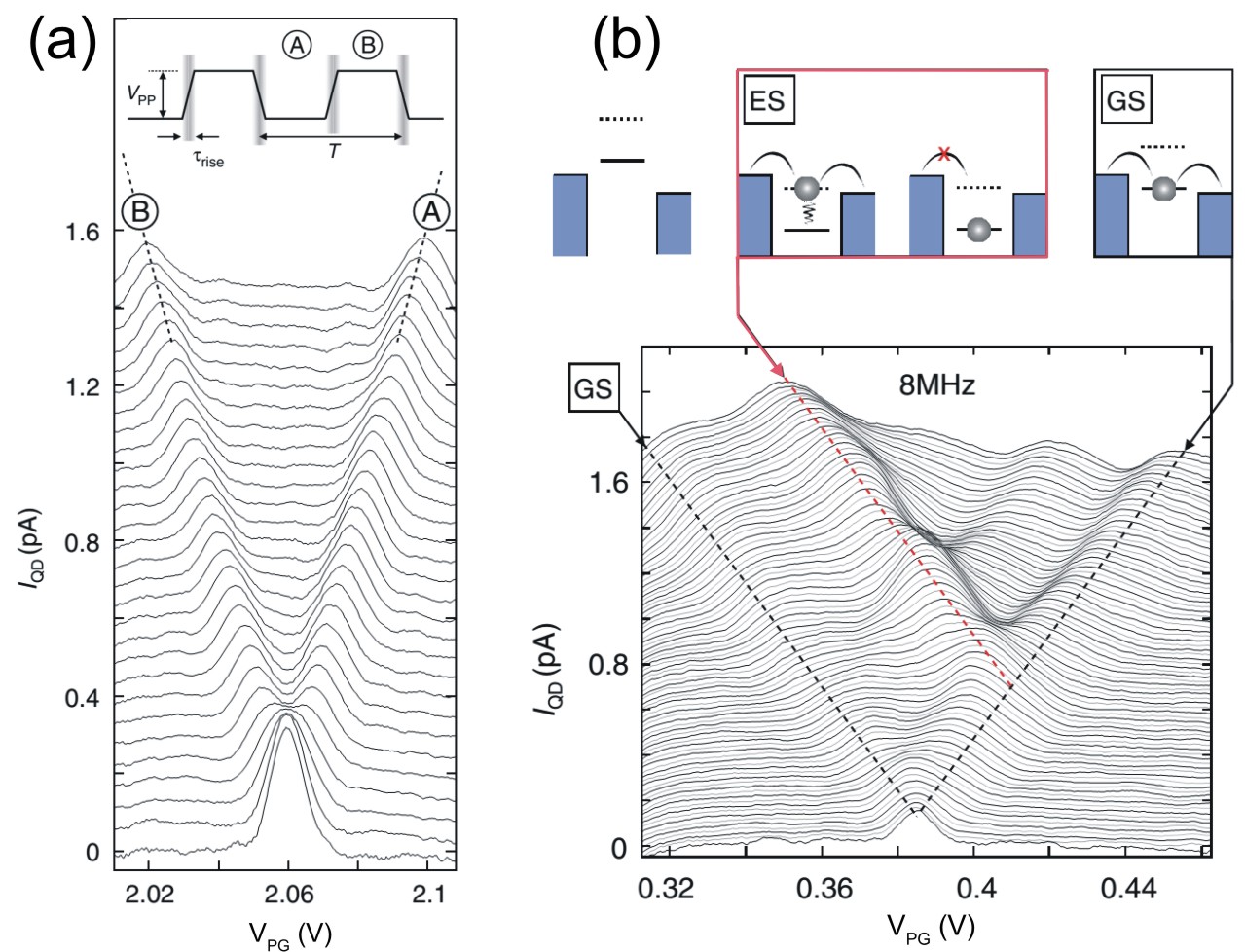}
\caption{(a) Current through the dot at V$_{SD}$=1.5 mV while applying a 100-kHz pulse. Different lines (from bottom to top) correspond to $V_{pp}$ being varied from 0 to 1.4 V in steps of 50 mV. Inset: Sketch of the pulse scheme employed in the measurements presented in this figure. Low and high pulse-level are labeled A and B, and T is the period of the pulse. (b) Top panel: Schematic of transport via GS, ES and, on the left, of a possible initialization stage. Bottom panel: Measurement similar to the ones shown in (a), but with a higher frequency of 8 MHz. $V_{pp}$ is varied from 0 to 2 V in steps of 25 mV (from bottom to top). Adapted with permission from ref. \cite{Volk2013}. Copyright 2013 Nature Publishing Group.}  
\label{Fig22}
\end{figure*}

Pulsed gating, in which a radio-frequency (RF) voltage is applied to the gates, is a powerful tool to manipulate electron spin and to study the spin relaxation time in 2DEG quantum dot systems \cite{Hanson2007}. In this section, we will describe how pulse gating can be used to investigate the charge relaxation dynamics of excited states (ESs) in GSQDs \cite{Droscher2012,Volk2013}. In these measurements, a rectangular pulse $V_{pp}$ with a duration $T$ [inset in Fig. \ref{Fig22}(a)] is applied on top of a DC voltage ($V_{PG}$) to the plunger gate located in the vicinity of the GQD. If the frequency of the pulse is low (2/\textit{T} $\leq$ $\Gamma_{R}$, $\Gamma_{L}$, where $\Gamma_{R(L)}$ is the tunneling rates of the right (left) barrier), the square-wave modulation of the gate voltage results simply in the splitting of the Coulomb resonance into two peaks. Fig. \ref{Fig22}(a) shows such a behavior when pulses with increasing amplitude (from bottom to top) are applied to the plunger gate. These peaks [labeled A and B in Fig. \ref{Fig22}(a)] result from the QD ground state (GS) entering the bias-window at two different values of $V_{PG}$, one for the lower pulse-level (A) and one for the upper one (B). This situation changes dramatically at higher frequencies (2/\textit{T} $\geq$ $\Gamma_{R}$, $\Gamma_{L}$), as shown in the bottom panel of Fig. \ref{Fig22}(b), where the splitting is broadened due to the reduced electron tunneling probability set by $T$ (black dashed line), and a number of additional peaks appear due to transient transport through the excited states of GQD (red dashed line). Each of these additional resonances corresponds to a situation in which the QD levels are pushed well outside the bias-window in the first half of the pulse [Fig. \ref{Fig22}(b), top left panel], and then brought into a position where transport can occur only through the ESs in the second one [Fig. \ref{Fig22}(b), top middle panel]. When the ES lie within the bias-window, an electron occupying the GS, either because of tunneling from the leads or relaxation from the ES, will block the current. Therefore, the additional resonances can be resolved in the DC-current measurements only if the frequency of the pulse is higher than the characteristic rate $\gamma$ of the blocking processes. As both tunneling and ES relaxation lead to the occupation of the GS, $\gamma$ is approximately given by $\gamma$ $\approx$ $\Gamma$$ + $1/$\tau$, where $\Gamma$ is the tunneling rate from lead to dot and $\tau$ is the intrinsic relaxation time of the ES. Since the lowest frequency at which signatures of transport through ESs emerge provides an upper bound for $\gamma$, and $\Gamma$ can be determined by the fitting of peak current through the dot. This in turn gives a lower bound $\tau$ $\geq$ 78 ns for the charge relaxation time of the GQD ESs \cite{Volk2013}.

The ES relaxation timescale is related to the lifetime of charge excitations, which is limited by electron-phonon interactions. The main potential source that induces the charge relaxation in supported graphene is through coupling to the longitudinal-acoustic (LA) phonon $via$ deformation potential (due to an area change of the unit cell) \cite{Ando2005,Kaasbjerg2012,Volk2013}. The fact that the observed timescale is a factor 5-10 larger than what has been reported in III-V QDs \cite{Fujisawa2001,Fujisawa2002,Fujisawa2002a} indicates that the electron-phonon interaction in sp$^{2}$-bound carbons is relatively weak, which is likely due to the absence of piezoelectric phonons in graphene \cite{Volk2013}.

\subsection{3.2 Graphene double quantum dots on SiO$_{2}$/Si substrates}
Graphene double quantum dots (GDQDs) are formed when two graphene islands are located close enough such that they are capacitively coupled to each other and individually coupled to the adjacent gates. Double quantum dots (DQDs) in a wide range of semiconductors are a model system for investigating the spin dynamics of electrons \cite{Hanson2007,Wiel2002,Pfund2006,Schroer2011,Pecker2013}. For example, spin-to-charge conversion using the Pauli spin blockade phenomenon and measurements of spin decoherence time were pioneered in GaAs and later realized in carbon nanotube and silicon DQDs \cite{Hanson2007,Ono2002,Johnson2005,Johnson2005a,Petersson2010,Chorley2011,Zwanenburg2013}. Graphene has been predicted to be particularly suitable for preparing spin-based qubits because of its weak spin-orbit interaction and hyper-fine effect \cite{Trauzettel2007}, which should lead to a long spin decoherence time ($T_2^\ast$). However, although the energy levels in a GQD have shown the ability to distinguish spin (see section 3.1.3), spin-related transport phenomena such as the Kondo-effect \cite{Goldhaber-Gordon1998} and spin blockade have thus far not been observed \cite{Moriyama2009,Molitor2010,Liu2010,Volk2011,Wang2012,Wei2013,Chiu2015,Deng2015}. Although attempts to probe the spin dynamics in such a system have failed, the control of confined charges in GDQDs can still be achieved. These include gate-tunable interdot coupling \cite{Molitor2009,Liu2010,Volk2011,Wei2013} and charge pumping \cite{Connolly.2013a}, which are discussed in the following sections.

\subsection{3.2.1 Coulomb blockade and magneto-transport}

\begin{figure}[!t]	
\includegraphics[scale=1.33]{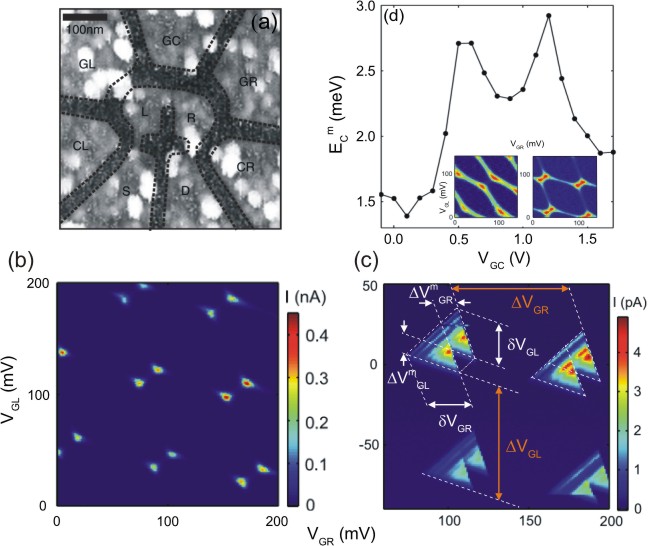}
\caption{(a) Atomic Force Microscope image of a GDQD etched by O$_{2}$ plasma. (b) Current through the GDQD in (a) as a function of $V_{GR}$ and $V_{GL}$ measured at a low bias voltage $V_{b}$ = 500 $\mu$V showing the triple points. (c) The same as (b) but at a higher bias voltage $V_{b}$ = 6 mV shows the bias triangles. (d) Mutual capacitive coupling between the two dots as a function of central plunger gate $V_{GC}$. All the data points correspond to the same triple point. Inset: Current as a function of $V_{GR}$ and $V_{GL}$ for two different central plunger gate voltages $V_{GC}$ = −1.9 V (left) and $V_{GC}$ = 0 V (right). Adapted with permission from ref. \cite{Molitor2010}. Copyright 2010 European Physical Society.}  
\label{Fig23}
\end{figure}

GDQDs can be fabricated lithographically by O$_{2}$ plasma etching out of a graphene flake or by defining the potential landscape using top gates on an etched GNR \cite{Moriyama2009,Molitor2010,Liu2010,Volk2011,Wang2012,Wei2013}. Fig. \ref{Fig23}(a) shows an AFM image of an etched GDQD device on SiO$_{2}$/Si substrate. Two plunger gates $V_{GR(GL)}$ are used to tune the energy levels in QD$_{R(L)}$ while three side gates ($V_{CL,GC,CR}$) are used to tune the tunnel barriers. Fig. \ref{Fig23}(b) shows the current through the device as a function of $V_{GR}$ and $V_{GL}$ at $V_{b}$= 500 $\mu$V, in which a honeycomb-like charge stability pattern typical for a double quantum dot device can be seen. In this low bias regime, transmission is only possible within small areas (known as triple points) in the stability diagram where the levels of two dots are aligned with a small bias window. When the applied bias is large, the current flow is possible over a wider range in gate space, resulting in current measured in the bias-dependent triangle-shaped regions (known as bias triangles), as shown in Fig. \ref{Fig23}(c). The dimensions of bias triangle allow the determination of the conversion factors between gate voltage and energy. The charging energies for the left dot $E^{L}_{C}$=$\alpha_{L}$$\cdot$$\Delta$V$_{GL}$=13.2 meV and for the right dot $E^{R}_{C}$=$\alpha_{R}$$\cdot$$\Delta$V$_{GR}$=13.6 meV are obtained using the voltage-energy conversion factor $\alpha_{L(R)}$=$eV_{b}$/$\delta$V$_{GL(GR)}$, which can be extracted from the bias triangles shown in Fig. \ref{Fig23}(c) (see section 2.2). The interdot coupling energy can also be determined from the splitting of the triangles [Fig. \ref{Fig23}(c)]: $E^{m}_{C}$=$\alpha_{L}$$\cdot$$\Delta V$$^{m}_{GL}$=$\alpha_{R}$$\cdot$$\Delta V$$^{m}_{GR}$=2.2 meV (see section 2.2). It is possible to modulate the interdot coupling strength by changing the voltage applied to the central gate, i.e., $V_{GC}$. The inset in Fig. \ref{Fig23}(d) shows examples of two charge stability diagrams recorded with exactly the same parameters, except for the voltage applied to the central plunger gate. This is also shown in Fig. \ref{Fig23}(d) where the interdot coupling energy $E^{m}_{C}$ extracted from the data is plotted as a function of $V_{GC}$. The oscillating behavior has been also reported in three different GDQD devices and was attributed to resonances induced by disorder states either in the middle GNR (connecting two dots) or in the graphene gate itself \cite{Molitor2009,Liu2010,Wei2013}. Since large gate-voltage ranges are used, the capacitive coupling of the gates to the disorder states can add or subtract charges discretely to these localized states, thus altering the entire environment abruptly and unpredictably. Consequently, the wavefunction in DQD needs to reconstruct itself, leading to the non-monotonic changes in the inter-dot coupling strength with gate voltage. 

\begin{figure}
\begin{center}
		\includegraphics[scale=0.53]{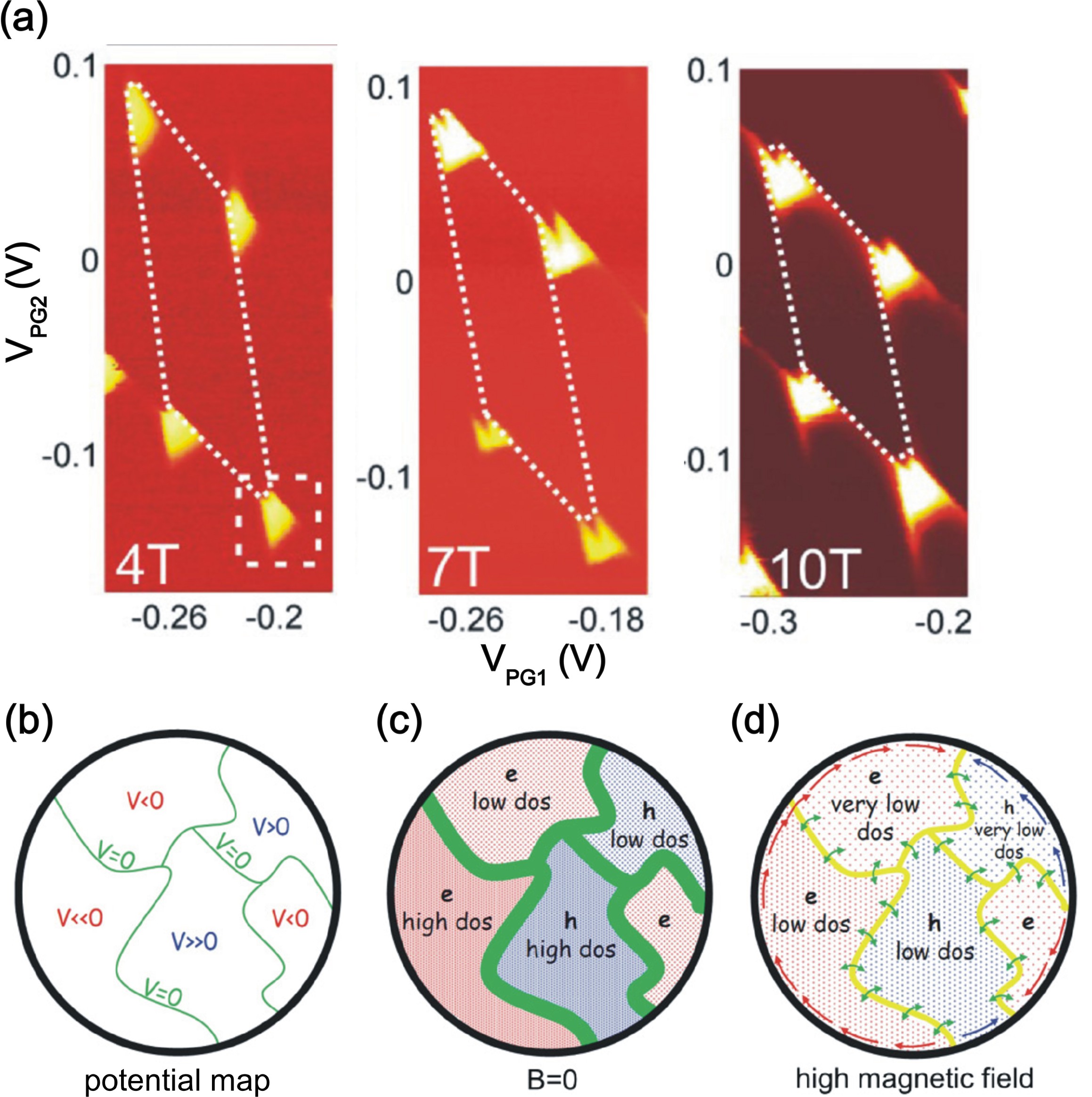}
\caption{(a) The evolution of the charge stability diagram of a large GDQD under the influence of perpendicular magnetic field from 4 to 10 Tesla, measured at $V_{b}$ = -1 mV. (b) Example of a potential distribution in a large disordered quantum dot. (c, d) Expected DOS distributions in the dot at zero magnetic field and high magnetic field, respectively. Adapted with permission from ref. \cite{Chiu2015a}. Copyright 2015 American Physical Society.}
\label{Fig24}
\end{center}	
\end{figure}

When a perpendicular magnetic field is applied to a large graphene dot in which substrate disorder plays an important role [meaning that the size of the QD is greater than the size of the disorder-induced charge puddles; see Fig. \ref{Fig3}(b)], it is possible to induce charge redistribution due to the merging of charge puddles in the dot. Charge stability diagrams of DQDs reveal a wealth of information about their charging energy, interdot coupling and cross gate coupling strength, making them an ideal way to probe charge rearrangements in QDs. Fig. \ref{Fig24}(a) shows the evolution of the charge stability diagram of a large GDQD (200 nm in diameter) for applied perpendicular magnetic fields ranging from 4 to 10 T \cite{Chiu2015a}. The field-dependent changes in the dimensions of the honeycomb (highlighted by the dotted hexagonal outlines) indicate the variations in the capacitances $C_{g1}$ and $C_{g2}$ and thus the changes in the charging energies of both dots [see Eqs. (\ref{eqn 36}), (\ref{eqn 37}), (\ref{eqn 44}) and (\ref{eqn 45})]. The QD charging energies vary from $E_{C1}$$\approx$3 meV and $E_{C2}$$\approx$6 meV at \textit{B}=4 T to $E_{C1}$$\approx$2.2 meV and $E_{C2}$$\approx$3.5 meV at 10 T. Note that the subscript 1(2) in $C_{g1(2)}$ and $E_{C1(2)}$ denotes the gate-dot capacitance and the charging energy for QD$_{1(2)}$. These results suggest that the `effective sizes' of both dots increase at high \textit{B}-fields, which is reflected on the decreasing charging energies. A schematic model shown in Fig. \ref{Fig24}(b), (c) and (d) serve as a qualitative explanation for this observation \cite{Chiu2015a}. Consider a varying background potential \textit{V} in a model QD, as shown in Fig. \ref{Fig24}(b), where \textit{V} fluctuates from positive (blue) to negative (red), passing through \textit{V}=0 (green). If \textit{V} varies slowly, in each region of a large dot, the energy bands will approximately correspond to the shifted energy bands of 2D graphene with the Fermi energy set to zero. A band gap is introduced to represent the quantum confinement effects of the dot, such that in the \textit{V}=0 (green) region, the density of states (DOS) is very low or 0, whereas in the \textit{V}$<<$0 (\textit{V}$>>$0) regions, it gives rise to the electron (hole) puddles with a high DOS, as shown in Fig. \ref{Fig24}(c). The DOS in the dot changes dramatically at high \textit{B}-fields, where the lowest LL (LL$_{0}$) is well developed, with the consequent closing of the band gap. Thus, in the \textit{V}=0 region, the DOS is expected to increase, resulting in the development of non-chiral channels connecting the puddles [the yellow region in Fig. \ref{Fig24}(d)], whereas in the \textit{V}$<<$0 (\textit{V}$>>$0) regions, the DOS decreases due to the more energetically separated LLs in high \textit{B}-fields. At the same time, the other LLs begin developing together with the chiral magnetic edge channel, as indicated in Fig. \ref{Fig24}(d) by the red (blue) arrows for the electron (hole) puddles. Since in this regime the DOS decreases in the bulk of the puddles while it increases at their edges, electron transport through the dot is not confined to a particular puddle but can be delocalized in the dot by flowing through both the chiral edge channels (red or blue arrows) and non-chiral channels (yellow region). In this sense, the current is delocalized in the dot, and charge rearrangement can be observed compared with the case of low \textit{B}-fields.

\subsection{3.2.2 Charge pumping}
\begin{figure}[!t]	
\includegraphics[scale=0.7]{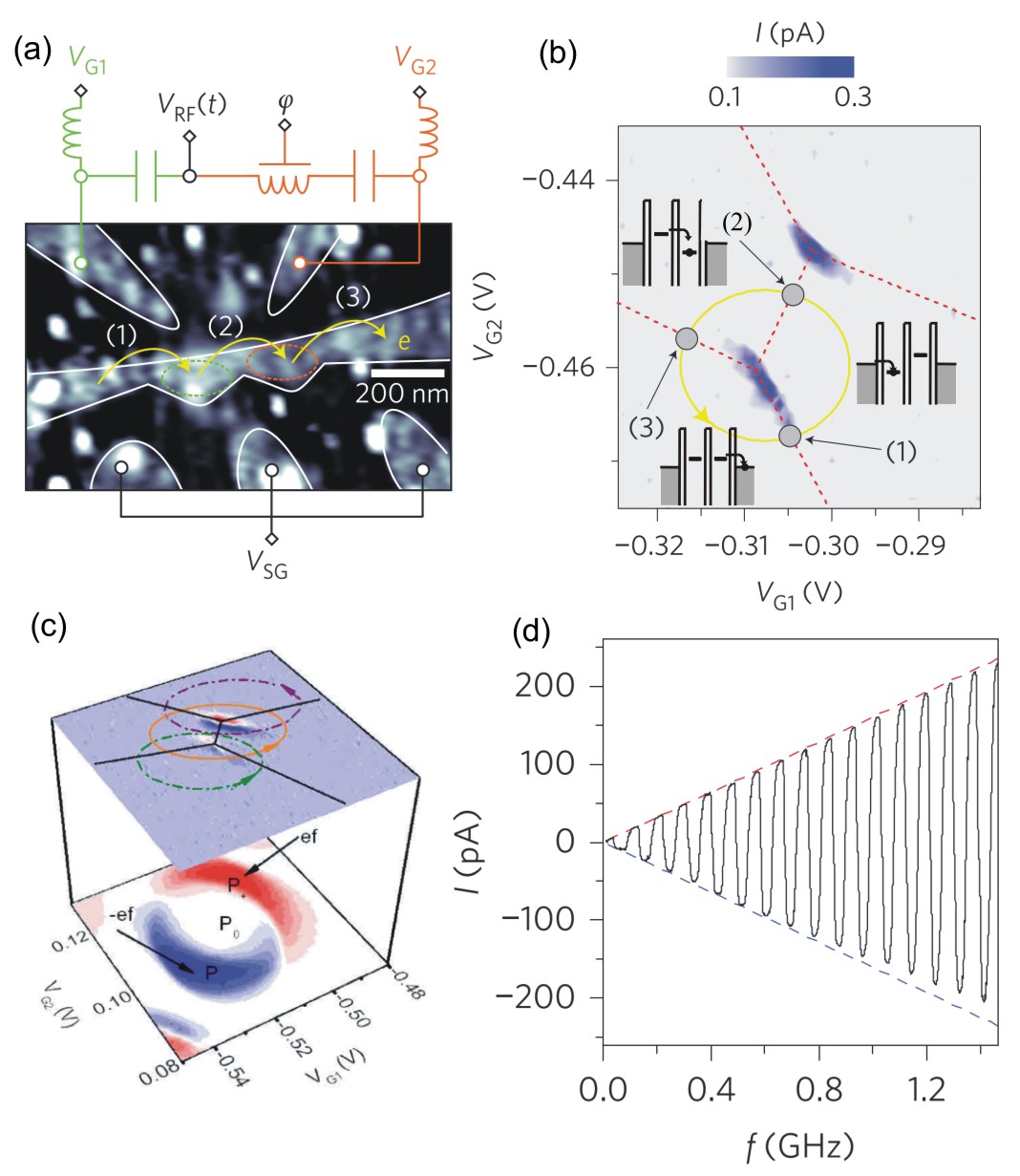}
\caption{(a) Atomic force micrograph of the device that shows the gates used to generate the pumped current in a GDQD device. An oscillating voltage $V_{RF}$($t$) is added to the DC voltages $V_{G1}$ and $V_{G2}$. A phase difference $\varphi$ is added to $V_{RF}$ before being added to one of the gates, which describes a circular trajectory (yellow circle) shown in (b). (b) Source-drain current as a function of $V_{G1}$ and $V_{G2}$ with an applied bias $\leq$1 $\mu$V. The trajectory (yellow) that encircles a triple point, passing through the sequence of transitions (1)$\rightarrow$(2)$\rightarrow$(3), as indicated in both (a) and (b). The insets denote different configuration of QD's energy level. (c) Plot showing a direct comparison between the DC (top) and AC (bottom) current behavior with $f$ = 12 MHz and $P$ = -25 dBm. Regions $P_{+}$, $P_{-}$ and $P_{0}$ refer to the positive, negative and zero pumped current, respectively. (d) Pumped current as a function of frequency at a power of $P$ = -15 dBm. Adapted with permission from ref. \cite{Connolly.2013a}. Copyright 2013 Nature Publishing Group.}  
\label{Fig25}
\end{figure}

Charge pumping, which refers to a device that can shuttle $n$ electrons per cyclic variation of control parameters to give the quantized current $I$ $\equiv$ $nef$, provides an exquisite way to link the electrical current to the elementary charge $e$ and frequency $f$ \cite{Geerligs1990,Kouwenhoven1991,Pothier1992}. Such quantized charge transport can be realized when out-of-phase RF signals are applied to the plunger gates of a DQD, as discussed in section 2.2 \cite{Fuhrer2007b,Chorley2012,Connolly.2013a}. Fig. \ref{Fig25}(a) shows a schematic of the measurement circuit and AFM image of a GDQD device used for charge pumping. The AC voltages $V_{RF}$($t$) on both plunger gates with a phase difference $\varphi$ between them drives the DQD into different charge states around the triple point. When $\varphi$ = 90, it effectively forms a circular pump loop through three charge states in the stability diagram: (1) loading an electron from source reservoir into the left dot, (2) electron transfer from left dot to the right dot and (3) unloading an electron from the right dot to the drain reservoir, as shown in Fig. \ref{Fig25}(a) and (b). When a cycle is complete, a single charge has been transferred from source to drain reservoir and establishes a current. The frequency $f$ of $V_{RF}$ determines the value of the quantized pumped current $I$ = $ef$, and the amplitude of $V_{RF}$ determines the size of area in gate space where pumped current is generated. Depending on the type of triple point that the pumping circle encloses, it generates a different direction of current. Thus, the current recorded around two nearby triple points will present a circular shape with equal values but different signs. Fig. \ref{Fig25}(c) shows a direct comparison of the locations in gate space around a pair of triple points without RF (top) and with RF (bottom) voltages applied to the plunger gates. If the pump loop only encloses one triple point (green and purple loop), it results in a flat regions, labeled $P_{+}$ and $P_{-}$, with a quantized pumped current $P_{+,-}$ = $\pm$$ef$ in the stability diagram. However, when the pump loop encloses a pair of triple points (orange loop), it leads to repeatedly increasing and decreasing the occupancy of each QD without any net transfer of electrons from source to drain. Thus, there is a central region (labeled $P_{0}$) where $I$ $\approx$ 0, giving rise to the crescent shape of pumped current as shown in Fig. \ref{Fig25}(c). Unambiguous confirmation of quantized charge pumping is shown in Fig. \ref{Fig25}(d), which plots the pumped current as a function of $f$ with the DC gate voltages fixed at the center of the $P_{+}$ region. The oscillatory behavior is introduced because of a frequency-dependent phase shift in the RF circuit. The pumped current follows the quantized value $I$ = $\pm$$ef$ over a range of frequencies up to gigahertz, an order of magnitude faster than the traditional metallic pump \cite{Keller1996}. The pumping frequency in graphene is characterized by the RC time constant of the tunnel barriers, where R and C are the effective resistance and capacitance of the GNRs. The two-dimensional nature of graphene leads to a small C and results in a large pump frequency set by the tunnel rate of tunnel barriers (GNRs) \cite{Connolly.2013a}. 

\ \\
We have reviewed the transport properties of graphene single dots and double dots fabricated on SiO$_{2}$ substrates. Additional relevant reviews of research on GQDs on SiO$_{2}$ substrates can be found in ref. \cite{Molitor2011,Stampfer2011,Guttinger2012,Neumann2013}. Next, we will review GQDs fabricated on hBN and discuss how the transport properties change with the reduced influence of substrate disorder.

\subsection{3.3 Graphene quantum dots on hBN}

While the behavior of graphene nanostructures fabricated on SiO$_{2}$ is clearly influenced by localized states, it remains an open question whether they originate predominantly from substrate disorder or edge roughness. Here, we review the studies of GQDs fabricated on hBN substrate. These devices, with reduced substrate disorder potential, are expected to enable the influence of substrate and edge disorder to be studied separately.

\begin{figure*}[!t]	
\includegraphics[scale=0.83]{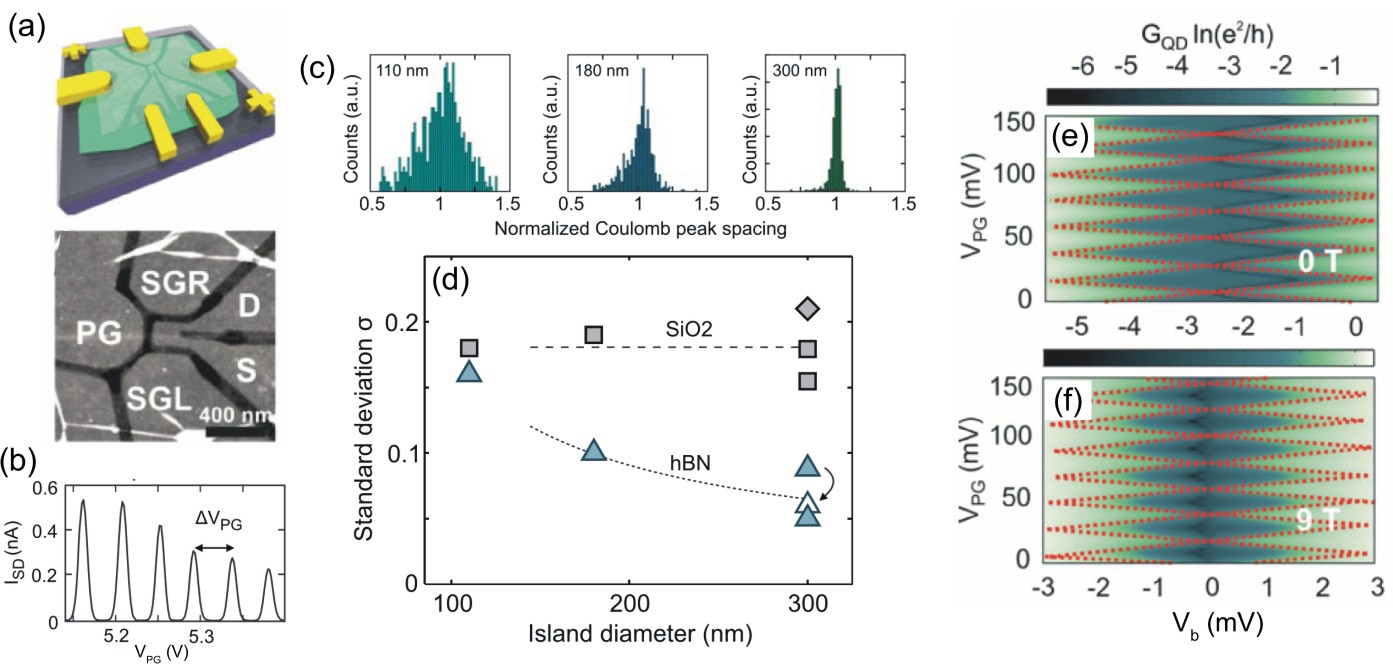}
\caption{(a) Top panel: Schematic illustration of a graphene SET on hBN. Bottom panel: Atomic force micrograph of an etched GQD on hBN with a diameter of 180 nm. (b) Source-drain current $I_{SD}$ as a function of $V_{PG}$ for the device shown in (a). (c) Normalized peak-spacing distribution for GQDs on hBN with diameters of $d$=110 nm (left panel), $d$=180 nm (middle panel), and $d$=300 nm (right panel). (d) Summary plot of the standard deviation $\sigma$ of the normalized peak-spacing distribution for different sized GQD on a SiO$_{2}$ (rectangular data points) and hBN (triangular data points) substrate. (e) Coulomb Diamond measurement of a GQD ($d$=300 nm) on hBN at a perpendicular B-field of 0 T. (f) The same measurement as (e) but at $B$=9 T. (a, b, c, d) adapted with permission from ref. \cite{Engels2013}. Copyright 2013 American Institute of Physics. (e, f) adapted with permission from ref. \cite{Epping2013}. Copyright 2013 John Wiley and Sons.}  
\label{Fig27}
\end{figure*}

GQDs with different diameters ranging from 100 to 300 nm have been fabricated on hBN substrates for transport characterization, as noted in a few literature \cite{Engels2013,Epping2013}. The sizes of the dots are close to the order of the expected size of charge puddles in bulk graphene on hBN ($\approx$ 100 nm in diameter) \cite{Xue2011}, so the substrate disorder is expected to play less important role. Fig. \ref{Fig27}(a) shows the schematic illustration of such a device (top panel) and the atomic force micrograph of an etched GQD on hBN with a diameter of 180 nm (bottom panel). The QD levels are tuned by a plunger gate (PG) while two side gate (SGR and SGL) are used to tune the resistance of the tunnel barrier GNRs. In a regime where the two barriers are pinched-off, the current $I_{SD}$ as a function of plunger gate voltage, as shown in Fig. \ref{Fig27}(b), confirms that the QD is operating in the Coulomb blockade regime. For a more detailed comparison between GQDs resting on hBN and SiO$_{2}$, the distribution of the Coulomb-peak spacing $\Delta V_{PG}$, i.e., the spacing between two subsequent Coulomb peaks, are statistically studied among dots with different sizes fabricated both on hBN and SiO$_{2}$ substrates. The normalized Coulomb peak spacings $\Delta V_{PG}$/$\overline{\Delta V_{PG}}$ for GQDs on hBN are reported as histograms in Fig. \ref{Fig27}(c), for QD diameter $d$=110 nm (left panel), $d$=180 nm (middle panel) and $d$=300 nm (right panel), respectively. The same type of measurements are also performed for GQDs on SiO$_{2}$, and the results are summarized in Fig. \ref{Fig27}(d), where the standard deviation of the normalized peak spacing distribution ($\sigma$) as a function of the QD diameter is presented for both hBN and SiO$_{2}$ substrates. A clear difference can be seen between these two cases. The standard deviation for GQDs on hBN shows a clear decreasing dependence from 0.16 for the dot with $d$=110 nm to 0.05 for the dot with $d$=300 nm, while in the case of GQDs on SiO$_{2}$ it is independent of $d$. The standard deviation $\sigma$, which can be considered as the strength of peak-spacing fluctuations, may result from (i) the fluctuations of single particle level spacing $\Delta$, (ii) fluctuations of the charging energy $E_{C}$ (i.e., fluctuations in the size of the dot), or (iii) fluctuations of the lever arm $\alpha$ (i.e., the position of the dot). The single-particle level spacing in GQD is $\Delta$($N$)=$\hbar v_{F}/(d\sqrt{N}$), where $N$ is the number of charge carriers on the dot and $v_{F}$ is the Fermi velocity \cite{Schnez2009}. If $N$ is the only variable, the single particle level spacing $\Delta$($N$) gives an upper limit at the order of 0.03 to $\sigma$ for $N$= 600 (the number of peaks studied), and should be independent of the dot size and substrate. This is not in agreement with the data shown in Fig. \ref{Fig27}(d) and leads to the assumption that the remaining two sources are responsible for the variability in peaks spacing. The standard deviation for GQDs on hBN can be represented as $\sigma$ $\approx$ $\sigma^{hBN}$ + $\sigma^{edge}$/$d$ $\approx$ 0.01 + 16/$d$ [nm], where $\sigma^{hBN}$ represents the substrate-induced disorder (independent of dot size) and $\sigma^{edge}$ represents the edge-induced disorder (scale with size as the edge-to-bulk ratio changes). Note that both values are obtained from the fit of the dotted line in Fig. \ref{Fig27}(d). By contrast, the standard deviation for GQDs on SiO$_{2}$ is independent of dot size and reads $\sigma^{SiO_{2}}$ $\approx$ 0.18. This suggests that the potential landscape in the dot on SiO$_{2}$ is dominated by substrate induced disorder, while contributions due to edge roughness, which are expected to scale with the size of the sample, play a minor role. These $\sigma$ values also lead to the conclusions that (i) the substrate-induced disorder in GQDs on hBN is reduced by roughly a factor 10 as compared to SiO$_{2}$ ($\sigma^{SiO_{2}}$=0.18 to $\sigma^{hBN}$=0.01), (ii) edge roughness is the dominating source of disorder for GQDs with diameters less than 100 nm. 

The reduced substrate disorder of GQDs on hBN can also reflect on the magneto-transport. If the magnetic length of the electrons on the GQDs is on the order of the disorder potential length scale, the electrons can accumulate in different charge puddles, leading to charge redistribution in the dot thus changing the charging energy (as discussed in 3.2.1). However, as a result of the reduced bulk disorder, this effect is assumed not to occur for GQDs on a hBN substrates. Fig. \ref{Fig27}(e)(f) show the comparison of Coulomb diamond measurements of a $d$=300 nm GQD on hBN at $B$=0 T and $B$=9 T. It can be seen by the similar Coulomb diamonds (with a charging energy $E_{C}$ $\approx$ 3 meV) for both magnetic fields, that the QD is stable and well-defined at 9 T, supporting the notion that the effective size of GQD is not affected in high magnetic fields.

\ \\
In summary, we have reviewed the transport properties of GQDs fabricated on both SiO$_2$ and hBN substrates. We conclude the main observations with the summary of references given in Table I. Here we note that, although the reduced substrate disorders for GQDs on hBN can in principle suppress some possible sources for fast spin relaxation, the spin-related transport phenomena (such as spin blockade) are still unreported \cite{Moriyama2009,Molitor2010,Liu2010,Volk2011,Wang2012,Wei2013,Chiu2015}. It has been reported that the spin relaxation time in monolayer graphene ranges from 100 ps to 2 ns, significantly shorter than theoretically predicted \cite{Tombros2007,Han2010a,Han2011,Maassen2012,Drogeler2014}. Two mechanisms have been proposed to explain this observation. One involves local magnetic moments, which enhance spin relaxation through the resonant scattering of electrons off magnetic moments. Adatoms, organic molecules, vacancies, or spin-active edges are the possible sources of such local magnetic moments \cite{Kochan2014}. The other mechanism is related to the interplay between the spin and pseudospin quantum degrees of freedom when disorder does not induce valley mixing \cite{Tuan2014}. Since graphene constrictions have been widely used as tunnel barriers in most GQDs reported thus far, resonant scattering of electrons off spin-active GNR edges can be inevitable, leading to enhanced spin relaxation that lifts the spin blockade. In fact, GQDs fabricated from lithographic etching are all expected to possess edge roughness and suffer from unwanted edge scattering. One possible solution to this problem is to use an electrical-field-induced bandgap in bilayer graphene to define GQDs \cite{Goossens2012,Allen2012}. However, the small induced energy gap ($\approx$ 200 meV \cite{Zhang2009b}) may limit the available energy range for quantum dot operation. The other approach is to use the tip induced deformation to define edge-free graphene quantum dot. When graphene is deformed, the strain in the membrane can induce a local pseudomagnetic field, which has been reported to be as high as a real magnetic field of 300 T \cite{Levy2010,PachecoSanjuan2014,Barraza-Lopez2013}. This strain-generated pseudomagnetic fields can introduce strong quantum confinement to electrons, from which a quantum dot can be formed. A few STM studies have been reported, but a novel method to probe spin dynamics in such a system is still lacking \cite{Klimov2012,Freitag2017}.

\ \\
\begin{center}
\begin{tabular}	  {|p{2cm}||p{6cm}|} 
 \hline
 \multicolumn{2}{|c|}{TABLE I. References for the main observations} \\
 \hline
 References& Main observations of graphene single quantum dots (GSQDs) \\
 
 \hline
 \cite{Schnez2009} & Observation of excited states \\
 \cite{Gutinger2008} & Charge detection \\
 \cite{Guttinger2009,Chiu2012} & Fock-Darwin spectrum in the few-electron and many-electron regimes    \\
 \cite{Guttinger2010} &  Zeeman splitting of spin states  \\
 \cite{Allen2012} & First suspended GQDs \\
 \cite{Puddy2013} & GQD defined by atomic force microscope cutting \\
 \cite{Goossens2012,Allen2012,Muller2014} & Bilayer GQDs defined by top gates \\
 \cite{Droscher2012,Volk2013} & High-frequency gate manipulation on GQDs \\
 \cite{Klimov2012} & GQDs defined $via$ strain engineering \\

 \hline

References& Main observations of graphene double quantum dots (GDQDs)\\
\hline
\cite{Molitor2010,Liu2010,Volk2011}    &Observation of excited states \\
\cite{Volk2011}    &Zeeman splitting \\
\cite{Volk2011,Fringes2012}    &Bilayer graphene double dot \\
\cite{Liu2010}    &GDQDs defined by gated GNRs \\
\cite{Roulleau2011}    &Electron-phonon coupling \\
\cite{Wei2013}    &Metal gate tuning \\
\cite{Connolly.2013a}&   Charge pumping  \\
\cite{Chiu2015a}&   Charge redistribution in magnetic fields \\
\cite{Deng2015}&   RF sensing of the number of charges \\
\hline

\hline

References& Main observations of GQDs on hBN\\
\hline
\cite{Epping2013}    & Size dependent mean Coulomb peak spacing fluctuation\\
\cite{Engels2013}    & Magnetotransport \\

\hline

\end{tabular} 
\end{center}

\maketitle
\section{4. Single Electron Transport in 2H-TMDs}

\begin{figure*}[!t]	
\includegraphics[scale=0.72]{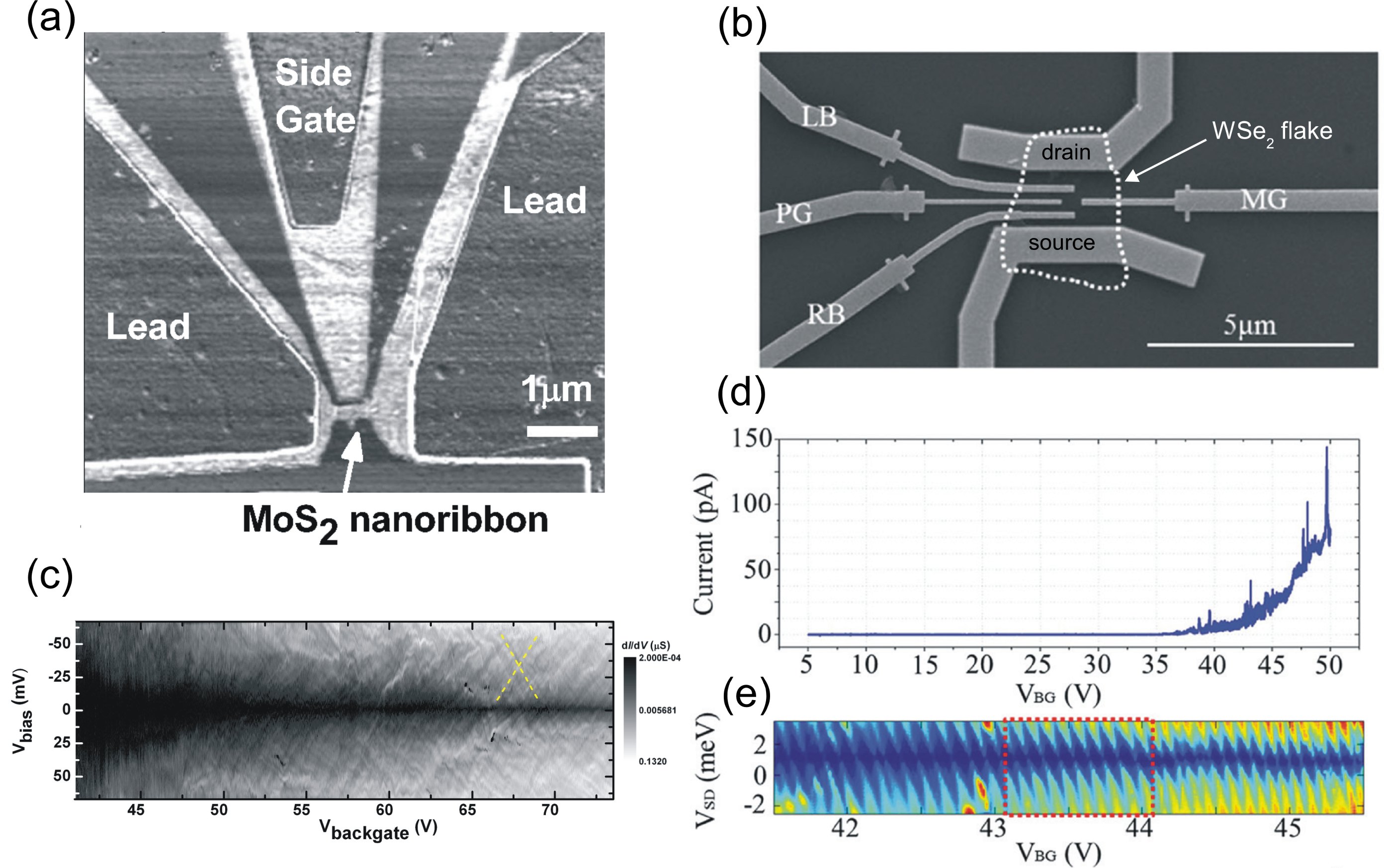}
\caption{(a) AFM image of the MoS$_2$ nanoribbon. (b) SEM image of the WSe$_2$ quantum dot device. The WSe$_2$ flake (4.5 nm in thickness) is highlighted by the white dotted line. The WSe$_2$ flake was directly contacted by the source/drain electrodes, and was separated from the four top gates (LB, PG, RB and MG) by a 40 nm layer of Al$_2$O$_3$ grown by atomic layer deposition (ALD). (c) Differential conductance of the MoS$_2$ nanoribbon as a function of DC bias voltage $V_{bias}$ and back-gate voltage $V_{backgate}$, with the side gate floated. (d) Source-drain current flow through the WSe$_2$ device as a function of back-gate voltage ($V_{BG}$). (e) Coulomb diamond measurements for WSe$_2$ QD from $V_{BG}$ = 41.5 V to $V_{BG}$ = 45.5 V, with all other top gate voltages fixed at -2 V. (a, c) adapted with permission from ref. \cite{Li2013}. (b, d, e) adapted with permission from ref. \cite{Song2015}. Copyright 2015 Royal Society of Chemistry.}  
\label{Fig28}
\end{figure*}

\begin{figure*}[!t]	
\includegraphics[scale=0.4]{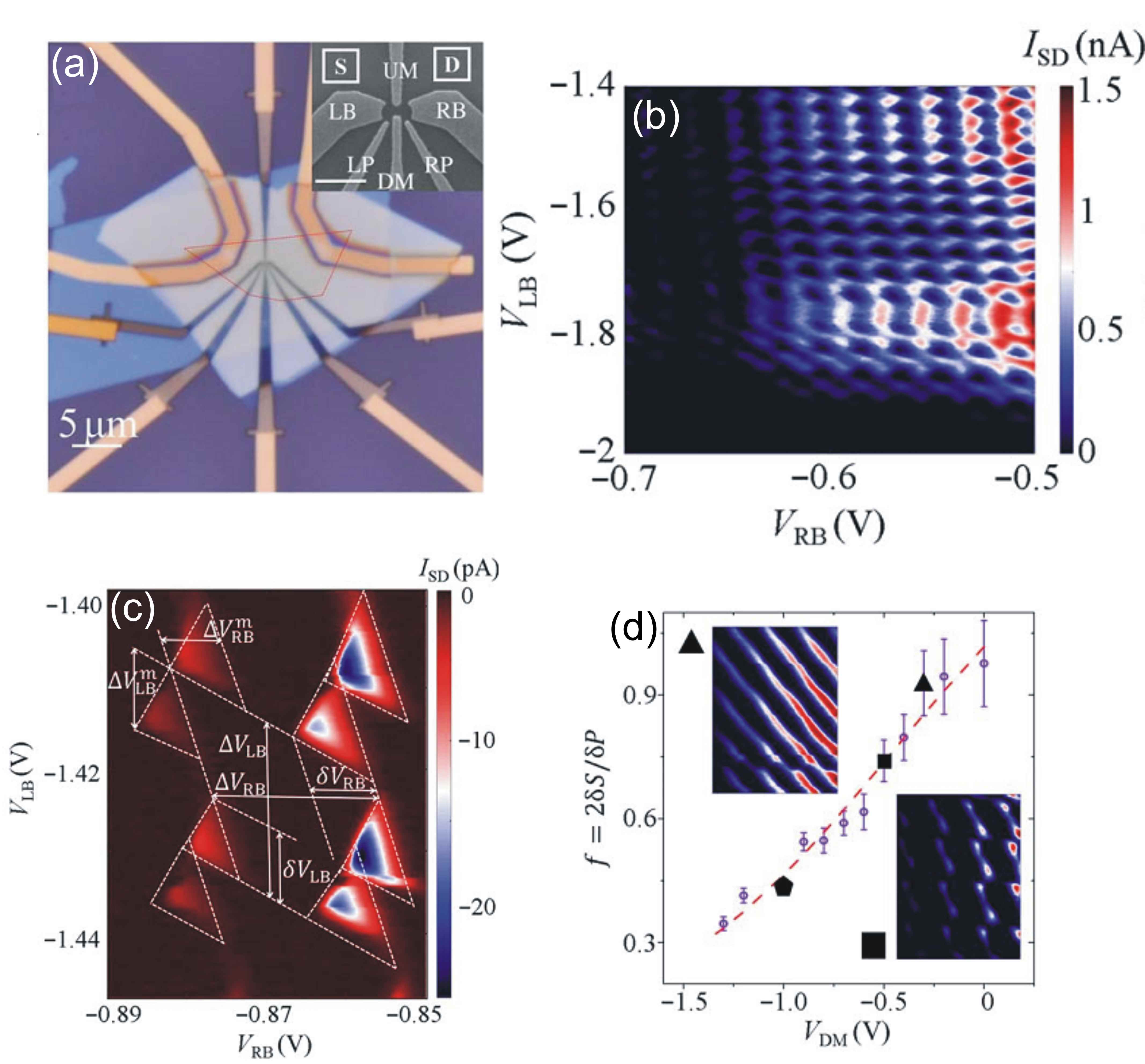}
\caption{Transport through a gate-defined MoS$_2$ double quantum dot. (a) Optical microscopy image of the device with a scale bar of 5 $\mu$m. The area enclosed in the red dashed lines indicates the location of the sandwiched MoS$_2$ flake. The inset with a scale bar of 500 nm shows a scanning electron microscopy image of the bottom gate structure taken before the stacks were transferred. The bottom gates are formed by 5 nm thick Pd. (b) DC current through the DQD versus V$_{LB}$ and V$_{RB}$ (plunger gates LB and RB), for V$_{BG}$ = 30 V, V$_{LP}$ = V$_{RP}$ = 0 V, V$_{UM}$ = -1.8 V, V$_{DM}$ = -1 V, and bias voltage V$_{SD}$ = 3 mV. (c) Charge stability diagram at V$_{BG}$ = 30 V, V$_{LP}$ = V$_{RP}$ = 0 V, V$_{UM}$ = -2.1 V, V$_{DM}$ = -1.2 V, and a bias voltage at V$_{SD}$ = -2 mV. The triple points expand into bias triangles, which allow the estimation of the lever arm between the gates and the dots and the charging energy of both dots. (d) $f$ = 2$\delta S$/$\delta P$ as a function of V$_{DM}$, where $\delta S$ is the diagonal splitting measured between vertices in a triangle and $\delta P$ is the distance between triple points. $f$ essentially represents the interdot coupling energy and shows a decreasing tendency when the middle gate V$_{DM}$ becomes more negative, demonstrating a monotonic change of $f$ by tuning V$_{DM}$. The insets indicate charge stability diagram at different values of V$_{DM}$, whereas other parameters are kept the same. The left inset corresponds to a SQD state while the right one is associated with a DQD state. Adapted with permission from ref. \cite{Zhang2017}. Copyright 2017 American Association for the Advancement of Science.}  
\label{Fig43}
\end{figure*}

The existence of band gaps close to the wavelengths of visible light has earned 2H-TMD nanostructures considerable attention in optical studies \cite{Gopalakrishnan2014,Lin2015,Yan2016}. However, it is also the sufficiently large bandgaps that distinguish 2H-TMDs from graphene and allows their nanostructures to be defined using electrical gating, as is commonly done in GaAs 2DEG systems. In this manner, the edge roughness created during the lithographic etching process, which is a common case in GQD fabrications, can be avoided. In this section, we will introduce a series of nanostructures defined from 2H-TMDs, including MoS$_2$ nanoribbons, WSe$_2$ single quantum dots and MoS$_2$ double quantum dots \cite{Li2013,Song2015,Zhang2017,Pisoni2018}. In the former case, the nanoribbons were fabricated using RIE etching [Fig. \ref{Fig28}(a)], whereas in the latter, the QDs were defined using metal top gates [Fig. \ref{Fig28}(b)]. Both devices were fabricated on SiO$_{2}$/Si substrates. The Coulomb diamond measurement for a MoS$_2$ nanoribbon is shown in Fig. \ref{Fig28}(c). The existence of diamonds confirms the presence of small localized sates (or QDs) in the nanoribbon. Moreover, the fact that larger diamonds are formed in the middle of the transport gap, whereas smaller diamonds are located away from the gap, indicating that the size of the localized state is strongly dependent on the Fermi energy, as is also observed for GNRs on SiO$_{2}$/Si substrates. This finding suggests that, in both cases, the potential in the nanoribbons can be described as a superposition of the substrate disorder potential and the confinement-induced energy gap, as already discussed in section 3. The back-gate sweep for the WSe$_2$ device is shown in Fig. \ref{Fig28}(d) and exhibits the characteristic behavior of an n-doped semiconductor with a transport gap for V$_{BG}$ $\leq$ 35 V. The Coulomb blockade regime for a single dot can be achieved by tuning the WSe$_2$ flake into the conducting regime (V$_{BG}$ $>$ 35 V) while keeping the area below the top gates in an insulating state (V$_{MG}$$=$V$_{PG}$$=$V$_{LB}$$=$V$_{RB}$ $=$ $-$2 V), such that the Coulomb diamonds can be measured as a function of V$_{BG}$, as shown in Fig. \ref{Fig28}(e). In this study, the charging energy $E_C$ was estimated to be approximately 2 meV, which corresponds to a QD radius of $r$ $=$ 260 nm (the plate capacitance model $E_{C}$=$e^{2}/8\epsilon\epsilon_{0}r$ is used, where $\epsilon$ is the relative permittivity of WSe$_2$) and is in reasonable agreement with the area defined by the top gates. Fig. \ref{Fig43}(a) shows the optical micrograph of a double quantum dot defined in MoS$_2$ using the same gating technique. MoS$_2$ and h-BN flakes were exfoliated and transferred onto the local bottom gates [labeled "UM", "LB", "LP", "DM", "RP", and "RB" in the inset of Fig. \ref{Fig43}(a)], which are prepatterned on SiO$_2$/Si substrate. After depositing the Ti/Au source and drain contacts [labeled "S" and "D"], another h-BN flake was transferred onto the whole structure to prevent the oxidation of MoS$_2$. By keeping the n-doped MoS$_2$ conductive (V$_{BG}$ = 30 V) while pinching off the two middle gates (V$_{LP}$ = V$_{RP}$ = 0 V, V$_{UM}$ = -1.8 V, and V$_{DM}$ = -1 V), a conductance map of the plunger gates V$_{RB}$ and V$_{LB}$ demonstrates a honeycomb-like pattern [Fig. \ref{Fig43}(b)], which is typical for DQD's charge stability diagram. Upon applying a suitable bias voltage, the triple points evolve into bias triangles, as shown in Fig. \ref{Fig43}(c), from which various parameters could be extracted (see section 2.2). The estimated charging energies for the left dot and for the right dot are $E_C^L$ = 4.8 meV and $E_C^R$ = 4.7 meV, while the interdot coupling energy is extracted to be $E_C^m$ = 1.2 meV. Furthermore, the estimated dot radius was $\approx$ 68 nm for both dots. These energy factors extracted from the gate-defined quantum dots are comparable with those obtained from the etched GQDs with similar size. One thing this DQD resembles the 2DEG QDs rather than GQDs is the capability of monotonic tuning on the interdot coupling, which allows a smooth transition from a single-dot state to a double-dot state in the same device. Such an effect is shown in Fig. \ref{Fig43}(d), where the the interdot coupling strength was monotonically tuned by changing the middle gate voltage V$_{DM}$, and the charge stability diagram for SQD and DQD were also observed accordingly [insets in Fig. \ref{Fig43}(d)]. Although the the 0D behaviors were demonstrated, no magnetic field dependence of the Coulomb resonances was reported in the above works. 

Despite a lack of experimental studies on the topic, here, we briefly discuss how the Coulomb resonances of 2H-TMD QDs should evolve with perpendicular magnetic fields \cite{Kormanyos2014}. The Hamiltonian for a QD in a two-dimensional semiconducting TMD defined by electrostatic gates under the influence of a perpendicular magnetic field can be written as follows (note that the model assumes that the system is n-doped and that the Hamiltonian describes the conduction band at the $K$ ($K'$) point) \cite{Kormanyos2014}:

\begin{eqnarray}
H_{dot} &=& H^{\tau,s}_{el} + H^{intr}_{SO} + H^{\tau}_{vl} + H_{sp} + V_{dot} \nonumber \\ 
&=& \frac{\hbar^2\hat{q}_+\hat{q}_-}{2m^{\tau,s}_{eff}} + \frac{1 + \tau}{2} sgn(B_z)\hbar\omega^{\tau,s}_c + \tau\Delta_{CB}s_z + \frac{\tau}{2}g_{vl}\mu_BB_z + \frac{1}{2}\mu_Bg^{\bot}_{sp}s_zB_z + V_{dot}
\label{eqn 55}
\end{eqnarray} The wave numbers $q_{\pm}$ = $q_{x}$ $\pm$ $iq_y$ are measured from the $K$ and $K'$ valley points of the TMD; therefore, the band dispersion is parabolic and isotropic (note that at zero $B$-field, $q_+q_-$ = $q^2_x$ + $q^2_y$). Here, the cyclotron energy $\hbar\omega^{\tau,s}_c$ = $e\left|B_z\right|/m^{\tau,s}_{eff}$ where $B_z$ is the applied magnetic field in z-axis and $m^{\tau,s}_{eff}$ denote the effective masses for bands with different valley ($\tau$ = $\pm$ 1) and spin ($s$ = $\pm$ 1) indices. $H^{intr}_{SO}$ = $\tau\Delta_{CB}s_z$ denotes the intrinsic spin-orbit coupling in the TMD, where $s_z$ is the spin Pauli matrix and $\Delta_{CB}$ determines the coupling strength. The next term, $H^{\tau}_{vl}$ = $\frac{\tau}{2} g_{vl}\mu_BB_z$, breaks the valley symmetry of the Landau levels ($g_{vl}$ is the valley $g$-factor, and $\mu_B$ is the Bohr magneton) and describes how the valley states move in the magnetic field. Finally, $g^{\bot}_{sp}$ = $g_e$ + $g^{\bot}_{so}$ is the total $g$-factor, where $g_e$ is the free-electron $g$-factor and $g^{\bot}_{so}$ is the out-of-plane effective spin $g$-factor addressing the SOC. $V_{dot}$ is the confinement potential for a QD of radius $R_d$ and describes the hard-wall boundary conditions: $V_{dot}$($r$) = 0 for $r \leq R_d$ and $V_{dot}$($r$) = $\infty$ for $r \geq R_d$. To solve the band dispersion for this Hamiltonian, we follow the route of solving the Fock-Darwin spectrum for the GQD, as depicted in section 2.1. We set $V_{dot}$ = 0 and find the eigenvalue and eigenfunction of Eq. (\ref{eqn 55}), and the bound-state solutions of the QD can then be determined from the condition that the wave function must vanish at $r$ = $R_d$. For further details on this formalism, refer to ref. \cite{Kormanyos2014}.

\begin{figure*}[!t]	
\includegraphics[scale=0.74]{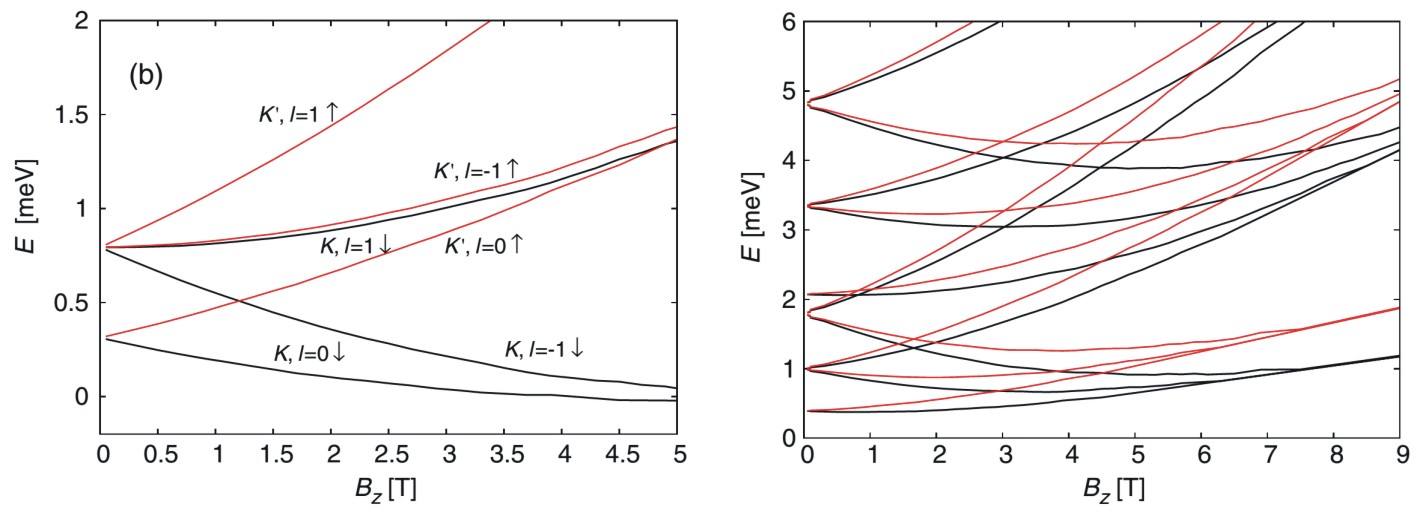}
\caption{(a) Fock-Darwin spectrum of a MoS$_2$ QD of radius $R_d$ = 40 nm. Labels show the valley ($K$ or $K'$), the orbital quantum number $l$, and the spin state ($\uparrow$ or $\downarrow$) for each level. (b) Fock-Darwin spectrum of a WS$_2$ QD of radius $R_d$ = 40 nm. Black (red) lines represent the spin $\uparrow$($\downarrow$) states from the $K$($K'$) valley. Adapted with permission from ref. \cite{Kormanyos2014}. Copyright 2014 American Physical Society.}  
\label{Fig29}
\end{figure*}

The numerically calculated spectra for QDs of $R_d$ = 40 nm in MoS$_2$ and in WS$_2$ are shown in Fig. \ref{Fig29}(a) and Fig. \ref{Fig29}(b), respectively. Note that the parameters $m^{\tau,s}_{eff}$, $g_{vl}$ and $g^{\bot}_{sp}$ used in the simulations for MoS$_2$ and WS$_2$ are different \cite{Kormanyos2014}. As can be seen in Fig. \ref{Fig29}(a), this spectrum mimics the one that we derived for a 2DEG QD [Fig. \ref{Fig12}(a)] because of the quadratic dispersion in the model. At zero magnetic field, states with an angular momentum of $\pm l$ within the same valley are degenerate, due to the effective time-reversal symmetry acting within each valley. For a finite magnetic field, all levels are both valley and spin split, which is different from Fig. \ref{Fig12}(a) where spin is not considered. For large magnetic fields, when $l_B$ $\leq$ $R_d$, the dot levels merge into Landau levels, as also shown in Fig. \ref{Fig12}(a) and (b). From Fig. \ref{Fig29}(a), one can see that the spin and valley degrees of freedom are locked together (meaning that spin $\downarrow$ ($\uparrow$) electrons only reside in the $K$ ($K'$) valley), suggesting that TMD QDs can be used as simultaneous valley and spin filters for single electrons. The spectrum for WS$_2$, as shown in Fig. \ref{Fig29}(b), is similar to that for MoS$_2$, but the level spacing at zero field is larger because of the smaller effective mass in WS$_2$ (the mean level spacing can be approximated as $\delta$ = $\frac{2\pi\hbar^2}{m_{eff}A}$, where $A$ is the area of the dot \cite{Kormanyos2014}). By contrast, at a finite $B$-field, the splitting between states belonging to different valleys (and also spins, as spin and valley are locked) is significantly larger for the MoS$_2$ than for the WS$_2$ because of the different signs of $\Delta_{CB}$ and, consequently, the different spin polarizations of the lowest levels in the two materials. The valley and spin pairs could, in principle, serve as valley or spin qubits, but the crossing between levels with different quantum numbers $l$ in finite magnetic fields may add complication to its realization. One way to circumvent this problem may be to use the lowest Kramers pairs [$\left|l = 0, K', \uparrow\right\rangle$ and $\left|l = 0, K, \downarrow\right\rangle$; see Fig. \ref{Fig29}(a)] at a low $B$-field as a combined spin-valley qubit. Note that the above results apply not only for monolayer 2H-TMDs but also for 2H-TMDs with an odd number of layers, in which the inversion symmetry is broken. 

After reviewed the relevant experimental studies related to developing spin qubits in 2D materials, in the next section, we will discuss the prospect of using 2D materials as an element in superconducting qubits.

\maketitle
\section{5. 2D material-based Josephson junction}

Josephson junction (JJ) is a core element in superconducting based quantum computing. When a Josephson junction is connected in parallel to a capaciator under the condition that Josephson energy is much larger than the charging energy ($E_J$ $\gg$ $E_C$), the eigen-energy of the quantum circuit forms anharmonic bound states with unequal energy spacing. The two lowest bound states of the quantum circuit can be used as a two-level system for qubit operation, and such qubit is generally referred as a transmon qubit. The anharmonic nature of the bounds states arises from the nonlinear phase relation in JJ's inductance ($L_J$), which is given by $V = L_J dI/dt$ and can be written as: 

\begin{eqnarray}
\ L_J = \frac{\Phi_0}{2\pi I_C cos\varphi},
\label{eqn 57}
\end{eqnarray} where $I = I_C sin\varphi$, $I_C$ is the critical current of JJ and $\varphi$ is the phase difference across the junction, $V = \frac{\Phi_0}{2\pi} \frac{d\varphi}{dt}$, and $\Phi_0$ is the magnetic flux quanta. The Josephson energy $E_J$ is given by $E_J = \hbar I_C /2e$ and the corresponding energy spacing between the lowest two bound states can be approximated as (when $E_J$ $\gg$ $E_C$) \cite{Larsen2015}:

\begin{figure}
\includegraphics[scale=0.38]{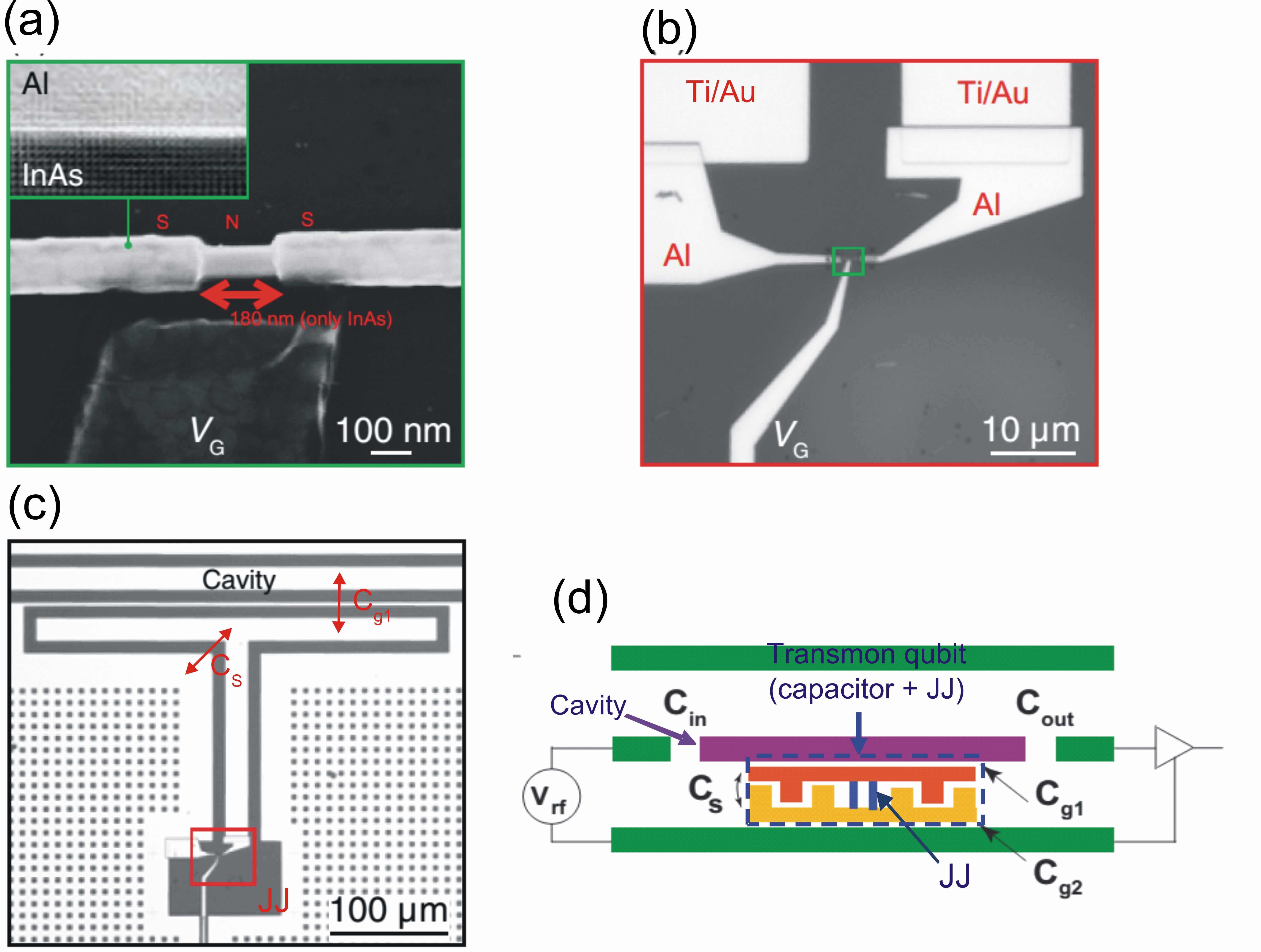}
\caption{InAs nanowire-based superconducting gatemon qubit. (a) Scanning electron micrograph of the Al-InAs-Al JJ. A segment (180 nm) of the epitaxial Al shell is etched to create a semiconducting weak link. Inset shows a transmission electron micrograph of the epitaxial InAs/Al interface. (b) Optical micrograph of the gatemon device. The green square indicate the region shown in (a). (c) Optical micrograph showing the gatemon defined between the T-shaped island and the surrounding ground plane. The circuit (JJ + capacitor) is coupled to a transmission line cavity $via$ C$_{g1}$. (d) Schematic of the entire circuit for readout and control of qubit. (a, b, c) adapted with permission from ref. \cite{Larsen2015}. Copyright 2015 American Physical Society. (d) adapted with permission from ref. \cite{Will2013}.}  
\label{Fig37}
\end{figure}

\begin{eqnarray}
\ E_{01} = \sqrt{8E_CE_J(V_g)},
\label{eqn 58}
\end{eqnarray} The critical current of a S-I-S (S stands for superconductor and I stands for insulator) Josephson junction is usually determined by its geometry, which is fixed after fabrications. Thus, the Josephson energy, hence the qubit frequency ($f_Q = E_{01}/h$), is fixed correspondingly. Although a configuration of multiple Josephson junctions can allow some tunability with magnetic flux \cite{Mooij1999,Orlando1999}, it usually requires complex fabrications which can be difficult for upscaling. On the other hand, a S-N-S Josephson junction, in which N stands for a semiconductor, allows the critical current to be mediated by the density of state of the middle semiconductor. By this manner, the qubit frequency can be readily tunable by an electric gate close to the semiconductor, leading to a gate-tunable transmon qubit, as known as gatemon. An example of a gatemon qubit is shown in Fig. \ref{Fig37}, where a InAs nanowire is contacted by two Al electrodes to form a Al-InAs-Al junction. The JJ is formed from a molecular beam epitaxy-grown InAs nanowire, 75 nm in diameter, with an $in$-$situ$ grown 30 nm thick Al shell. By wet etching away a 180 nm segment of the Al shell, followed by a deposition of a gate 100 nm away from it, a gateable junction is fabricated as shown in Fig. \ref{Fig37} (a) and (b). The JJ is bridging between a T-shape Al island and the surrounding Al ground plane, leading to a junction shunted by a capacitor $C_S$ [see the arrow next to $C_S$ in Fig. \ref{Fig37} (c)], which effectively forms a transmon qubit (JJ + capacitor). Fig. \ref{Fig37} (d) illustrates the measurement scheme for the transmon qubit. The on-chip cavity is capacitively couple to the qubit through $C_{g1}$ [also indicated in Fig. \ref{Fig37} (c)], and the $C_{in}$ ($C_{out}$) defines the the capacitive coupling between the incoming (outgoing) transmission line and the cavity. When the incoming RF signal resonates with the qubit ($f_{RF} = f_{Q}$), the transmitted signal will undergo a significant loss, from which qubit frequency can be determined. The read out of qubit state is operated in a dispersive regime, where the resonant frequency of the cavity will depend on the qubit state. For readers who are interested in the relevant techniques, please refer to \cite{Reed2014}.  

It is natural to consider 2D materials as a good intermediate medium in a S-N-S junction owing to their 2D (allowing different JJ sizes) and gateable nature. In order to do so, a gateable JJ based on 2D materials must be demonstrated. We will review a selection of experimental studies in the subsequent sections.    

\subsection{5.1 Graphene-based Josephson junction}

\begin{figure}
\includegraphics[scale=0.6]{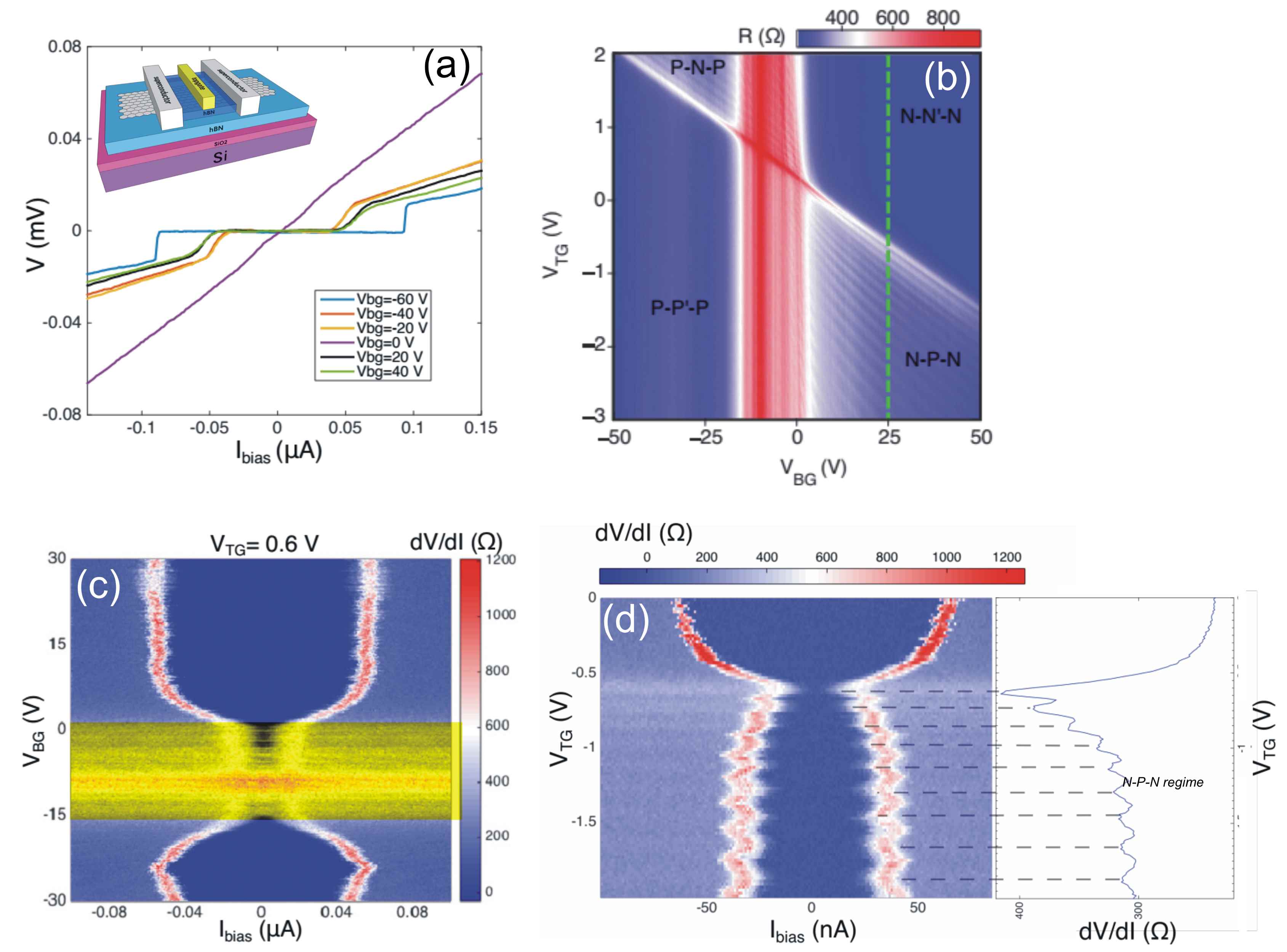}
\caption{Electrical characterization on a dual-gated S-Graphene-S junction. (a) $I$-$V$ characteristic of the junction measured at different back gate voltage showing the gate-tunable critical current. The inset shows a 3-D schematic of the dual-gated S-G-S junction. (b) 2-D normal state resistance map measured as a function of $V_{TG}$ and $V_{BG}$. Note that the Fabry-Pérot oscillations are observed in the N-P-N and P-N-P configurations. (c) Differential resistance $dV$/$dI$ as a function of backgate voltage $V_{BG}$. Topgate voltage $V_{TG}$ is set at 0.6 V. (d) Fabry-Pérot oscillation of Josephson effect, where the critical current $I_C$ oscillates out-of-phase with the normal state resistance $R_N$. Adapted with permission from ref. \cite{Wang2016}.} 
\label{Fig38}
\end{figure}

When a graphene flake is in contact with a superconductor, it acquires superconducting properties because of the proximity effect, a well-known phenomenon that can be described as the leakage of Cooper pairs of electrons from a superconductor (S) into a normal-type conductor (N). The proximity effect takes place only if the S-N interface is highly transparent to electrons, which makes graphene a good candidate as it can form Schottky barrier-free contacts with metals. Graphene-based Josephson junctions have been widely studied over the past decade, from the early demonstration of bipolar supercurrent \cite{Heersche2007} to the more recent studies on Andreev reflection \cite{Dirks2011,Efetov2015,Bretheau2017}, current phase relation \cite{Nanda2017}, edge-mode superconductivity \cite{Allen2015,Lee2017} and supercurrents in the quantum Hall regime \cite{Amet2016}. The inset in Fig. \ref{Fig38} (a) illustrates an example of a dual-gated graphene Josephson junction, where graphene is contacted by two superconducting electrodes (Al) to form a S-G-S junction. The graphene flake can also be electrically gated through a global Si back gate and a local Ti/Au finger-like top gate using a piece of hBN as a dielectric layer. The graphene device exhibits ohmic behavior with a normal state resistance, $R_N$, in the range from 0.2 to 1 k$\Omega$ within the gate tunable range. Fig. \ref{Fig38} (b) shows $R_N$ measured as a function of top gate $V_{TG}$ and back gate $V_{BG}$ by applying a current excitation much larger than the critical Josephson current of the device (hence no Josephson effect is expected). Since the top gate in the S-G-S junction can tune the carrier density in the locally gated region, the polarity of the entire device can be determined by the four distinct regions in the 2D resistance map, i.e., P-N-P, N-P-N, P-P'-P, and N-N'-N, as indicated in Fig. \ref{Fig38} (b). In particular, the P-N interface in graphene can be highly transparent for charge carriers (as the chiral Dirac fermion can pass the potential barrier $via$ Klein tunneling), while their trajectories could resemble that of refracted light at the interface of metamaterials with negative refractive index. Therefore, in analogy to wave optics, the locally gated region can be regarded as a Fabry-Pérot cavity for the charge carriers and result in the interference pattern as observed in the P-N-P and N-P-N regions in Fig. \ref{Fig38} (b). This interference pattern can be correlated with the supercurrent as will be discussed below.

When the temperature is below the critical temperature of the superconductor ($T_C$=1.1 K for Al), the pseudo four-probe current bias measurement, in which the voltage across the graphene is measured while sweeping the bias current through the same source-drain electrodes, was performed at different back gate voltages, as shown in Fig. \ref{Fig38} (a). The proximity effect in graphene manifests itself through the appearance of a dissipationless supercurrent, i.e., no voltage drop accompanies the finite current flow, and the switching from superconductive to dissipative conduction occurs when $I$ approaches a critical current, $I_C$, leading to the abrupt appearance of a finite voltage. Similar measurements can be also performed by a standard AC+DC technique, in which a small AC modulation is added on top of the DC bias current, and the resulting $dV$/$dI$ was detected by a lock-in amplifier. The $I_c$, defined by the boundary of the zero $dV$/$dI$ region, displays a strong gate dependence as shown in Fig. \ref{Fig38} (c). The gate-dependence of $I_C$ can be attributed to the $I_C R_N$ relation in a S-N-S junction \cite{Likharev1979}. The product of critical current and normal state resistance for a S-N-S junction can be roughly characterized by $I_C R_N \approx \Delta_0/e$, where $\Delta_0$ is the superconducting energy gap of the contact electrodes (note that the exact form of the formula may vary depending on whether the junction is in the diffusive or ballistic regime \cite{Doh2005,Titov2006}). Since the 2D and semiconducting nature of the graphene allows its carrier density, hence the normal state resistance, to be controlled by the voltage applied to the back-gate, the critical current also shows a strong gate-dependence. As expected, $I_C$ nearly diminishes at the Dirac point ($V_{BG}$ = -13 V), where $R_N$ is at its maximum [see Fig. \ref{Fig38} (b)], as can be observed in Fig. \ref{Fig38} (c). The strong correlation between $I_C$ and $R_N$ can be further investigated in Fig. \ref{Fig38} (d), where the left panel displays the $dV$/$dI$ v.s. $I_{bias}$ as a function of $V_{TG}$ in the Fabry-Pérot oscillation regime, and the right panel plots normal resistance in the same regime [linecut along the green dashed line in Fig. \ref{Fig38} (b)] for comparison. It can be clearly seen that $I_c$, defined by the boundary of the zero $dV$/$dI$ region, oscillates out-of-phase with the Fabry-Pérot oscillation of normal state resistance $R_N$, in agreement with the $I_C R_N$ relation mentioned above. The observed $I_C$ varies from 10 nA at $V_{BG}$ = -13 V to 80 nA at $V_{BG}$ = -60 V, demonstrating a good gate-tunability readily for use in graphene-based gatemon qubits.

\begin{figure}
\includegraphics[scale=0.75]{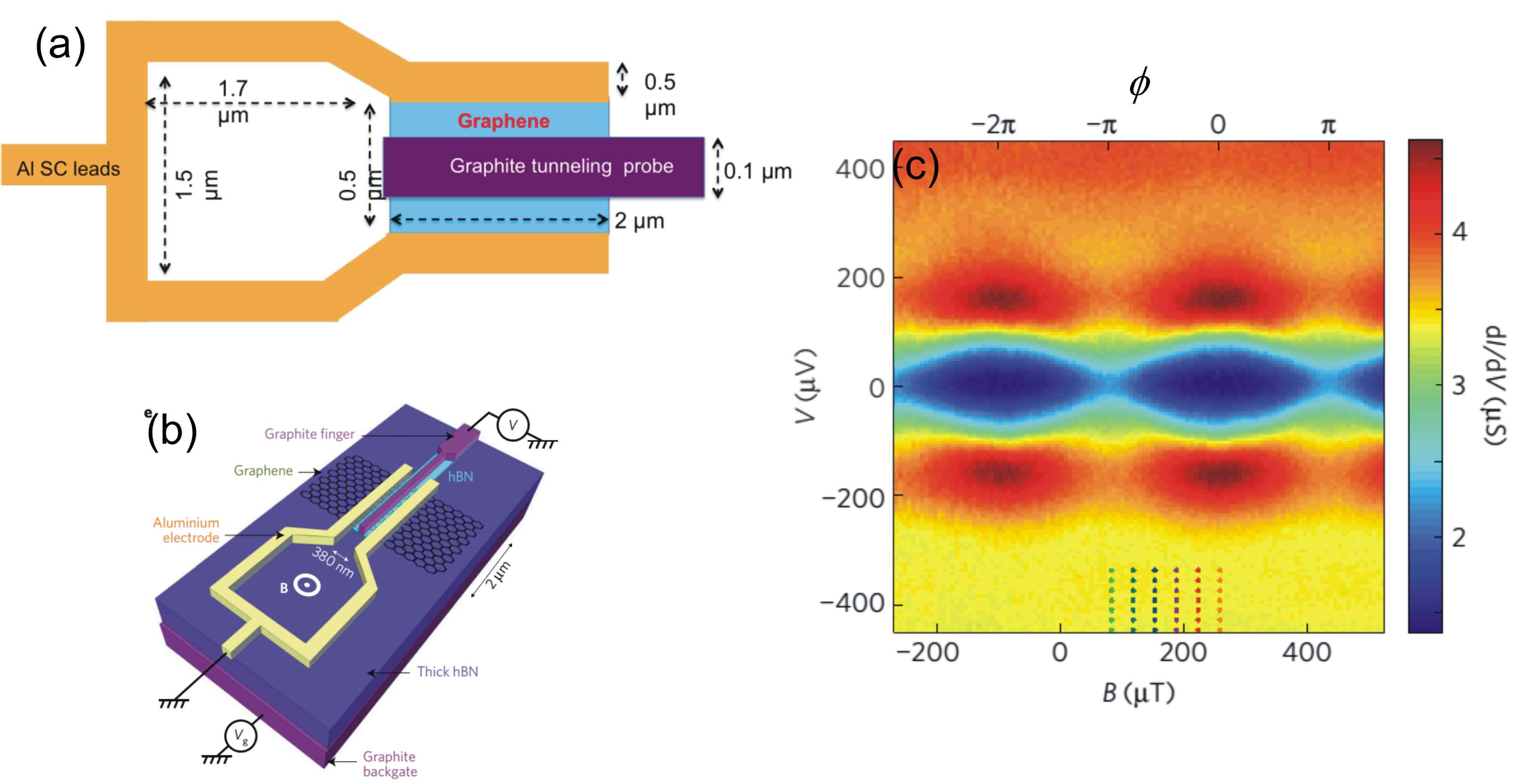}
\caption{ABS oscillations in a SQUID-like graphene Josephson junction (S-G-S). (a) Top view of the SQUID-like S-G-S junction, showing the relevant dimensions of the device. (b) Schematic side view of the SQUID-like S-G-S junction. An encapsulated graphene flake is connected to two superconducting electrodes. (c) A 2-D map of $dI$/$dV$ measured on the device with the geometry shown in (b), showing the ABS oscillations within the induced superconducting gap in graphene. (a) adapted with permission from ref. \cite{Wang2016}. (b, c) adapted with permission from ref. \cite{Bretheau2017}. Copyright 2017 Nature Publishing Group.}  
\label{Fig40}
\end{figure}

In section 2.3, we have discussed how Andreev reflection, and hence the corresponding ABS, can be related with proximity effect. To manifest the phase dependence of ABS energy, here we introduce a SQUID-like graphene Josephson junction, in which the phase difference across the weak link can be well-controlled by applying an external magnetic field \cite{Bretheau2017}. The top and side view of this SQUID-like S-G-S junction is shown in Fig. \ref{Fig40} (a) and (b), respectively. A monolayer graphene sheet is encapsulated between two hBN sheets, where the bottom one is 15 nm thick and the top one is one atom thick (0.3 nm). The bottom hBN isolates graphene from a graphite local backgate, which enables the electrostatic control of the Fermi energy of graphene. On top of the top hBN sits a 150-nm-wide metallic probe made of a thin graphite. The use of a graphite probe (instead of metal) is to limits the doping in the underneath graphene, and allow the low carrier density regime to be accessed. The graphene sheet, with a width $W$ = 2 $\mu$m, is connected to two superconducting Al electrodes, with a spacing $L$ = 500 nm. The extension of the electrodes is patterned in a loop that enables the control of phase difference $\phi$ = $2\pi\Phi$/$\Phi_0$ across the graphene weak link by applying a magnetic flux $\Phi$ = $B \cdot Area$ through the loop ($\Phi_0$ is flux quantum). In such a way, one could study the phase dependence of ABS energy discussed in section 2.3. The tunneling measurement to study the DOS of proximitized graphene was performed by applying a voltage modulation $dV$ = 10 $\mu V$ to the graphite probe and measured the differential conductance $dI$/$dV$ through the Al lead using a standard lock-in technique. Fig. \ref{Fig40} (c) shows $dI$/$dV$ measured as a function of DC bias voltage $V_b$ and magnetic field $B$ for a device with the geometry shown in Fig. \ref{Fig40} (b). The observed oscillation within the induced superconducting gap in graphene (-160 $\mu V$ $\leq$ $V_b$ $\leq$ 160 $\mu V$) strongly indicates the phase dependence of ABS in the short junction limit, where only a pair of ABS is modulated by phase \cite{Bretheau2017}. However, it worth noting that there exists non-vanishing DOS inside the gap at all fields, suggesting there are ABS whose energies span the full spectrum within the gap. These ABS can be regarded effectively as contributed by a set of long junction modes, which are less sensitive to phase modulation compared with the short junction modes \cite{Bretheau2017}. Their origin may be due to the weak coupling between graphene and the superconductor or impurity scattering within the normal region. The oscillation in Fig. \ref{Fig40} (c) possess a periodicity $\delta B$ = 360 $\mu$T, which corresponds to one flux quantum $\Phi_0$ threading the loop of enclosed area $A$ = 5.7 $\mu$m$^2$ (taking into account the Meissner effect). The corresponding phase $\phi$ = $2\pi A(B-B_0)$/$\Phi_0$ is shown on the top axis of Fig. \ref{Fig40} (c), where $B_0$ = 250 $\mu$T is an offset magnetic field in this experiment. The gap and side peaks are most pronounced at $\phi$ = 0 and get reduced when the phase is swept towards $\phi$ = $\pi$, as expected from Eq. (\ref{eqn 64}).

\subsection{5.2 TMDs-based Josephson junction}
Apart from graphene, the S-N-S junctions made of other layered materials are relatively limited. Although there are works using superconducting layered materials as a component, they are used to provide superconductivity not as an intermediate layer in a S-N-S junction \cite{Yabuki2016,Kim2017}. Owing to the intrinsic band gap in semiconducting 2H-TMDs, the critical current in S-TMD-S junctions is expected to be highly tunable. However, it is also this semiconducting nature that leads to a difficulty to make good ohmic contacts on them, as high Schottky barriers are usually present in the metal/TMD interfaces. The resistive Schottky barrier tends to weaken the induced superconductivity, while the transport in 2H-TMDs is usually diffusive owing to the low mobility, both effects result in a very poor weak link for the supercurrent to be observed. On the other hand, the T$_d$-phase TMDs (which is three-dimensional stack of 1T'-phase monolayers), such as T$_d$-WTe$_2$ and T$_d$-MoTe$_2$, are highly conducting semimetal and could be readily used as a weak link to transmit supercurrent. In this section, we will introduce experiments that have been performed to study the Josephson effects in junctions made of 2H-MoS$_2$ \cite{Island2016} and T$_d$-WTe$_2$ \cite{Shvetsov2018}.

\begin{figure}
\includegraphics[scale=0.57]{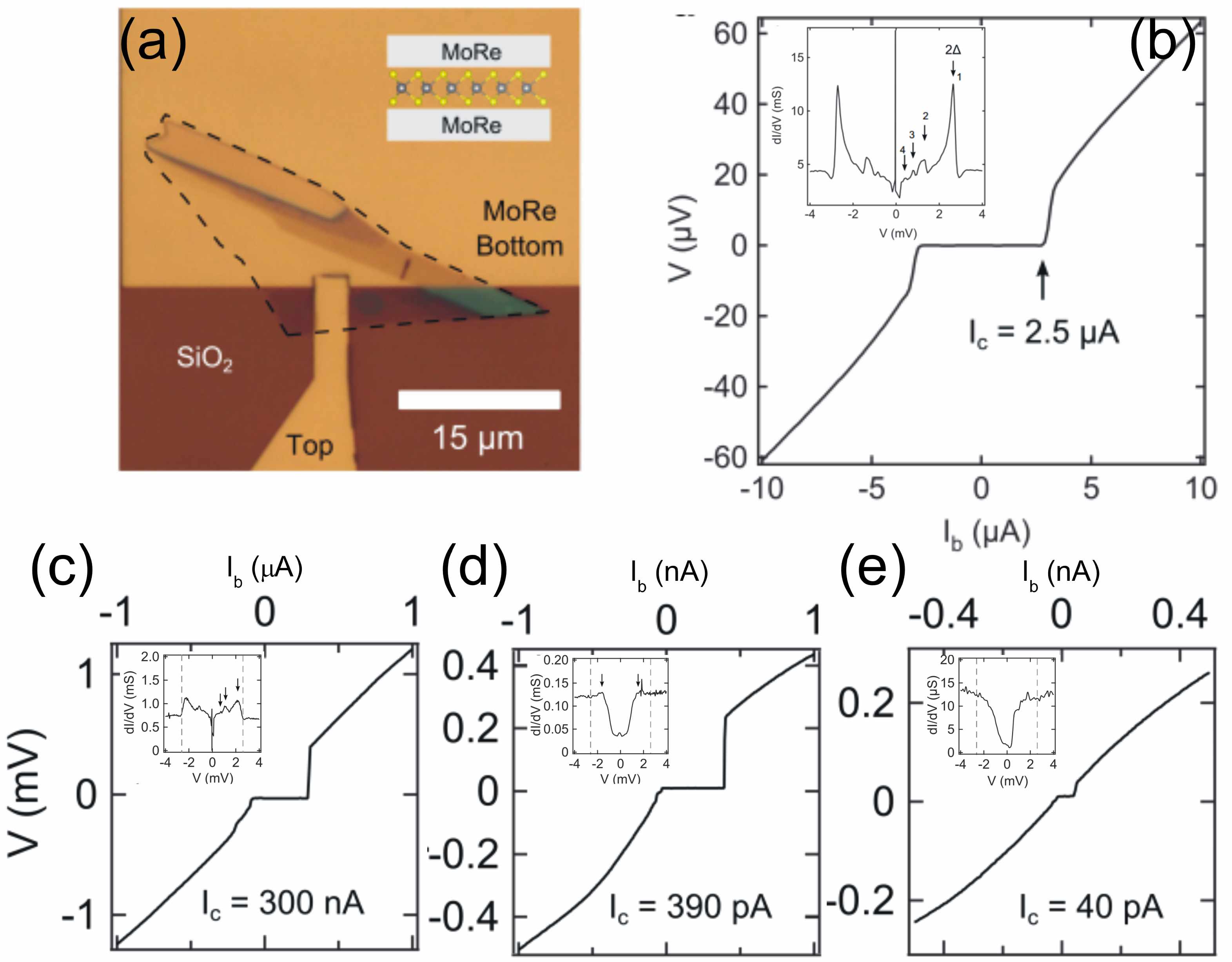}
\caption{Layer dependent Josephson effect in vertical S-MoS$_2$-S junctions. (a) Optical image showing a monolayer 2H-MoS$_2$ is sandwiched between two MoRe electrodes to form a vertical junction. (b) $V$-$I_b$ curve of the monolayer MoS$_2$ junction presenting a critical current of 2.5 $\mu$A. The inset shows the differential conductance ($dI_b$/$dV$) as a function of $V$. The arrows with integer number mark the conductance peaks arising from Andreev reflections. (c, d, e) Current bias sweeps at $T$ = 1.2 K for vertical junctions having 2-4 layers of MoS$_2$, respectively. Insets show the differential conductance ($dI_b$/$dV$) vs. $V$ for each junction, taken at 30 mK. Figures adapted with permission from ref. \cite{Island2016}. Copyright 2016 IOP Publishing Ltd.}  
\label{Fig41}
\end{figure}

Fig. \ref{Fig41} (a) shows the optical image of the vertical S-MoS$_2$-S junction, where a monolayer 2H-MoS$_2$ is sandwiched between two MoRe electrodes. The use of superconducting MoRe ($T_c$ $\approx$ 10 K) as electrodes is to provide Schottky barrier-free contact to MoS$_2$ as a result of their work functions match. The successfully induced superconductivity in the monolayer weak link can be seen in Fig. \ref{Fig41} (b), in which a critical current $I_C$ of 2.5 $\mu$A was observed in the $V$-$I_b$ curve. The differential conductance measured as a function of bias voltage was also performed [inset in Fig. \ref{Fig41} (b)], which shows symmetric peaks at $\pm$2.6 mV marking the quasiparticle gaps. In addition, sub-gap conductance peaks indicating multiple Andreev reflections (MAR) were also observed symmetrically in voltage axis, suggesting the transport is in the diffusive regime despite the junction is short \cite{Octavio1983}. The positions of the MAR peaks in energy are given by $eV_n$ = 2$\Delta$/$n$, where n is a positive integer \cite{Octavio1983}, as indicated by the arrows and number in the inset of Fig. \ref{Fig41} (b). Linear fitting of the MAR peaks gives the superconducting gap of $\Delta$ = 1.3 meV, which is expected from MoRe thin film \cite{Island2016}. The thickness dependence of the vertical junction reveals more information on the Josephson coupling between the top and bottom MoRe electrodes. Fig. \ref{Fig41} (c) - (e) show the measured $V$-$I_b$ (main panels) and $dI_b$/$dV$-$V$ (insets) curves for three devices consisting of 2, 3 and 4 layers of MoS$_2$, respectively. As can be seen in Fig. \ref{Fig41} (b) - (e), the critical current decreases with increasing thickness, with high critical currents for the single and bilayer devices and several orders of magnitude lower values for the three and four layer junctions. The observed MAR in monolayer and bilayer junctions [inset in Fig. \ref{Fig41} (b) and (c), see arrows] indicates a highly transparent and metallic weak link \cite{Blonder1982}, possibly due to the doping of the MoS$_2$ from direct contact with the MoRe electrodes. On the other hand, the trilayer and four layer device shows the tunneling behavior through an undoped semiconductor, which is supported by the formation of a reduced quasiparticle gap as shown in the inset of Fig. \ref{Fig41} (d) and (e). These results indicated that, the uncoupled layers (layers that are not directly in contact with the MoRe electrodes) in the three and four layer devices provide a tunnel barrier, which reduces the critical current densities and results in the more well-defined quasiparticle transport gaps.

\begin{figure}
\includegraphics[scale=0.53]{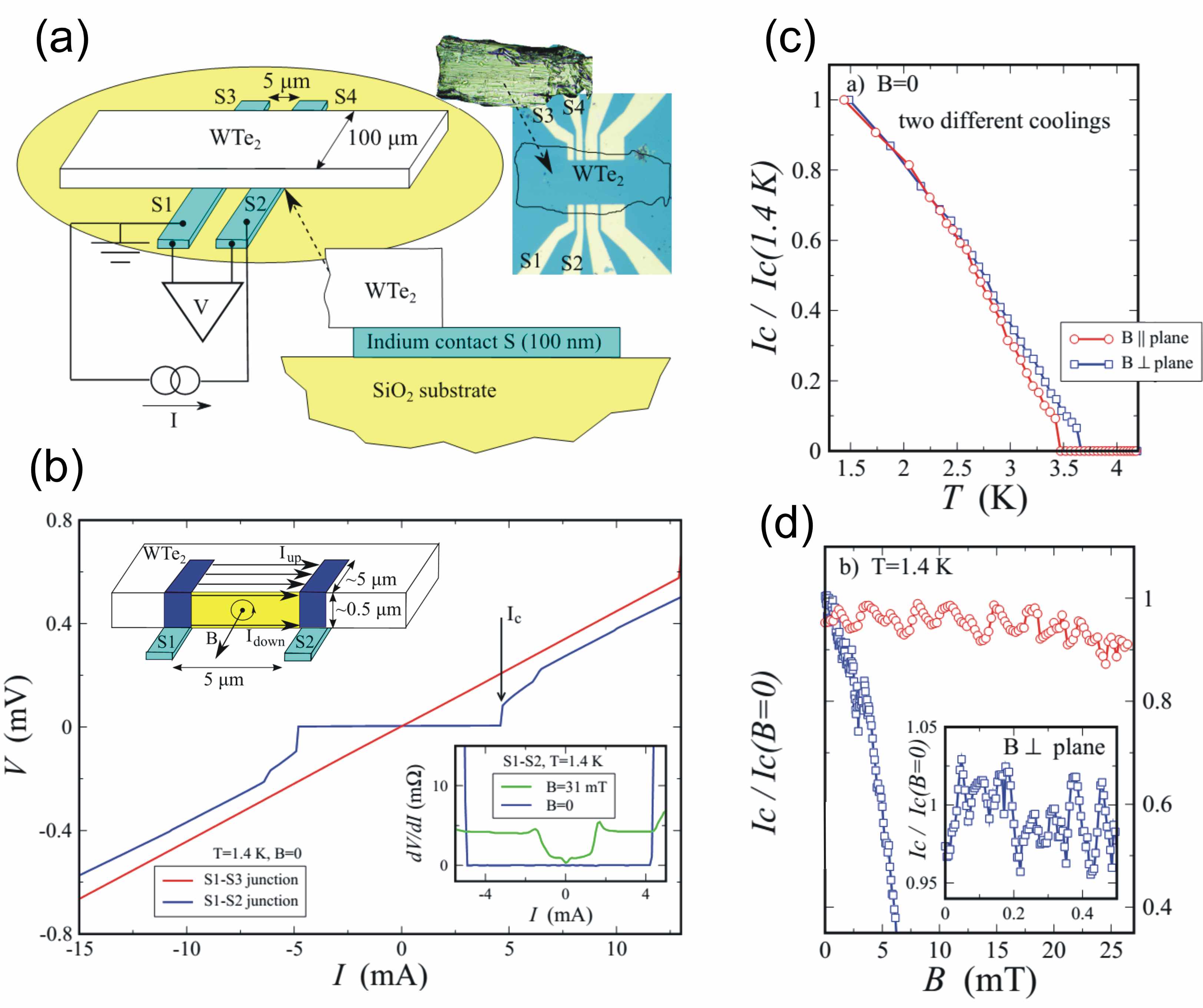}
\caption{Josephson effects in a lateral In-WTe$_2$-In junction. (a) Sketch of the sample with indium contacts to the bottom surface of a T$_d$-WTe$_2$ crystal (not to scale). Right inset shows the top-view image of the indium leads and WTe$_2$ crystal. (b) $V$-$I$ characteristics in two different experimental configurations in zero magnetic field at $T$ = 1.4 K. The blue curve is obtained for 5 $\mu$m long In-WTe$_2$-In junction between the superconducting leads S1 and S2, as depicted in (a). In contrast, the resistance is always finite between 80 $\mu$m separated S1 and S3 indium leads, see the red curve. Right inset: $dV$/$dI$-$I$ characteristics for the S1-WTe$_2$-S2 junction at $T$ = 1.4 K, obtained in zero field (the blue curve) and for the critical field $B$$_c$ = 31 mT (the green one). Left inset: Schematic diagram of a double-slit SQUID geometry, realized by Weyl surface states in WTe$_2$ semimetal. (c, d) Suppression of the critical current $I_C$ by temperature (c) and perpendicular (blue) and parallel (red) magnetic fields (d). Figures adapted with permission from ref. \cite{Shvetsov2018}.}  
\label{Fig42}
\end{figure}

Weyl semimetals are three-dimensional phases of matter with gapless electronic excitations that are protected by topology and symmetry \cite{Armitage2018}. The Weyl fermions, which are used to describe the low-energy excitations in Weyl semimetals, can be viewed as a three-dimensional analogs of graphene, i.e., dispersing linearly along all the three momentum directions across the Weyl points (WPs). The WPs always appear in pairs with opposite chirality, and there exists open curve-like Fermi surfaces with nondegenerate spin texture, as known as surface Fermi arcs, connecting them \cite{Armitage2018}. The Weyl semimetals can be further classified into two types by whether the Lorentz symmetry is respected: type-I hosts point-like Fermi surfaces and its WPs respect the Lorentz symmetry, whereas type-II breaks the Lorentz symmetry and contains electron and hole pockets in its Fermi surfaces \cite{Weng2015,Bulmash2014}. Recently, type-II Weyl semimetals have been proposed to exist in the orthorhombic T$_d$-phase TMDs (such as WTe$_2$ and MoTe$_2$ \cite{Soluyanov2015,Chang2016}), which is later supported by several experimental studies \cite{Jiang2017,Li2017}. As an analogy to the Dirac fermions in graphene, the Weyl fermion carriers in T$_d$-TMDs are expected to transmit electrical currents effectively, and therefore could provide a good weak link in Josephson junctions. Fig. \ref{Fig42} (a) illustrate a lateral S-N-S junction made of superconducting indium leads and T$_d$-WTe$_2$. The device is fabricated by weakly pressing a WTe$_2$ single crystal ($\approx$ 0.5 mm $\times$ 100 $\mu$m $\times$ 0.5 $\mu$m dimensions) onto the indium leads pattern (100 nm thick), so that planar In-WTe$_2$ junctions [5 (W) $\times$ 5 (L) $\mu$m$^2$] are formed at the bottom surface of the WTe$_2$ crystal. The In-WTe$_2$-In junctions were measured by a quasi-four probe technique [as illustrated in Fig. \ref{Fig42} (a)] at a temperature around 1.4 K, which is well below the critical temperature of indium ($T_c$ $\approx$ 3.4 K). The measured $V$-$I$ curve between two 5 $\mu$m spaced contacts S1 and S2 shows a supercurrent with $\pm I_C$ $\approx$ 4 mA [blue curve in Fig. \ref{Fig42} (b)], while that between the 80 $\mu$m separated S1 and S3 leads always presents a finite resistance [red curve in Fig. \ref{Fig42} (b)]. The observed supercurrent between S1 and S2 indicates an unprecedentedly long junction 5 $\mu$m $\gg$ $\xi_{In}$, where $\xi_{In}$ $\approx$ 300 nm is the indium phase coherence length. The temperature and $B$-field dependence of critical current $I_C$ were performed on the junction in order to reveal more information for Josephson effect. To analyze $I_C$ ($B$,$T$) behavior, $dV$/$dI$ vs. $I$ was measured by adding an additional AC modulation (100 nA, 10 kHz) on top of the DC current, followed by an AC part of $V$ ($\approx$ $dV$/$dI$) detected by a lock-in amplifier. The right inset in Fig. \ref{Fig42} (b) shows an example of the measured $dV$/$dI$-$I$ curve, from which $I_C$ can be determined by the edge of the gap. The obtained $I_C$ ($B$,$T$) was summarized in Fig. \ref{Fig42} (c) and (d). The temperature dependence $I_C$ ($T$) in Fig. \ref{Fig42} (c) does not present an exponential decay that is expected for long S-N-S junctions ($L$ $\gg$ $\xi_{In}$). Instead, the $I_C$ ($T$) dependence is even slower than the linear function of $T$, which is usually the case for a short junction limit \cite{Kulik1969}. The slow temperature dependence of $I_C$ ($T$) in this long device may be associated with the topological Fermi arc surface states in Weyl semimetal \cite{Shvetsov2018}. Because of the topological protection, they can efficiently transfer the Josephson current and lead to a slow temperature dependence. The field dependence of $I_C$ ($B$) also reveals a scenario of surface transport, which can respond differently to magnetic field orientation, as shown in Fig. \ref{Fig42} (d). Strong suppression of $I_C$ ($B$) is observed when the applied field is perpendicular to the junction plane (as expected for standard Josephson effect), while $I_C$ ($B$) diminishes very slowly with the parallel magnetic field. For both orientations, $I_C$ ($B$) oscillates with $B$ within 5 percents of $I_C$ magnitude, with a period of $\Delta$$B$ = 2 mT for the parallel field and $\Delta$$B$ = 0.1 mT for the perpendicular one. The observed $I_C$ ($B$) suppression in parallel magnetic fields resembles double-slit SQUID behavior, indicating a possible surface transport as illustrated in the left inset of Fig. \ref{Fig42} (b). Since the thickness of WTe$_2$ (500 nm) is comparable with indium phase coherence length (300 nm), it is likely the regions of proximity-induced superconductivity couples two opposite sample surfaces near the In leads [blue regions in the inset of Fig. \ref{Fig42} (b)] and essentially forms a SQUID configuration. Therefore, parallel magnetic field induces the magnetic flux threading the loop area enclosed by the surfaces (highlighted in yellow in the inset) and results in the oscillation of $I_C$ ($B$$_{||}$). Since $A \cdot \Delta$$B$ $\approx$ $\Phi_0$, $\Delta$$B$ = 2 mT gives $A$ $\approx$ 10$^{-8}$ cm$^2$, which corresponds to 300 nm sample thickness for a 5 $\mu$m long junction and is in good agreement with the device dimensions. The perpendicular field dependence of $I_C$ ($B$$_{\bot}$) reflects homogeneous supercurrent distribution within the surface state in the top and bottom S-N-S junctions \cite{Shvetsov2018}. The observed period of $\Delta$$B$ = 0.1 mT corresponds to an effective junction area of $A$ $\approx$ 2 $\times$ 10$^{-7}$ cm$^2$ ($\approx$ 5 $\mu$m $\times$ 5 $\mu$m), again well corresponds to the sample dimensions.

T$_d$-TMDs have more exotic topological properties in the monolayer form, which will be introduced in the next section.

\maketitle
\section{6. Quantum spin Hall edge states in 1T'-TMDs}

Majorana fermions (MFs) or Majorana bound states (MBS) in condensed matter systems are a special type of excitation that is predicted to exist in 1D or 2D p-wave superconductors \cite{Kitaev2001,Alicea2012,Leijnse2012}. They obey non-Abelian (non-commutative) exchange statistics, which effectively means particle exchange with different routes will lead to different end states. This is the essential property for utilizing them as an element for quantum computing. By braiding a few MFs (computing) and annihilating them (readout), one could encode bits of information \cite{Nayak2008}. What stands out from conventional quantum computing (based on a two-level system) is that, the Majorana-based qubit is expected to have a very long decoherence time. Another way to say this is that Majorana states are topologically protected, i.e. as long as they are well separated from each other, any local perturbation will not be able to affect them simultaneously. This non-local property distinguishes them from qubits made of two-level systems, which often suffer from local couplings to charges, phonons or magnetic moments that induce decoherence. Engineering MFs in condensed matter system requires several important ingredients. They are believed to exist in the vortices of a 2D p-wave superconductor or in an s-wave superconductor proximatized 2D topological insulator (TI) or semiconducting nanowire with a large g-factor and strong spin-orbit coupling \cite{Alicea2012,Leijnse2012}. The last has been realized experimentally in several InSb and InAs nanowires since 2012 \cite{Mourik2012,Albrecht2016,Deng2016}. In the former case, the spin polarized edge states in TIs provide the spin nondegenerate states while the s-wave superconductors provide Cooper pairing strength, the combination of the two essentially forms a "spinless" superconductor that is capable of hosting MFs \cite{Alicea2012}. Experimental search for MBS in 3D TI platforms has been focused on using scanning tunneling microscopy (STM) to probe the center of vortices formed on the surfaces of proximatized 3D TIs \cite{Xu2015,Sun2016}. As for 2D TIs, although possible Majorana-related physics have been revealed in HgTe-based Josephson junctions \cite{Bocquillon2016,Wiedenmann2016}, definite experimental evidence for MBS in such a system remains elusive up to date. 1T'-TMDs as nature 2D TIs with large bulk gap and narrow edge states (compared with HgTe quantum wells, see section 1.3) thus provide another interesting platform to probe MBS. 

\begin{figure}
\includegraphics[scale=0.57]{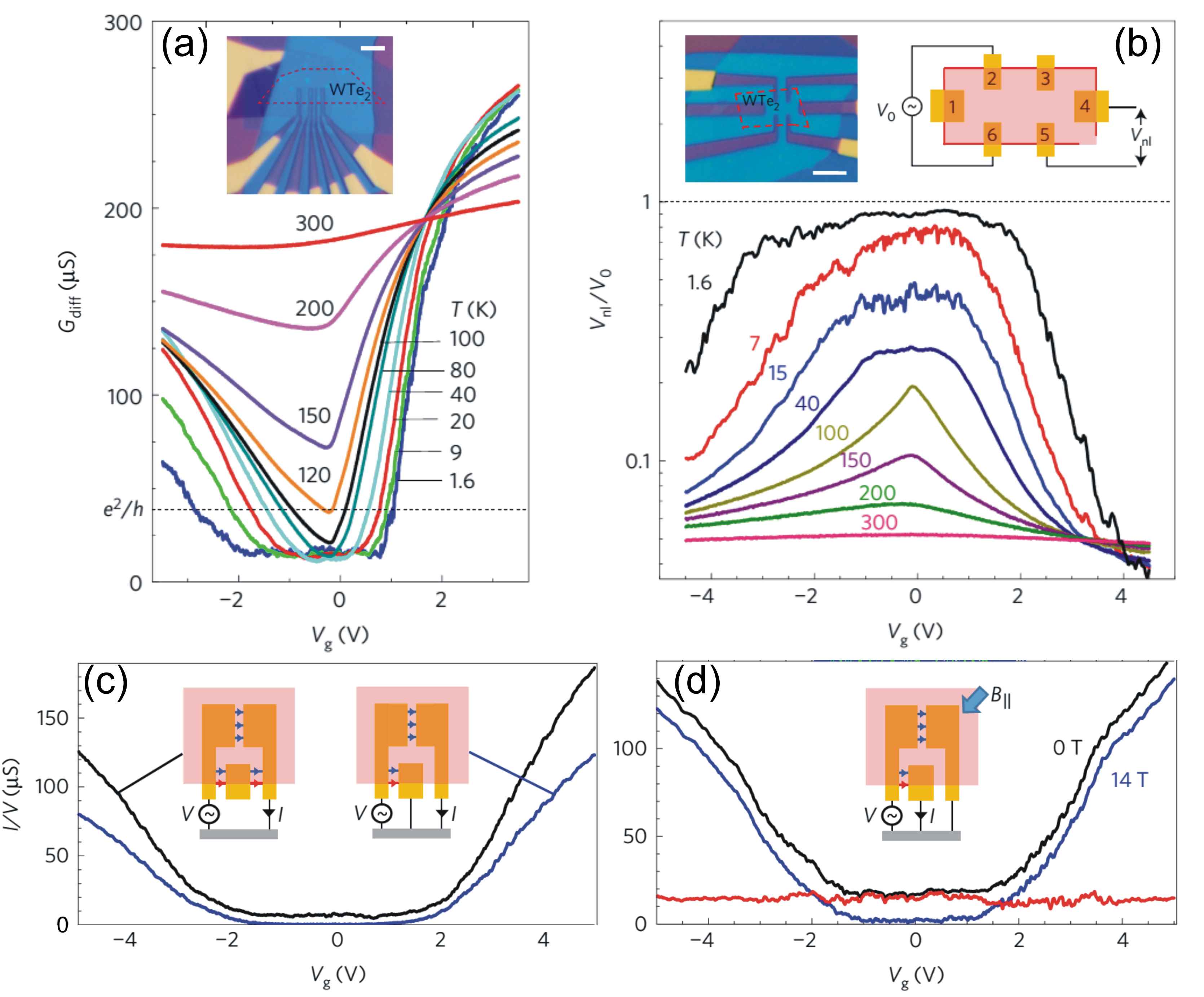}
\caption{(a) Temperature dependence of the characteristics for Pd contacts on an edge of a monolayer 1T'-WTe$_2$ ($L$ = 0.24 $\mu$m, $W$ = 3.3 $\mu$m) device. Inset shows the optical image of monolayer 1T'-WTe$_2$ device. Scale bar, 5 $\mu$m. (b) Nonlocal measurements on a device shown in the left inset. The measurement scheme is illustrated in the right inset, where a voltage $V_0$ ($V_0$ = 100 $\mu$V at 11.3 Hz) is applied across contacts 2 and 6 and the voltage drop $V_{nl}$ between contacts 4 and 5 is detected. (c) Measurements on a device with pincer-shaped contacts overlapping one straight edge (the contact separation along the edge is 0.22 $\mu$m; the pincer spacing is 0.28 $\mu$m). Measurements were performed at $T$ = 10 K. (d) Effect of in-plane magnetic field $B_{||}$ = 14 T on $I$/$V$ between adjacent contacts ($T$ = 10 K). The red trace is the magnitude of the decrease. Figures adapted with permission from ref. \cite{Fei2017}. Copyright 2017 Nature Publishing Group.}  
\label{Fig45}
\end{figure}

In section 1.3, we have briefly introduced the theoretical background for 1T'-TMDs (monolayer form of T$_d$-TMDs), whose band structures indicate that they are 2D TIs with quantum spin Hall edge states. Quantum transport through a 2D TI should exhibit the following characteristics \cite{Wu2018}: (i) helical edge modes protected by time-reversal symmetry (TRS), characterized by an edge conductance that is approximately the quantum value of $e^2/h$ per edge; (ii) saturation to the conductance quantum in the short-edge limit; and (iii) suppression of conductance quantization upon application of a magnetic field, owing to the breakdown of TRS. In the following, we will present experimental evidence that at low temperatures monolayer 1T'-WTe$_2$ exhibits an insulating bulk and conducting edge, and describe the properties of the edge conduction, including its dependence on gate voltage, magnetic field, temperature, and contact separation. All the properties eventually indicate that 1T'-WTe$_2$ meet the above criteria for being a 2D TI. Fig. \ref{Fig45} (a) shows the two-terminal differential conductance measurements of an encapsulated monolayer 1T'-WTe$_2$ device, which has a row of bottom Pd contacts along one edge [see the inset in Fig. \ref{Fig45} (a)] and a graphite top gate $V_g$ to tune the Fermi energy of the whole flake. On cooling from 300 K, the monolayer shows metallic behavior but develops a strong $V_g$ dependence with a wide minimum near $V_g$ = 0. However, below $T$ = 100 K the minimum stops dropping and instead broadens into a plateau of conductance (at $\approx$ 16 $\mu$S) reminiscent of quantized edge state conductance. We will show below that the plateau is due to edge conduction remaining when the bulk becomes insulating below $T$ $\approx$ 100 K. Fig. \ref{Fig45} (b) shows the nonlocal measurements that were used to detect edge conduction in another monolayer 1T'-WTe$_2$ device [left inset of Fig. \ref{Fig45} (b)]. In the measurement setup shown in the right inset of Fig. \ref{Fig45} (b), a small excitation $V_0$ between contacts 2 and 6 on opposite edges was applied, and the nonlocal voltage $V_{nl}$ induced between contacts 4 and 5 (which are far out of the normal current path between contacts 2 and 6) was detected. At low $T$ and small $V_g$, $V_{nl}$/$V_0$ grows large, suggesting that in this regime most of the current propagates along the edge. At higher $T$ or larger $V_g$, $V_{nl}$/$V_0$ falls off as more current takes the direct path through the bulk. Although this measurement indicates that the current follows the edge, in this geometry it is difficult to quantitatively separate edge contributions form bulk conduction. To address this, another geometry which employs a series of pincer-shaped contacts overlapping one straight edge of a monolayer flake, is adopted and shown schematically in the insets of Fig. \ref{Fig45} (c) and (d). The gate dependence of conductance between a pair of pincers [black trace in Fig. \ref{Fig45} (c)] behaves similarly to a pair of adjacent contacts as shown in Fig. \ref{Fig45} (a). However, if the smaller rectangular contact interposed between them is grounded [right inset in Fig. \ref{Fig45} (c)], so that any current flowing near the edge is shorted out, the measured $I/V$ between pincers is suppressed nearly to zero around $V_g$ = 0 [blue trace in Fig. \ref{Fig45} (c)]. This indicates that around $V_g$ = 0 most of the current flows near the edge while bulk is insulating. In contrast, with the middle contact still grounded but $V_g$ larger than about $\pm$2 V, the appearance of non-zero $I/V$ indicates the conduction through the two-dimensional bulk (directly across the gap between the pincers). 

The effect of a in-plane $B$-field on edge conduction is shown in Fig. \ref{Fig45} (d), where the black trace is a measurement at $B_{||}$ = 0 T between two adjacent contacts (see the configuration shown in the inset), and the blue trace is the same measurement done with an in-plane field of $B_{||}$ = 14 T. As can be seen in blue trace, near $V_g$ = 0, where the bulk is insulating, $I/V$ drops nearly to zero, implying that the edge conduction is strongly suppressed by the magnetic field due to the breakdown of TRS. In addition, the magnitude of the drop, plotted in red in Fig. \ref{Fig45} (d), is similar at all $V_g$, indicating that the edge makes a roughly constant contribution to the conductance, independent of gate voltage and bulk conductivity. 

\begin{figure}
\includegraphics[scale=0.57]{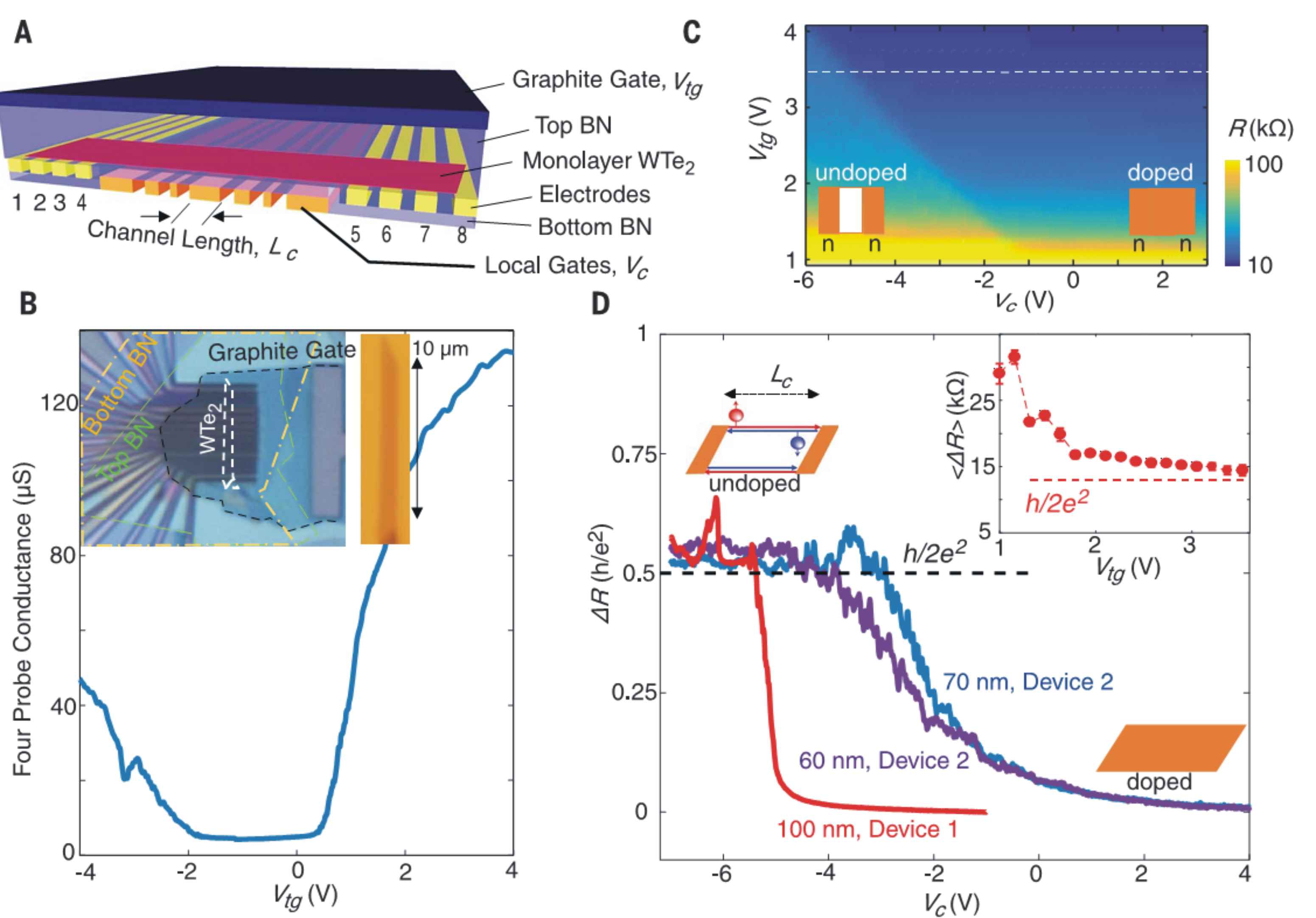}
\caption{Length-dependence study on edge conduction. (a) Schematic of the device structure. (b) Four-terminal conductance
measurement at 4 K of the device (see inset) as a function of $V_{tg}$ across all the local gates, which are floating. Inset shows the optical image of the device and the corresponding monolayer WTe$_2$ flake before fabrication. (c) 2D map of the resistance tuned by $V_{tg}$ and the 100 nm-wide local gate $V_c$ at 4 K. Two regions are separated by a step in the resistance distinguishing the doped and undoped local channel, as depicted by the inset schematics. (d) $\Delta R$ versus $V_c$ for the 100-nm-wide gate on the device shown in (b) at $V_{tg}$ = 3.5 V [white dashed line in (c)] and the 60- and 70-nm wide gates on another device at $V_{tg}$ = 4.1 V (taken at 5 K). Inset shows the average step height $\Delta R$, extracted from (c), as a function of $V_{tg}$, showing a clear saturation toward $h/2e^2$ for large $V_{tg}$. Figures adapted with permission from ref. \cite{Wu2018}. Copyright 2018 American Association for the Advancement of Science.}  
\label{Fig46}
\end{figure}

The temperature, gate, and magnetic-field dependence on the edge conduction have been checked. However, we note that the edge conductance never reaches the expected quantum value of $e^2/h$ = 38.7 $\mu$S, as can be seen in Fig. \ref{Fig45} (a). This could be due to the imperfect transmission between the metal contacts and the edge or the backscattering from multiple magnetic impurities, which all prevent observation of the intrinsic edge conductance \cite{Fei2017}. To address this, a study adopting another device geometry is shown in Fig. \ref{Fig46} (a), in which the device consists of eight contact electrodes (four on each side), a top graphite gate, and a series of in-channel local bottom gates with width $L_c$ varying from 50 nm to 900 nm. The monolayer flake is carefully selected to have a long strip shape (a few $\mu$m wide and about ten $\mu$m long) and is fully encapsulated between two hBN flakes. The goal of the design was to minimize the effect of contact resistance and to enable a length-dependence study on a single device. Fig. \ref{Fig46} (b) shows a typical measurement of the four-terminal conductance across all the local gates ($\approx$ 8 $\mu$m long) as a function of top gate voltage, $V_{tg}$. Like previous study [Fig. \ref{Fig45} (a)], a finite conductance plateau develops around $V_{tg}$ = 0 V, indicating a regime of edge conduction. In order to study the length-dependence of edge conduction, a short transport channel with length $L_c$ was selectively defined by a local gate voltage $V_c$, whereas the rest of the flake is highly doped by $V_{tg}$ to ensure good contact to the electrodes [see the left inset of Fig. \ref{Fig46} (c)]. Fig. \ref{Fig46} (c) shows the 2D resistance map as a function of $V_{tg}$ and $V_c$ for a local gate with $L_c$ = 100 nm. The step in conductance (color contrast) indicates a transition from a bulk-metallic state (doped) to a bulk-insulating state (undoped) within the locally gated region, from which the offset resistance (resistance change from the value in the highly doped limit) can be defined as: $\Delta R$ = $R$($V_c$) $-$ $R$($V_c$ = -1V). The saturated value of $\Delta R$ at large negative $V_c$, as shown in Fig. \ref{Fig46} (d), thus measures the resistance of the undoped channel, which can only originate from the edges because the monolayer interior is insulating. Notably, $\Delta R$ for different channel length (100-nm channel and the 60- and 70-nm channels) all saturate at $h/2e^2$. Given that the sample has two edges, the observed conductance per edge is therefore $e^2/h$, pointing to helical edge modes as the source of the conductance and confirming the abovementioned criteria (i) and (ii). The observed quantum spin Hall edge states in 1T'-TMDs can be readily integrated with s-wave superconductors for testing Majorana-related physics, such as the 4$\pi$-periodicity current phase relation \cite{Fu2009} and doubling of the voltage spacing for Shapiro steps \cite{Bocquillon2016}.

\maketitle
\section{7. Summary}
In summary, we have provided the theoretical background and a number of experimental studies relevant to developing various qubits in 2D materials. The single-electron transport properties of graphene nanodevices (GNRs, GSQDs and GDQDs) fabricated on SiO$_{2}$ and hBN substrates were both reviewed. GSQDs fabricated on SiO$_{2}$ and hBN show a distinct difference in their Coulomb blockade peak-spacing fluctuations, indicating that edge roughness is the dominant source of disorder for QDs with diameters of less than 100 nm. For 2H-TMD nanostructures, despite Coulomb blockade has been demonstrated in both SQDs and DQDs, $B$-field dependence of the dot levels is still lacking, presumably due to the intrinsic high disorder in the materials. Nevertheless, we discussed their Fock-Darwin spectra for the potential use in spin and valley qubits. We also reviewed the transport properties of Josephson junctions made of various 2D materials. The $I_C$$R_N$ correlation and current-phase relation in graphene-based lateral junctions have been studied, while the layer dependence of critical current was demonstrated in vertical junctions made of 2H-MoS$_2$. In addition, Weyl semimetal T$_d$-WTe$_2$-based junctions have shown an unusual temperature dependence of critical current, suggesting the existence of topologically protected surface states. Finally, we reviewed recent experimental studies showing that monolayer 1T'-WTe$_2$ is a natural 2D TI with quantum spin Hall edge states which are protected by TRS.   

The absence of spin blockade in GDQDs \cite{Moriyama2009,Molitor2010,Liu2010,Volk2011,Wang2012,Wei2013,Chiu2015} and the fact that the measured spin relaxation times in 2D graphene flakes are shorter than expected \cite{Tombros2007,Han2010a,Han2011,Maassen2012,Drogeler2014} suggest that there are extrinsic effects that govern the spin relaxation dynamics in graphene. These effects may be related to the scattering of electrons off magnetic impurities originating from carbon atom vacancies or graphene edge roughness \cite{Chiu2017}. Since graphene constrictions have been used as tunnel barriers in most GQDs reported thus far, alteration of the electron spin by the constriction edge during transport could be inevitable. Possible solutions to circumvent this issue could be the followings: (i) using an electrical-field-induced bandgap in bilayer graphene to define GQDs \cite{Goossens2012,Allen2012,Overweg2018,Overweg2018a}; (ii) functionalization of graphene, such as fluorinated graphene (FG), to simultaneously define the quantum confinement and passivate the graphene edges \cite{Withers2011,Lee2011}; and (iii) choosing magnetic inert materials, such as hBN, as a tunnel barrier to build vertical graphene nanodevices \cite{Bischoff2015,Bischoff2015a}. In the first two cases, nanodevices defined in gated bilayer and functionalized graphene have been achieved, but no DQD structures (useful for $T_1$ and $T_{2}^{*}$ measurements) have been reported. Vertical tunneling \cite{Ashoori1993} to GQDs using hBN as a tunnel barriers may also serve as a promising solution for minimizing the edge effect from graphene constrictions, despite the edge of GQDs still exists. It is also possible to probe spin relaxation time in vertical QDs using pulse gating techniques \cite{Hanson2007}. However, spin qubits made in vertical geometry are difficult for upscaling, as currently they are mostly made in lateral configurations in GaAs and Si systems \cite{Hanson2007,Zwanenburg2013}. The large bandgaps of few-layer 2H-TMDs can be advantageous for defining QDs $via$ electrical gating. But the disorder nature of the material might lead to a strong decoherence that shorten $T_1$ and $T_{2}^{*}$. Nevertheless, spin blockade or spin relaxation time measurements have not been reported from this material either. 

On the other hand, the graphene-based Josephson junctions have shown promising prospects for use in superconducting qubits owing to its ballistic nature and gate-tunable critical current \cite{Kroll2017,Wang2018}. The lateral 2H-TMDs Josephson junctions have not been reported so far, perhaps due to their diffusive nature that is difficult to preserve the induced superconductivity. On the contrary, T$_d$ phase TMDs-based junctions have shown great conductivity for sustaining large Josephson current, but they are not gateable except for the few-layer form ($\leq$ 3 layers). The few-layer T$_d$-TMDs are reported to oxidize very quickly in air \cite{Fei2017,Wu2018}, which may add difficulty for making T$_d$-TMDs-based gatemons. Apart from graphene, experimental reports on other 2D material-based gatemon qubits are still lacking up to date.

The pursuit of MFs in various solid state systems has been a hot topic over the past decade, from the early tunneling experiments on semiconducting nanowires \cite{Mourik2012,Albrecht2016,Deng2016} to the more recent studies on proximatized TIs \cite{Xu2015,Sun2016,Bocquillon2016,Wiedenmann2016} and quantum anomalous Hall insulator \cite{He2017}. Later studies of STM experiments further confirm their location at the end of a chain of magnetic atoms \cite{Nadj-Perge2014}. However, whether these unconventional quaisparticles obey non-Abelian exchange statistics will not be known until braiding and readout are performed. More sophisticated devices, such as QD chains and T-junctions \cite{Heck2012,Fulga2013,Alicea2011,Sau2011,Hyart2013,Flensberg2011,Schrade2015}, that allow exchange of particles are needed for further investigations. 1T'-TMDs with the proposed gate-tunable topological phases are particular interesting to test these theoretical proposals and other potential computing schemes \cite{Hoffman2016}.

\end{document}